\documentclass{JHEP3}
\usepackage[english]{babel}
\usepackage{cite}

\usepackage{epsf}
\usepackage{amssymb}
\usepackage{amsmath}
\usepackage{amsfonts}
\usepackage{psfrag,epsfig,graphicx,graphics}

\usepackage{float}

\def\slashchar#1{\setbox0=\hbox{$#1$}
   \dimen0=\wd0
   \setbox1=\hbox{/} \dimen1=\wd1
   \ifdim\dimen0>\dimen1
      \rlap{\hbox to \dimen0{\hfil/\hfil}}
      #1
   \else
      \rlap{\hbox to \dimen1{\hfil$#1$\hfil}}
      /
   \fi}

\def\bei{\begin{itemize}}
\def\ei{\end{itemize}}

\def\beeq{\begin{eqnarray}} 
\def\beqa{\begin{eqnarray}}
\def\bea{\begin{eqnarray}}

\def\eea{\end{eqnarray}}
\def\eqa{\end{eqnarray}}
\def\eeeq{\end{eqnarray}}

\def\eqar{\end{array}}
\def\beqar{\begin{array}}

\def\beas{\begin{eqnarray*}}
\def\beqas{\begin{eqnarray*}}

\def\eqas{\end{eqnarray*}}
\def\eeas{\end{eqnarray*}}

\def\beq{\begin{equation}} 
\def\be{\begin{equation}}

\def\ee{\end{equation}}
\def\eq{\end{equation}}
\def\eeq{\end{equation}}

\def\beqd{\begin{displaymath}}
\def\eeqd{\end{displaymath}}
\def\eqd{\end{displaymath}}

\def\beeq{\begin{eqnarray}} \def\eeeq{\end{eqnarray}}

\newcommand{\fin}{\end{document}}

\newcommand{\pom}{\mathbb{P}}

\newcommand{\veck}{{\bf k}}
\newcommand{\vecki}{{\bf k}_i}
\newcommand{\veckone}{{\bf k}_1}
\newcommand{\vecktwo}{{\bf k}_2}
\newcommand{\veckj}{{\bf k}_{J}}
\newcommand{\veckji}{{\bf k}_{J,i}}
\newcommand{\veckjone}{{\bf k}_{J,1}}
\newcommand{\veckjtwo}{{\bf k}_{J,2}}

\newcommand{\deins}[1]{{\rm d}#1\,}
\newcommand{\dzwei}[1]{{\rm d}^2#1\,}
\newcommand{\dk}{\dzwei{\veck}}

\newcommand{\dkone}{\dzwei{\veckone}}
\newcommand{\dktwo}{\dzwei{\vecktwo}}

\newcommand{\dsigma}{\deins{\sigma}}
\newcommand{\dsigmahat}{\deins{{\hat\sigma}_{\rm{ab}}}}
\newcommand{\dnu}{\deins{\nu}}

\newcommand{\dx}{\deins{x}}
\newcommand{\dxone}{\deins{x_1}}
\newcommand{\dxtwo}{\deins{x_2}}
\newcommand{\dyjetone}{\deins{y_{J,1}}}
\newcommand{\dyjettwo}{\deins{y_{J,2}}}
\newcommand{\dphij}{\deins{\phi_{J}}}
\newcommand{\dphijone}{\deins{\phi_{J,1}}}
\newcommand{\dphijtwo}{\deins{\phi_{J,2}}}
\newcommand{\dtwojets}{{\rm d}|\veckjone|\,{\rm d}|\veckjtwo|\,\dyjetone \dyjettwo}

\newcommand{\dtwojetsK}{{\rm d}|\veckjone|\,{\rm d}|\veckjtwo|}

\newcommand{\dtwojetsY}{{\rm d}|\veckjone|\,{\rm d}|\veckjtwo|\,dY}

\newcommand{\shat}{{\hat s}}
\newcommand{\non}{\nonumber\\}
\newcommand{\asbar}{{\bar{\alpha}}_s}

\newcommand{\chihat}{{\omega}}

\def\PS{d({\rm P.\,S})}

\def\intbin{\int_{\rm bin} \PS}

\title{Confronting Mueller-Navelet jets in NLL BFKL with
LHC experiments 
at 7 TeV}
\author{B. Duclou\'e\\
LPT, Universit{\'e} Paris-Sud, CNRS, 91405, Orsay, France\\
Email: \email{bertrand.ducloue@th.u-psud.fr}}
\author{ L. Szymanowski\\
National Center for Nuclear Research (NCBJ), Warsaw, Poland\\
Email: \email{Lech.Szymanowski@fuw.edu.pl}}
\author{S. Wallon\\
LPT, Universit{\'e} Paris-Sud, CNRS, 91405, Orsay, France {\em \&} \\
UPMC Univ. Paris 06, facult\'e de physique, 4 place Jussieu, 75252 Paris Cedex 05, France\\
Email: \email{wallon@th.u-psud.fr}}

\abstract{More than 25 years ago, Mueller Navelet jets were proposed as a decisive test of BFKL dynamics at hadron colliders. We here study this process at NLL BFKL accuracy, taking into account NLL corrections to the Green's function and to the jet vertices. We present detailed predictions for various observables that can be measured at LHC in ongoing experiments like ATLAS or CMS at $\sqrt{s}=7$ TeV: the cross-section, the azimuthal correlations and the angular distribution of these jets. For this purpose, we apply realistic kinematical cuts and binning, and study the dependence of our results with respect to several parameters. We then compare our results with those that can be obtained in a fixed order NLO treatment, and propose specific observables which could actually be used as a probe of BFKL dynamics.
}

\date{\today}
\begin{document}

\pagestyle{empty}
\newpage

\mbox{}

\pagestyle{plain}

\setcounter{page}{1}
\section{Introduction}
\label{Sec:Intro}

One of the important longstanding theoretical questions raised by QCD is its behaviour in  the
perturbative Regge limit $s \gg -t$.
Based on theoretical grounds,   
one should identify and test suitable observables in order to test these peculiar dynamics.

First, one should select processes in which the presence of a hard scale justifies the use of perturbative QCD.
At high energy, QCD is a massless theory with vector bosons, and it has two kinds of
 infrared (IR) divergences, namely the soft and the collinear divergences.
For sufficiently inclusive quantities, both kinds of divergences cancel. Still, they are responsible for large logarithms, which may compensate the smallness of the strong coupling. At leading order, the soft singularities manifest themselves as powers of $\alpha_S \, \ln s/|t|$, resummed by the leading logarithmic (LL) BFKL $\pom$omeron~\cite{Fadin:1975cb, Kuraev:1976ge, Kuraev:1977fs, Balitsky:1978ic}. The collinear singularities are responsible for large logarithms of ratios of the transverse scales, which are resummed 
at leading logarithmic order (LLQ) 
by the DGLAP equation~\cite{Gribov:1972ri, Lipatov:1974qm, Altarelli:1977zs, Dokshitzer:1977sg}.
 
The Regge limit is expected to be governed by the soft perturbative dynamics of QCD, which we want to reveal, and not by its collinear dynamics. 
The key point is thus to select processes in which the hard collinear scales are of similar magnitude, in such a way that the difference between a fixed 
order calculation and a collinear resummed result should be tiny, while 
a BFKL type of resummation should modify the predictions dramatically.

During the last 25 years, there have been many attempts to see manifestations 
of BFKL resummation effects. In inclusive DIS at HERA~\cite{Askew:1993rn,Navelet:1996jx,Munier:1998vk} or in total $\gamma^* \gamma^*$ cross-section at $e^+ e^-$ colliders~\cite{Bartels:1996ke,Brodsky:1996sg,Bialas:1997eq,Boonekamp:1998ve,Kwiecinski:1999yx,Brodsky:1998kn,Brodsky:2002ka}, the hard scale is the 
$\gamma^{*}$ virtuality. Exclusive processes have also been proposed and studied, 
either for heavy meson production ($J/\Psi$,  $\Upsilon$), the hard scale being provided by the meson mass~\cite{Ryskin:1992ui,Frankfurt:1997fj,Enberg:2002zy,Ivanov:2004vd},
or meson electroproduction at large $t$~\cite{Enberg:2002zy,Ivanov:2000uq}, for which HERA data seems to favour a BFKL picture~\cite{Enberg:2003jw,Poludniowski:2003yk}. At future high energy and high luminosity colliders like ILC, processes like $\gamma^{(*)} \gamma^{(*)} \to \rho \, \rho$ could be a realistic 
exclusive test of the hard $\pom$omeron~\cite{Pire:2005ic,Enberg:2005eq,Segond:2007fj,Ivanov:2005gn,Ivanov:2006gt,Caporale:2007vs}, with the planned 
detectors designed to cover the very forward region.

\begin{figure}[htbp]
\centering
\includegraphics[height=8cm]{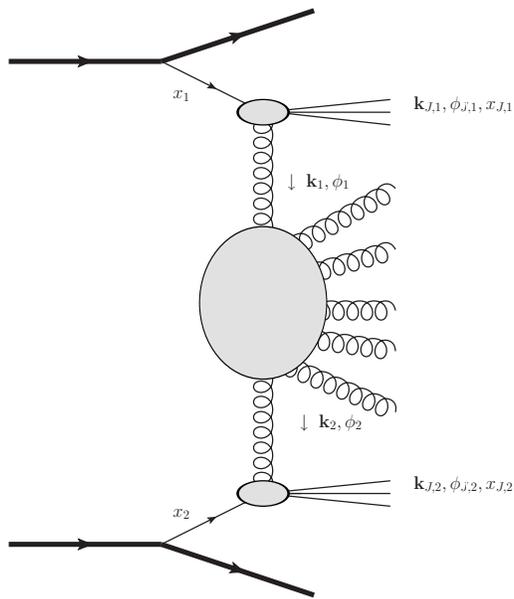}
\caption{kinematics}
\label{Fig:kinematics}
\end{figure}

Jets have been proposed as a powerful tool in order to study BFKL dynamics, like
diffractive high energy double jet production~\cite{Cox:1999dw, Enberg:2001ev, Chevallier:2009cu}  as well as central jets \cite{Ostrovsky:1999kj,Bartels:2006hg} in hadron-hadron collisions. In this paper, we focus on Mueller-Navelet jets~\cite{Mueller:1986ey}. 
This test of BFKL  is based on 
the measure of two jets at large $p_T$ (hard scale), such that $s \gg  
p_T^2 \gg \Lambda_{\rm QCD}^2$, separated by a large rapidity $Y$, including possible activity between the two  observed jets, as illustrated in figure~\ref{Fig:kinematics}. The idea is to consider two jets of similar $p_T$ in order to minimize the effect of collinear resummation. 
From a lowest order treatment it is clear that these two jets should be back-to-back, in the very forward and very backward 
regions.

On the other hand, the expectation is that the large value of $Y=\ln (s/p_T^2)$ should examplify the effect of BFKL dynamics, due to possible emission of gluons between them (thus the Pomeron contributes there at $t=0$ at the level of the cross-section), leading to enhanced  terms  which sum up as
$\sum (\alpha_s  Y)^n$ (LL),  $\alpha_s\sum (\alpha_s  Y)^n$
(NLL)~\cite{Ciafaloni:1998gs, Fadin:1998py}, etc..., leading to a power--like rise for the cross-section.

Besides the cross-section  a more exclusive observable within this process drew the attention, namely the azimuthal correlation between these jets \cite{DelDuca:1993mn, Stirling:1994zs}. 
 The signal of a BFKL dynamics is a decorrelation of relative azimuthal angle between emitted jets when increasing  $Y\,.$ Indeed, while a fixed order calculation implies that the two jets would be emitted back-to-back, the fact that more and more (untagged) gluons can be emitted between them when increasing their relative rapidity should lead to a decorrelation of this relative azimuthal angle.
Studies were made at LL  \cite{DelDuca:1993mn,Stirling:1994zs, DelDuca:1994ng}, which overestimates this decorrelation by far. A better agreement with the data \cite{Abbott:1999ai} could be obtained in the LL scenario using an event generator which takes into account in an exact way the energy-momentum conservation, which is a subleading effect in a pure BFKL approach \cite{Orr:1997im}. On the other hand, the (kinematically) modified LL BFKL approach \cite{Kwiecinski:2001nh},   again  based on LL jet vertices, could also provide some better agreement with the data.

At the same time, an exact fixed NLO $(\alpha_s^3)$ Monte Carlo calculation using 
the program JETRAD~\cite{Giele:1993dj} lead to a too low estimate of the decorrelation, while the Monte Carlo NLO program HERWIG~\cite{Marchesini:1991ch} was in 
perfect agreement with the data. It should be noted that this last treatment includes some Sudakov resummation effects, which might be important.
The inclusion of such effects within a BFKL approach in an open problem 
which might be of interest for phenomenology. We leave this issue for further studies.

Starting from first principles from the point of view of Regge and Quasi-multi-Regge kinematics,
NLL  \cite{Vera:2006un, Vera:2007kn} 
and collinear resummed NLL \cite{Marquet:2007xx} studies (with LL jet vertices) have been performed, improving the situation
with respect to pure LL BFKL, but still leading to a much stronger decorrelation than the one seen by the data.  

In a previous work, we showed, based on a full NLL analysis \cite{Colferai:2010wu}, that contrarily to the expectation, the NLL corrections to the Green's function
and to the jet vertices~\cite{Bartels:2001ge,Bartels:2002yj} are of similar magnitude, based on a \textsc{Mathematica} code. We focused there on a center-of-mass energy $\sqrt{s}=$14 TeV, and considered jets with fixed values of transverse momenta $\veckji$. 

In the present paper, we pursue this study and make detailed predictions for observables to be extracted  in the ongoing experiments ATLAS and CMS. Since experimental data are given in bins, it thus requires that we integrate
$\veckji$ over a finite range. We also fix the 
center-of-mass energy to be
$\sqrt{s}=$7 TeV.  In the study of the azimuthal correlations, 
we extend the use of the collinear resummation method to non-zero conformal spins. 
We then
study in detail the azimuthal distribution of jets, which is directly experimentally accessible.
Finally, we make a detailed comparison (with the same set of parameters) of our predictions with the NLO fixed order results based on the code used in ref.~\cite{Aurenche:2008dn}.

Our numerical predictions are based on a new \textsc{Fortran} code which allows us to perform more detailed studies of the dependency on various parameters (PDFs, renormalization/factorization scale, choice of $s_0$ scale). 
To check the consistency of our results, a
detailed comparison has been made, in the mixed case of NLL BFKL Green's function and LL jet vertices $V^{(0)}$, with the previous 
studies of ref.~\cite{Vera:2007kn,Marquet:2007xx}.

\section{Basic formulas for LL and NLL calculation}
\label{sec:NLLcalculation}

\subsection{Kinematics and general framework}
\label{sec:kinematics}

We consider two hadrons (in practice protons) which collide at a center-of-mass energy $\sqrt{s}$ producing two very forward jets, whose transverse momenta  are labeled by Euclidean two dimensional vectors $\veckjone$ and $\veckjtwo$, and by their azimuthal angles  $\phi_{J,1}$ and $\phi_{J,2}$. The jet rapidities  $y_{J,1}$ and $y_{J,2}$  are related to the longitudinal momentum fractions of the jets via $x_J = \frac{|\veckj|}{\sqrt{s}}e^{y_J}$. The two partons produced by each of these two hadrons, which initiate the hard process, are treated in a collinear way.
For large $x_{J,1}$ and $x_{J,2}$, collinear factorization leads to
a differential cross-section which reads
\begin{equation}
  \frac{\dsigma}{\dtwojets} = \sum_{{\rm a},{\rm b}} \int_0^1 \dxone \int_0^1 \dxtwo f_{\rm a}(x_1) f_{\rm b}(x_2) \frac{\dsigmahat}{\dtwojets},
\label{diff-cross-section}
\end{equation}
where $f_{\rm a,b}$ are the parton distribution functions~(PDFs) of a parton a (b) in the according proton, characterized by their longitudinal momentum fraction $x_i$. The hard process is then described using $k_T$-factorization. The  logarithmically enhanced contributions are taken care of by convoluting, in transverse momentum space, the BFKL Green's function $G$
with the two jet vertices, according to
\begin{equation}
  \frac{\dsigmahat}{\dtwojets} = \int \dphijone\dphijtwo\int\dkone\dktwo V_{\rm a}(-\veckone,x_1)\,G(\veckone,\vecktwo,\shat)\,V_{\rm b}(\vecktwo,x_2)\,,
\label{eq:bfklpartonic}
\end{equation}
where the Mandelstam variable  $\shat=x_1 x_2 s$ refers to the hard subprocess. The jet vertices
 $V_{a,b}$ were  calculated at
 NLL order in ref.~\cite{Bartels:2001ge,Bartels:2002yj}.
Combining the PDFs with the jet vertices, we can thus write
\beqa
  &&\frac{\dsigma}{\dtwojets} \nonumber\\
&=& \int \dphijone\dphijtwo \! \!\int \! \dkone\dktwo \Phi(\veckjone,x_{J,1},-\veckone)\,G(\veckone,\vecktwo,\shat)\,\Phi(\veckjtwo,x_{J,2},\vecktwo) \,,
\label{eq:sigma-2-jets}
\eqa
where
\beqa
\Phi(\veckji,x_{J,i},\vecki) = \int dx_i\, f(x_i)\, V(\vecki,x_i).
\label{eq:def-phi}
\eqa
In order to deal both with the cross-section and with the 
azimuthal decorrelation, it is convenient to define the  coefficients
\beqa
  &&\mathcal{C}_m(|\veckjone|, |\veckjtwo|, Y) \equiv \int dy_1 \, dy_2 \, \delta(y_1+y_2-Y) \int  \dphijone\dphijtwo\cos\big(m(\phi_{J,1}-\phi_{J,2}-\pi)\big) \nonumber \\
&&\times \int \dkone\dktwo \Phi(\veckjone,x_{J,1},-\veckone)G(\veckone,\vecktwo,\shat)\Phi(\veckjtwo,x_{J,2},\vecktwo) \,.
\label{eq:mathcalc}
\eqa
The differential cross-section then corresponds to $\mathcal{C}_0$ which reads
\begin{equation}
 \int dy_1 \, dy_2 \, \delta(y_1+y_2-Y) \frac{\dsigma}{\dtwojets}  = \frac{\dsigma}{\dtwojetsY}= \mathcal{C}_0\,,
\label{eq:dsigma}
\end{equation}
while the azimuthal decorrelation for fixed $(|\veckjone|, |\veckjtwo|, Y)$ is given by
\beq
\label{eq:def_cor}
  \langle\cos(m\varphi)\rangle \equiv \langle\cos\big(m(\phi_{J,1}-\phi_{J,2}-\pi)\big)\rangle = \frac{\mathcal{C}_m}{\mathcal{C}_0} \,.
\eq

\subsection{LL order}
\label{sec:LL}

The jet vertex $V$ at lowest order just 
implements the fact that the jet is made of a single parton, of the same nature as the collinear parton initiating the hard process. It
reads~\cite{Bartels:2001ge,Bartels:2002yj}:
\begin{align}
  V_{\rm a}^{(0)}(\veck,x) =& \, h_{\rm a}^{(0)}(\veck)\mathcal{S}_J^{(2)}(\veck;x) , & h_{\rm a}^{(0)}(\veck) =& \, \frac{\alpha_s}{\sqrt{2}}\frac{C_{A/F}}{\veck^2} , \label{def:V0}\\
 & & \mathcal{S}_J^{(2)}(\veck;x) =& \, \delta\left(1-\frac{x_J}{x}\right)|\veckj|\delta^{(2)}(\veck-\veckj).
\end{align}
In the definition of $h_{\rm a}^{(0)}$, $C_A=N_c=3$ is to be used for initial gluon and $C_F=(N_c^2-1)/(2N_c)=4/3$ for initial quark.

In the LL approximation, 
 the BFKL kernel, because of its conformal invariance, is diagonalized by the eigenfunctions
\begin{equation}
  E_{n,\nu}(\vecki) = \frac{1}{\pi\sqrt{2}}\left(\vecki^2\right)^{i\nu-\frac{1}{2}}e^{in\phi_i}\,,
\label{def:eigenfunction}
\end{equation}
with an eigenvalue given by
\beq
  \chihat_{LL}(n,\nu) = \asbar\chi_0\left(|n|,\frac{1}{2}+i\nu\right)\,,
\label{def:omega_0}
\eq
with $\asbar = N_c\alpha_s/\pi$ and
\beq
\chi_0(n,\gamma) = 2\Psi(1)-\Psi\left(\gamma+\frac{n}{2}\right)-\Psi\left(1-\gamma+\frac{n}{2}\right)\,,
\label{def:chi_0}
\eq
 where $\Psi(x) = \Gamma'(x) /\Gamma(x)$.
Using this basis for both the Green's function and the jet vertices, one thus obtains, introducing the arbitrary (at LL) scale $s_0$,
\begin{equation}
  \mathcal{C}_m = (4-3\delta_{m,0})\!\int dy_1 \, dy_2 \, \delta(y_1+y_2-Y)\!\!\!\int \dnu C_{m,\nu}(|\veckjone|,x_{J,1})C^*_{m,\nu}(|\veckjtwo|,x_{J,2})\left(\frac{\shat}{s_0}\right)^{\chihat(m,\nu)},
\label{eq:cm2}
\end{equation}
where
\beq
C_{m,\nu}(|\veckj|,x_{J})
= \int\dphij\dk \dx f(x) V(\veck,x)E_{m,\nu}(\veck)\cos(m\phi_J) \,.
\label{eq:mastercnnu}
\eq

\subsection{NLL order}
\label{sec:NLL}

At NLL, the jet can be made of either a single or two partons. The collinear singularities can be absorbed consistently in the renormalized PDFs, as was shown in refs.~\cite{Bartels:2001ge,Bartels:2002yj}, for a given infrared-safe
jet algorithm. These jet vertices read symbolically
\begin{equation}
 \label{eq:VLO+NLO}
  V_{\rm a}(\veck,x) = V^{(0)}_{\rm a}(\veck,x) + \alpha_s V^{(1)}_{\rm a}(\veck,x).
\end{equation}
The explicit form for the NLL $V^{(1)}_{\rm a}$ are rather lengthy and are intimately dependent on the jet algorithm.
They will not be reproduced here\footnote{They can be found in ref.~\cite{Colferai:2010wu}, as extracted from refs.~\cite{Bartels:2001ge,Bartels:2002yj} after correcting a few misprints of ref.~\cite{Bartels:2001ge}. They have been recently reobtained in ref.~\cite{Caporale:2011cc}.}\!.
In our study we will use the cone algorithm with a size of $R_{\rm cone}=0.5$.  Note that other jet algorithms can be used, which do not affect significantly our obtained results.

The main issue when dealing with NLL corrections is to treat the NLL BFKL kernel. The point here is to avoid dealing explicitly with two convolutions in transverse momentum space, between the jet vertices and the Green's function. In general this kind of convolution is very difficult to handle with for numerical evaluations. Instead we prefer  to mimic the treatment used for LL studies and work in the ($n$, $\nu$) space.
One is thus looking for a convenient basis in order to deal
with the NLL BFKL kernel. 
The functions (\ref{def:eigenfunction}) cannot be used in principle, since conformal invariance is now broken.
Anyway, the action of the NLL BFKL kernel on these LL eigenfunctions has been calculated in ref.~\cite{Kotikov:2000pm}, and it turns out that $E_{n,\nu}$ are still eigenfunctions in an extended sense, if one now promotes the eigenvalue to become an operator containing a derivative with respect to $\nu$ \cite{Ivanov:2005gn,Vera:2006un,Vera:2007kn}. When convoluting  with jet vertices, this derivative acts on them, thus leading to a contribution to the eigenvalue which now depends on the jet vertices \cite{Ivanov:2005gn,Vera:2006un,Vera:2007kn,Schwennsen:2007hs}
\begin{multline}
   \label{eq:omegaNLO}
   \chihat_{NLL}(n,\nu) = \asbar \chi_0\left(|n|,\frac{1}{2}+i\nu\right)   + \asbar^2 \Bigg[\chi_1\left(|n|,\frac{1}{2}+i\nu\right)\\
-\frac{\pi b_0}{2N_c}\chi_0\left(|n|,\frac{1}{2}+i\nu\right) \left\{-2\ln\mu_R^2-i\frac{\partial}{\partial\nu}\ln\frac{C_{n,\nu}(|\veckjone|,x_{J,1})}{C_{n,\nu}(|\veckjtwo|,x_{J,2})}\right\}\Bigg],
\end{multline}
where
\begin{align}
  \chi_1(n,\gamma) =& \phantom{+}\mathcal{S}\chi_0(n,\gamma) + \frac{3}{2}\zeta(3)-\frac{\beta_0}{8N_c}\chi_0^2(n,\gamma)\non
& +\frac{1}{4}\left[\psi''\left(\gamma+\frac{n}{2}\right)+\psi''\left(1-\gamma+\frac{n}{2}\right)-2\phi(n,\gamma)-2\phi(n,1-\gamma)\right]\non
&- \frac{\pi^2\cos(\pi\gamma)}{4\sin^2(\pi\gamma)(1-2\gamma)}\Bigg\{\left[3+\left(1+\frac{N_f}{N_c^3}\right)\frac{2+3\gamma(1-\gamma)}{(3-2\gamma)(1+2\gamma)}\right]\delta_{n,0}\non
&\hspace{2cm}-\left(1+\frac{N_f}{N_c^3}\right)\frac{\gamma(1-\gamma)}{2(3-2\gamma)(1+2\gamma)}\delta_{n,2}\Bigg\},
\label{eq:nlokernel}
\end{align}
with the constant ${\mathcal S} = (4 - \pi^2 + 5 {\beta_0}/{N_c})/12$. $\zeta(n)=\sum_{k=1}^\infty k^{-n}$ is the Riemann zeta function while the function $\phi$ reads
\begin{multline}
  \phi(n,\gamma) = \sum_{k=0}^\infty \frac{(-1)^{k+1}}{k+\gamma+\frac{n}{2}}\Bigg(\psi'(k+n+1)-\psi'(k+1)\\
+ (-1)^{k+1}\left[\beta'(k+n+1)+\beta'(k+1)\right]
 +\frac{\psi(k+1)-\psi(k+n+1)}{k+\gamma+\frac{n}{2}}\Bigg),
\end{multline}
with
\begin{equation}
  \beta'(\gamma) = \frac{1}{4}\left[\psi'\left(\frac{1+\gamma}{2}\right)-\psi'\left(\frac{\gamma}{2}\right)\right].
\end{equation}

At NLL accuracy, only the leading order vertex coefficients 
enter in the derivative term of \eqref{eq:omegaNLO}, so that
\begin{equation}
   \label{eq:dnuerivative}
-2\ln\mu_R^2-i\frac{\partial}{\partial\nu}\ln\frac{C^{\rm (LO)}_{n,\nu}(|\veckjone|,x_{J,1})}{\left(C^{\rm (LO)}_{n,\nu}(|\veckjtwo|,x_{J,2})\right)^*}
= 2\ln\frac{|\veckjone|\cdot|\veckjtwo|}{\mu_R^2} .
\end{equation}

\subsection{Strong coupling, renormalization scheme and PDFs at NLL}

In this paper we will mainly use the MSTW 2008 PDFs~\cite{Martin:2009iq}. We will make comparisons with several other sets of PDFs, as provided by the Les Houches Accord PDF Interface (LHAPDF)~\cite{Whalley:2005nh}.

We use the two-loop strong coupling constant in the form
\begin{equation}
  \label{eq:runningcoupling}
  \alpha_s(\mu_R^2) = \frac{1}{b_0 L}\left(1+\frac{b_1}{b_0^2}\frac{\ln L}{L}\right),
\end{equation}
with $L=\ln \mu_R^2/\Lambda_{\rm QCD}^2$, and
\begin{align}
  b_0 =& \frac{33-2N_f}{12\pi} ,  & 
  b_1 =& \frac{153-19N_f}{24\pi^2} .
\end{align}
In the following, $\alpha_s$ or $\asbar$ without argument is to be understood as $\alpha_s(\mu_R^2)$ or $\asbar(\mu_R^2)$ respectively.
The MSTW 2008 PDFs assume $\mu_R$ and $\mu_F$ to be equal. Therefore, we make the same identification everywhere in our analysis.
The renormalization scale $\mu_R$ is chosen to be $\mu_R=\sqrt{|\veckjone|\cdot |\veckjtwo|}$.

\subsection{Choice of scale $s_0$}

At NLL,
one should also pay attention to the choice of scale $s_0$. The choice $s_0 =\sqrt{s_{0,1} \, s_{0,2}}$   with 
$s_{0,i}= \frac{x_{i}^2}{x_{J,i}^2}\veckji^2$ which we adopt is natural, since it does not depend on the momenta $\veck_{1,2}$ to be integrated out. Besides, the dependence with respect to $s_0$ of the whole amplitude can be studied,
when taking into account the fact that both the NLL BFKL Green's function and the vertex functions are $s_0$ dependent. We refer to Sec.~3.2.2 of ref.~\cite{Colferai:2010wu} for a detailed discussion.

\subsection{Collinear improvement}

Several methods have been developed to improve the NLL BFKL Green's function for $n=0$, by imposing compatibility with the DGLAP equation \cite{Gribov:1972ri,Lipatov:1974qm,Altarelli:1977zs,Dokshitzer:1977sg} in the collinear limit \cite{Salam:1998tj,Ciafaloni:1998iv,Ciafaloni:1999yw,Ciafaloni:2003rd}.
This is only required by the Green's function. Indeed, the collinear improvement deals with poles in the  $\gamma$ plane ($\gamma$ being a variable conjugated to transverse momentum in the Mellin transform). We have checked,  based on a numerical study~\cite{Colferai:2010wu}, that the 
jet vertices are free of $\gamma$ poles and thus do not call for any collinear improvement.
In order to study the effect of such possible collinear improvement \cite{Salam:1998tj,Ciafaloni:1998iv,Ciafaloni:1999yw,Ciafaloni:2003rd}, 
a first attempt was performed in ref.~\cite{Colferai:2010wu}, for $n=0$, using the scheme 3 of ref.~\cite{Salam:1998tj}. Focusing on  $n=0$ is enough for the study of the cross-section. 

In view of the study of azimuthal correlation, a consistent treatment requires to take into account these collinear improvements also for $n \neq 0\,.$
This has been investigated for the NLL BFKL Green's function in refs.~\cite{Vera:2007kn,Schwennsen:2007hs,Marquet:2007xx}. We take into account these effects in this paper, thus improving the study of our previous work~\cite{Colferai:2010wu}.

\section{Binning in $|\veckj|$}

\subsection{Integration over $|\veckj|$}
\label{SubSec:integration-kJ}

The experimental binning imposes that the values of $|\veckji|$
should be integrated in a given range. 
Each rapidity $y_i$ varies in the range
$y_{\rm min} \le y_i \le y_{\rm max}$. In practice, we take $y_{\rm min}=0$ and $y_{\rm max}=4.7\,.$
The total relative rapidity $Y=y_1+y_2$ which is experimentally accessible varies between 0 and 9.4. We will restrict ourselves to the region $Y \gtrsim  4$ (see discussion at the end of this section).

\begin{figure}[htbp] 
\psfrag{80LL}{\scalebox{.80}{$ 80\%$}}
\psfrag{80NLL}{\scalebox{.80}{$ 80\%$}}
 \psfrag{sigma}{\hspace{0.1cm}$\left(\frac{d\sigma}{dY}\right)_{\rm bin}$ \scalebox{0.9}{[nb]}}
 \psfrag{kjmax}{$k_{J\, {\rm max}}$ \scalebox{0.9}{[GeV]}}
 \psfrag{90}{\scalebox{0.7}{$\hspace{-0.5cm} 90\%$} \scalebox{0.8}{$\hspace{-0.05cm} \sigma_{max}$}}
 \centerline{\includegraphics[width=10cm]{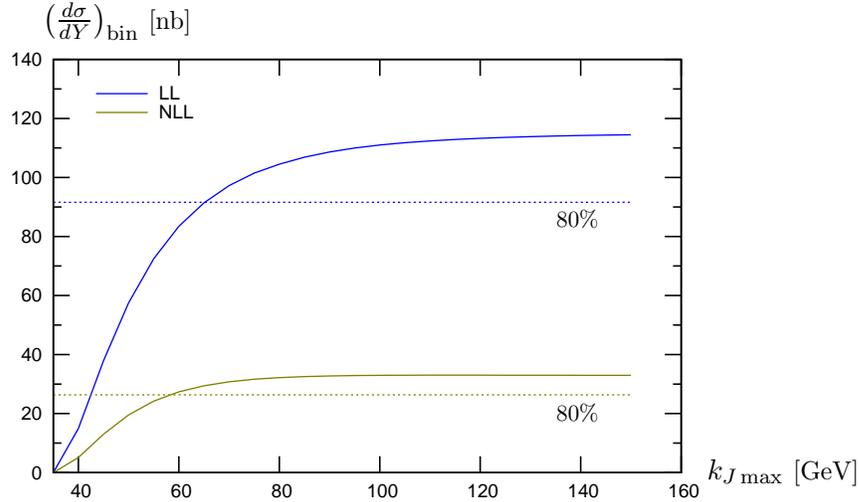}}
 \caption{Growth of the cross-section with $k_{J\, {\rm max}}=k_{J\, {\rm max},1}=k_{J\, {\rm max},2}\,,$ for $Y=6.6$ and $k_{J\, {\rm min}}=k_{J\, {\rm min},1}=k_{J\, {\rm min},2}=35$~GeV.}
 \label{Fig:ubound}
\end{figure}

The phase space, at fixed $|\veckjone|$, $|\veckjtwo|$, $Y$, is defined as
\beqa
\PS \equiv d |\veckjone| \,d |\veckjtwo| 
\, dy_1 \, dy_2
\, \delta(y_1+y_2-Y)\,.
\label{def-PS}
\eqa
Correspondingly, the integration over the bin phase-space is
defined as 
\beqa
\intbin \equiv 
\int\limits_{k_{J {\rm min},1}}^{k_{J\, {\rm max},1}} d |\veckjone|
\int\limits_{k_{J {\rm min},2}}^{k_{J\, {\rm max,2}}} d |\veckjtwo|
\int\limits_{y_{\rm min}}^{y_{\rm max}} dy_1
\int\limits_{y_{\rm min}}^{y_{\rm max}} dy_2
\, \,\delta(y_1+y_2-Y)\,.
\label{def-dPSbin}
\eqa
For a given observable ${\cal O}\,,$
we thus define
\beqa
{\cal O}_{\rm bin} = \intbin \,\, {\cal O}\,,
\label{def-Obin}
\eqa
which equivalently can be written as
\beqa
&&\hspace{-.95cm}{\cal O}_{\rm bin} = \!\!\int\limits_{k_{J {\rm min},1}}^{k_{J {\rm max},1}} 
\! d |\veckjone| \!\!
\int\limits_{k_{J {\rm min},2}}^{k_{J {\rm max},2}} 
d |\veckjtwo| \!\!
\int\limits_{y_{\rm min}}^{y_{\rm max}} dy_1
\int\limits_{y_{\rm min}}^{y_{\rm max}} dy_2 \,
\,\delta(y_1+y_2-Y)\, {\cal O}(|\veckjone|, |\veckjtwo|, y_1, y_2) 
\label{def-dbin1}
\\
&&\hspace{-.95cm}= \!\!\!\!\!
\int\limits_{k_{J {\rm min},1}}^{k_{J\, {\rm max},1}} \!\!
\! d |\veckjone| \!\!
\int\limits_{k_{J {\rm min},2}}^{k_{J\, {\rm max},2}} \!\!
d |\veckjtwo| \!\!
\int\limits_{y_{\rm min}}^{y_{\rm max}} dy_1 \!
\,\,\Theta(y_{\rm min} \le Y-y_1 \le y_{\rm max})\, {\cal O}(|\veckjone|, |\veckjtwo|, y_1, Y-y_1)\,.
\label{def-dbin2}
\eqa
The resulting cross-section
\beqa
\left(\frac{d\sigma}{dY}\right)_{\rm bin}= 
\intbin \frac{\dsigma}{\dtwojets}
\label{def_sigma_intk}
\eqa
is in practice numerically evaluated by sampling each $y_i$ with a $(y_{\rm max}-y_{\rm min})/10=0.47$ binning. It thus means that in eq.~(\ref{def-dbin2}), the $y_1$ integration is replaced by a discrete sum, which is then multiplied by a $0.47$ width.
This cross-section is shown in figure~\ref{Fig:ubound}, in the pure LL approximation as well as in the full NLL treatment. This figure shows that a very significant fraction ($\sim 80 \%$) of the cross-section is obtained for $k_{J\, {\rm max}} \sim 60$ GeV.
We will further discuss this in the next subsection in relation with energy-momentum conservation issues.

As long as the jet vertices are treated in the LL approximation, the integration with respect to $|\veckji|$ can be performed analytically. As a consistency check, we compare, in the Tevatron kinematics used in ref.~\cite{Vera:2007kn}, the integration with respect to $|\veckji|$ with boundaries
$k_{J\, {\rm min},1}=20$ GeV, $k_{J\, {\rm min},2}=50$ GeV and $k_{J\, {\rm max},1}=k_{J\, {\rm max},2}=\infty\,.$ 
Numerically, due to numerical instabilities when evaluating
the Green's function for fixed values of $|\veckji|$ at low $Y$, the comparison with data is expected to be rather poor for 
$Y < \frac{\pi}{2 \alpha_s N_c}$, i.e. typically for
$Y \lesssim 4\,.$ A detailed study of this issue will be made elsewhere~\cite{small-Y}. 
In the rest of this paper, we will restrict ourselves to the region
$Y \gtrsim 4\,.$

\begin{figure}[htbp]
 \psfrag{sigma}{\hspace{0.1cm}$\sigma_{\rm bin}$ \scalebox{0.9}{(nb)}}
 \psfrag{kjmax}{$k_{J\, {\rm max}}$ \scalebox{0.9}{(GeV)}}
 \psfrag{90}{\scalebox{0.7}{$\hspace{-0.5cm} 90\%$} \scalebox{0.8}{$\hspace{-0.05cm} \sigma_{max}$}}
\vspace{.3cm}\centerline{\hspace{-4.8cm}\includegraphics[width=10cm]{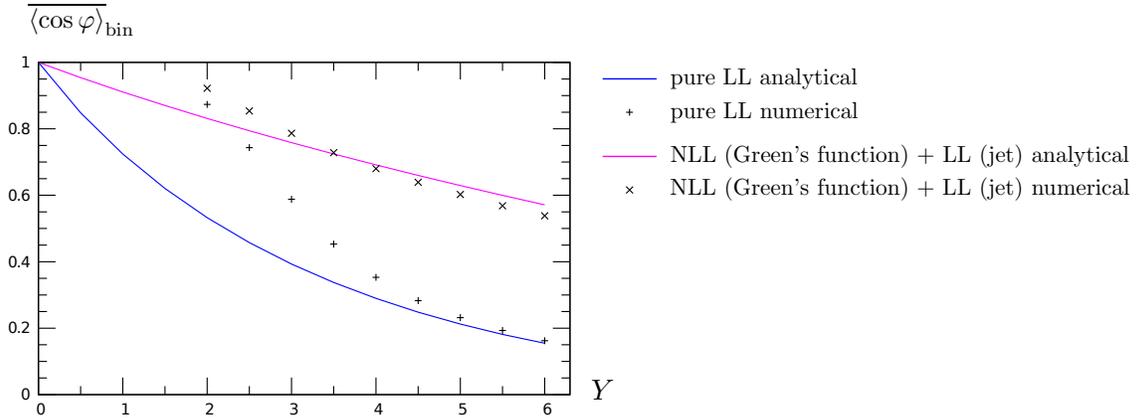}}
 \caption{Comparison of $\overline{\langle \cos \varphi \rangle}_{\rm bin}$, either using  the numerical integration  over $|\veckjone|, |\veckjtwo|$, based on our \textsc{Fortran} code (crosses), or the analytical integration (solid) as shown in ref.~\cite{Vera:2007kn}, for the Tevatron kinematics, for both pure LL BFKL and mixed NLL BFKL treatments.}
 \label{Fig:integrated_k-Tevatron}
\end{figure}

To perform this comparison, we use the observable $\overline{\langle \cos \varphi \rangle}_{\rm bin}\,,$ which is the average of $\langle \cos \varphi \rangle$ on the experimental bin, that is
defined here as (see eq.~\ref{eq:def_cor}))
\beqa
\overline{\langle \cos (m \varphi) \rangle}_{\rm bin} 
&=& \frac{\displaystyle
\int\limits_{k_{J {\rm min},1}}^{k_{J\, {\rm max},1}} 
\int\limits_{k_{J {\rm min},2}}^{k_{J\, {\rm max},2}} 
%
%
 \dtwojetsK \, {\cal C}_m}
{\displaystyle
\int\limits_{k_{J {\rm min},1}}^{k_{J\, {\rm max},1}} 
\int\limits_{k_{J {\rm min},2}}^{k_{J\, {\rm max},2}} 
%
   \dtwojetsK
 \, {\cal C}_0}  \nonumber \\
&=& 
\frac{\displaystyle
\intbin
\langle \cos (m \varphi) \rangle\, \displaystyle \frac{\dsigma}{\dtwojets}}{ \displaystyle \intbin  \, \displaystyle\frac{\dsigma}{\dtwojets}    }\,.
\label{def-cosphi-bin}
\eqa
This comparison is shown in figure~\ref{Fig:integrated_k-Tevatron}, showing the consistency of our numerical results\footnote{This comparison focuses on the Green's function. It thus assumes that the PDFs are equal to 1, with $x_{J, i}=1$. The scales $\mu_F$ and $\sqrt{s_0}$ are taken to be $\sqrt{k_{J {\rm min,\,1}} \,\, k_{J {\rm min,\,2}}}\,,$ with $k_{J {\rm min,\,1}}=20$ GeV and $k_{J {\rm min,\,2}}=50$~GeV\,.}\!.
 

\def\scalecos{.9}
\psfrag{C1C0}{\scalebox{\scalecos}{$\overline{\langle \cos \varphi \rangle}_{\rm bin}$}}
\psfrag{Y}{$Y$}
\psfrag{LLmath}[l][l][.8]{pure LL analytical}
\psfrag{LLfortran}[l][l][.8]{pure LL numerical}
\psfrag{NLLmath}[l][l][.8]{NLL (Green's function) + LL (jet) analytical}
\psfrag{NLLfortran}[l][l][.8]{NLL (Green's function) + LL (jet)  numerical}

\subsection{Energy-momentum conservation issues}
\label{SubSec:energy-momentum}

It is well known that the BFKL equation does not preserve energy-momentum conservation.
However, this violation is expected to be smaller at higher order in perturbation theory, i.e. when comparing NLL BFKL versus LL BFKL.
In practice, one should thus avoid to use all the available collider energy.
This means that one should satisfy the constraint
\beq
y_{J,i}  \ll \cosh^{-1} \frac{{ x_i} \, E}{ k_{J,i}}\,.
\label{Yconstraint}
\eq
This implies that taking a lower $k_J$ provides a larger validity domain. This justifies in our opinion a strong experimental effort to extract low $k_J$ data.

In practice, with only a lower cut on $k_J$, one has to integrate over regions where the BFKL approach may not be valid anymore. For example, $k_{J}=60$ GeV leads to  a constraint $y_{J,i}  \ll 7.3\,.$
For this reason it would be nice to have a measurement with bins including an upper cut on transverse momentum, $k_{J\, {\rm min}} \le k_J \le k_{J\, {\rm max}}\,.$ Since the cross-sections are expected to be large, we believe that the statistics should be large enough
to allow for a narrow binning in $k_J$, which should thus be mainly a  detector issue\footnote{At CMS,  a measurement with  $k_{J \,{\rm min}}=35$ GeV seems to be possible, while 
going down to $20$ GeV would probably require a dedicated trigger.}\!.
To conclude this section, we  note that the
$k_J$ integration reduces the $Y$ domain between jets. This 
$Y$ domain is also reduced by the 
 $x_i$ integration, which is weighted by PDFs, strongly peaked at small  $x_i\,.$

 \section{Results: symmetric configuration}
 \label{Sec:sym}
 
 In this section, we consider a symmetric configuration as planned to be studied by the CMS collaboration. We thus consider bins with cuts
 \beqa
  35\,{\rm GeV} < &|\veckjone|,|\veckjtwo|& < 60 \,{\rm GeV} \,, \nonumber\\
 0 < &y_1, \, y_2& < 4.7\,.
\label{sym-cuts} 
\eqa
We consider several kinds of scenarios, starting from a pure LL approximation up to full NLL and collinear improved NLL approximations. The convention for colors is the same in the whole paper:
\beq
\hspace{-.3cm}
\begin{tabular}{ll}
blue: &  pure LL result \\
magenta: &   combination of LL vertices with  pure NLL Green's function \\
green: &  combination of LL vertices with collinear improved NLL Green's function \\
brown: &  pure NLL result \\
red: &  full NLL vertices with  collinear improved NLL Green's function.
\end{tabular}
\label{def:colors}
\eq

\subsection{Cross-section}

\begin{figure}[htbp]
\centerline{\includegraphics[width=10cm]{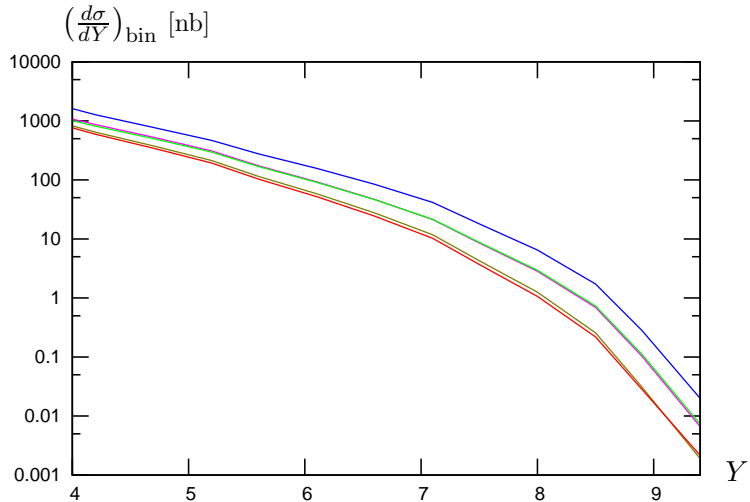}}
 \caption{Differential cross-section as a function of the jet rapidity separation $Y$, integrated over bins $35\,{\rm GeV} < |\veckjone|,|\veckjtwo| < 60 \,{\rm GeV}$ and $0 < y_1, \, y_2 < 4.7$, for the 5 scenarios described in the text, see (\ref{def:colors}).}
  \label{Fig:sigma-7-35-35}
\end{figure}

We first consider the cross-section.
The obtained results are displayed in figure~\ref{Fig:sigma-7-35-35}. Note that
the Monte Carlo integration leads to a precision of the order of $2\%$ to $5\%$, which is too small to be seen in this figure.

This result confirms the fact that NLL corrections to the jet vertices are huge, of the same order of magnitude as the NLL corrections to the Green's function. The full NLL result leads to a cross-section which is significantly smaller than the one based on LL vertices combined with the  pure NLL Green's function.

The curves obtained when combining  the LL vertices with the pure NLL Green's function and when combining
 the LL vertices with the collinear improved NLL Green's function are almost indistinguishable. Similarly, the curves obtained when combining  the NLL vertices with the pure NLL Green's function and when combining
 the NLL vertices with the collinear improved NLL Green's function are very close.
 
 \psfrag{sigma}{$\hspace{0cm}\mathcal{C}_0  = \frac{d\sigma}{d Y}\left[{\rm nb}\right]$}
\psfrag{Y}{$Y$}

\psfrag{LL}[l][r][.8]{\qquad \qquad \footnotesize pure LL}

\psfrag{LLmix}[l][r][.8]{\qquad \qquad \footnotesize LL vertex + NLL Green fun.}
\psfrag{LLplus}[l][r][.8]{\qquad \qquad \footnotesize LL vertex + NLL resum. Green fun.}
\psfrag{NLL}[l][r][.8]{\qquad \qquad \footnotesize NLL vertex + NLL Green fun.}
\psfrag{NLLplus}[l][r][.8]{\qquad \qquad \footnotesize NLL vertex + NLL resum. Green fun.}

\begin{figure}[htbp]
\vspace{.4cm}
 \begin{minipage}{0.49\textwidth}
      \includegraphics[width=7.7cm]{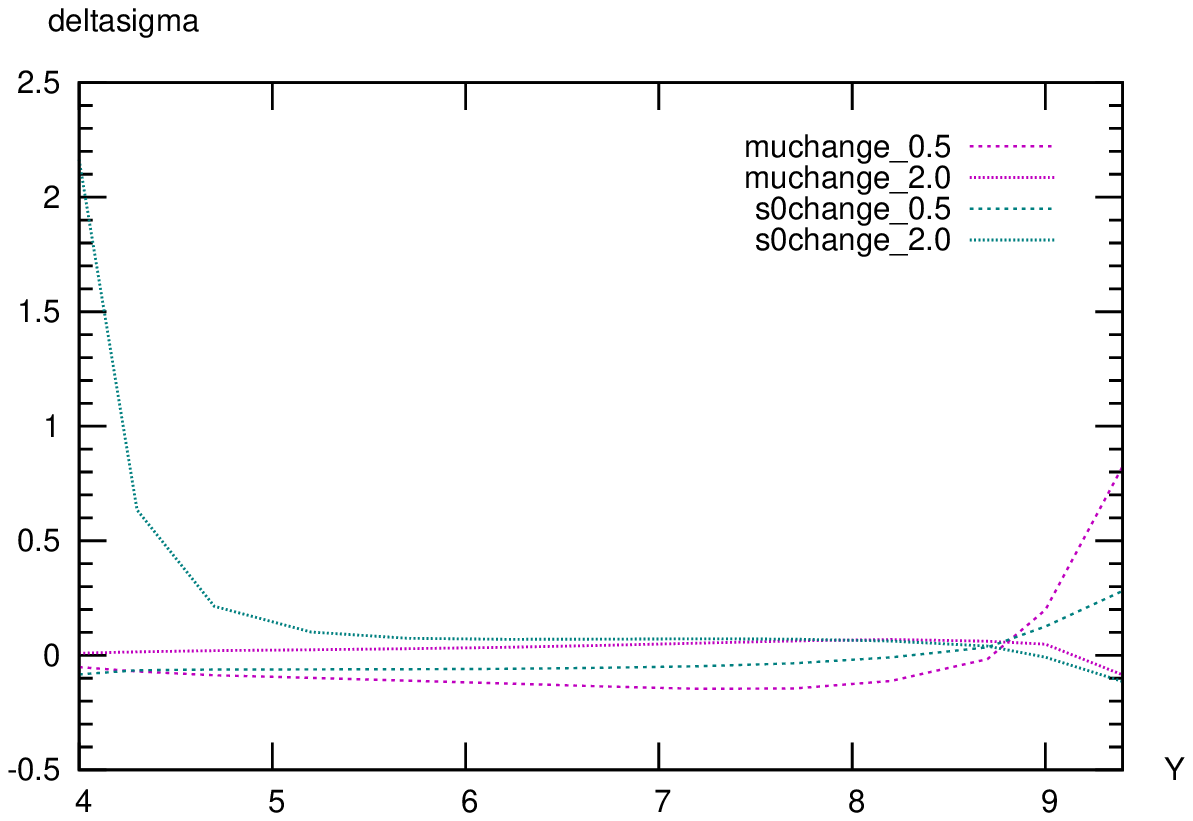}
  \end{minipage}
\begin{minipage}{0.49\textwidth}
      \psfrag{ecartrel}{\raisebox{-0.1cm}{$\frac{\Delta \sigma}{\sigma}$}}
      \psfrag{Y}{\scalebox{0.9}{$Y$}}
     \hspace{-.2cm} \includegraphics[width=7.72cm]{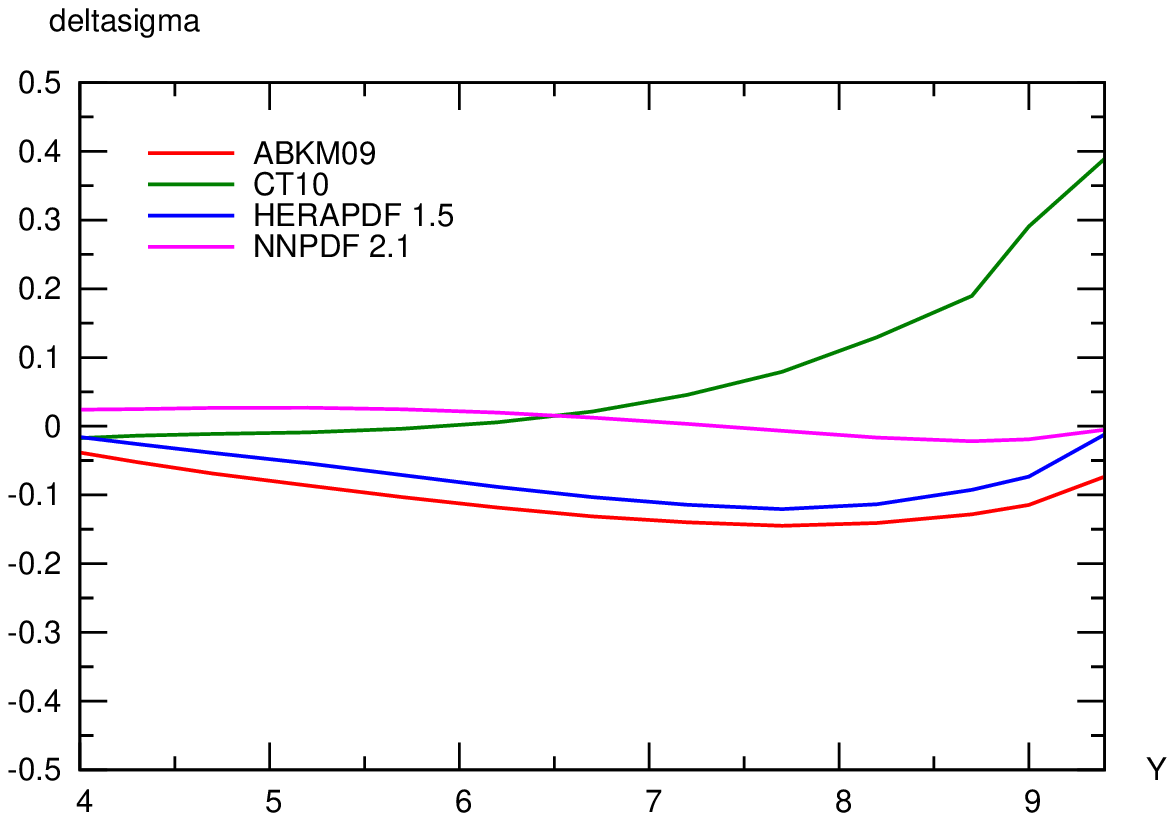}
  \end{minipage}
\caption{
    Left:
     Relative variation of the cross-section  when varying $\sqrt{s_0}$ and $\mu_F$ with a factor 2.
    Right: Relative variation of cross-section with respect to MSTW PDFs due to the replacement by other sets of PDFs, as indicated.}
\label{Fig:sigma-7-35-35-parameters}
\end{figure}

In figure~\ref{Fig:sigma-7-35-35-parameters}, in the pure NLL case, we display the uncertainties  due to the changes of the various involved parameters. 
The first effect which we study is the variation of the scales $s_0$ and $\mu_F$, as shown in figure~\ref{Fig:sigma-7-35-35-parameters}~(L). 
The large uncertainty at very low $Y$ is related to the specific
instabilities of NLL Green's function mentioned at the end of section~\ref{SubSec:integration-kJ}, while the large $Y$ uncertainty is related to kinematical boundary effect (the cross-section almost vanishes). 

The second effect, due to the dependency on the set of PDFs, is shown in figure~\ref{Fig:sigma-7-35-35-parameters}~(R). Note that we only display the pure NLL case, although the trend is similar for other scenarios.

Both of these effects are much smaller than the changes due to the NLL corrections to the jet vertices.

\def\sca{.7}

\psfrag{muchange_0.5}[l][r][\sca]{\qquad  \  \footnotesize $\mu_F \to \mu_F/2$}
\psfrag{muchange_2.0}[l][r][\sca]{\qquad \ \footnotesize $\mu_F \to2 \mu_F$}
\psfrag{s0change_0.5}[l][r][\sca]{\qquad \ \footnotesize $\sqrt{s_0} \to \sqrt{s_0}/2$}
\psfrag{s0change_2.0}[l][r][\sca]{\qquad \ \footnotesize $\sqrt{s_0} \to 2 \sqrt{s_0}$}

\def\scat{.8}
\psfrag{sigma}{\hspace{0.1cm}$\left(\frac{d\sigma}{dY}\right)_{\rm bin}$ \scalebox{0.9}{[nb]}}

\def\scalecos{.9}
\psfrag{deltasigma}{\raisebox{0.2cm}{\scalebox{\scalecos}{$\displaystyle\frac{\Delta \sigma_{\rm bin}}{\sigma_{\rm bin}}$}}}

\def\sca{.6}
\psfrag{muchange_0.5}[l][r][\sca]{\hspace{-1.9cm}\footnotesize $\mu_F \to \mu_F/2$}
\psfrag{muchange_2.0}[l][r][\sca]{\hspace{-2.05cm} \footnotesize $\mu_F \to2 \mu_F$}
\psfrag{s0change_0.5}[l][r][\sca]{\hspace{-2.2cm} \footnotesize $\sqrt{s_0} \to \sqrt{s_0}/2$}
\psfrag{s0change_2.0}[l][r][\sca]{\hspace{-2.2cm} \footnotesize $\sqrt{s_0} \to 2 \sqrt{s_0}$}
  
  \def\sca{.7}
  
\subsection{Azimuthal correlations}

\begin{figure}[htbp]
\def\scalecos{.9}
\psfrag{cos}{\raisebox{0cm}{\scalebox{\scalecos}{$\overline{\langle \cos \varphi \rangle}_{\rm bin}$}}}
\centerline{\includegraphics[width=10cm]{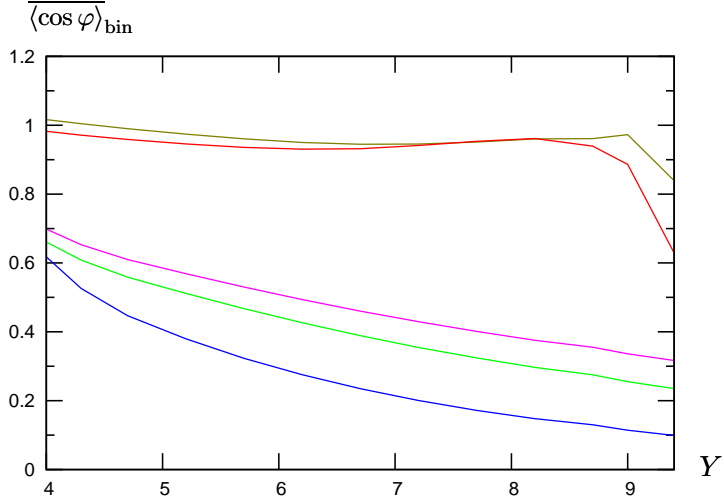}}
  \caption{The bin averaged $\overline{\langle \cos \varphi \rangle}_{\rm bin}$ as a function of the jet rapidity separation $Y$, integrated over bins $35\,{\rm GeV} < |\veckjone|,|\veckjtwo| < 60 \,{\rm GeV}$ and $0 < y_1, \, y_2 < 4.7$, for the 5 scenarios described in the text, see (\ref{def:colors}).}
  \label{Fig:cos-7-35-35}
\end{figure}

We now consider the azimuthal correlations.
The obtained results are displayed in figure~\ref{Fig:cos-7-35-35}, again using the color conventions (\ref{def:colors}). Note that
the Monte Carlo integration leads to a precision of the order of a few $\%$ when using the NLL vertices, the numerical uncertainty being negligible in the case of LL jet vertices. We do not show it on this figure.

Comparing on one hand the pure LL scenario with the mixed LL vertex combined with the NLL Green's function, and  
the mixed LL vertex combined with the NLL Green's function with the full NLL treatment on the other hand, we see that the correction due to the jet vertex produces the largest correction. 
On the same plot, one can see the effect of collinear improvement.
When including this effect for the whole set of the conformal spins $n$, we obtain very close results for the pure NLL and the collinear improved  NLL approaches. This can be compared with the mixed LL jet combined with either a pure NLL Green's function or collinear improved NLL Green's function scenarios: the resulting modification is much smaller at NLL. This is in our opinion a sign of convergence of the perturbative series.
One can see on figure~\ref{Fig:cos_n0-to-ngen} this effect already at the level of fixed $|\veckjone|=|\veckjtwo|=35$ GeV, when passing from $n=0$ collinear improved NLL Green's function (left) to 
all $n$ collinear improved NLL Green's function (right)\footnote{We plot these curves for $\sqrt{s}=7$ TeV, with the same choice of parameters for $y_i$ as in ref.~\cite{Colferai:2010wu}, i.e. $3<y_i <5$ and thus $6< Y< 10$.}\!.
Note that this resummation affects only the green and the red curves in figure~\ref{Fig:cos_n0-to-ngen}, which are thus the only modified ones when passing from figure~\ref{Fig:cos_n0-to-ngen}~(L) to figure~\ref{Fig:cos_n0-to-ngen}~(R). Furthermore, we see that including the collinear resummation for all $n$ does not lead anymore to $\langle \cos \varphi \rangle$ potentially
above 1 (except in the very large $Y$ domain, due to the kinematical boundary effect).

We have made a similar check, in the three scenarios with  LL jet vertices, between the results obtained with our code and the one used in  ref.~\cite{Marquet:2007xx}. These results are in very good agreement, taking into account the slightly different numerical treatments and the fact that we use scheme 3 of ref.~\cite{Salam:1998tj} while the code of ref.~\cite{Marquet:2007xx} is based on scheme 4.


\def\scalecos{.9}
\begin{figure}[htbp]
  \psfrag{cos}{\raisebox{.1cm}{\scalebox{\scalecos}{$\langle \cos \varphi\rangle$}}}
  \psfrag{Y}{\scalebox{0.9}{$Y$}}
  \begin{minipage}{0.49\textwidth}
      \includegraphics[width=7cm]{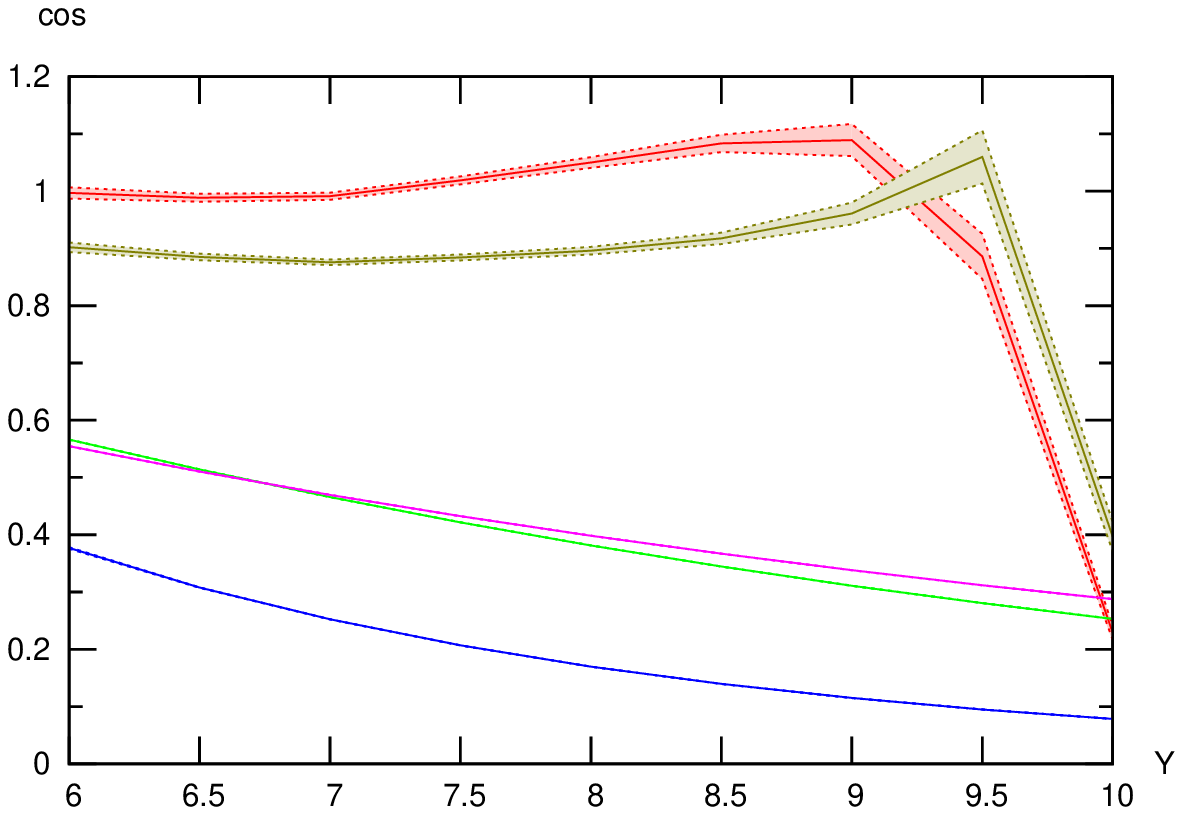}
  \end{minipage}
  \begin{minipage}{0.49\textwidth}
      \includegraphics[width=7cm]{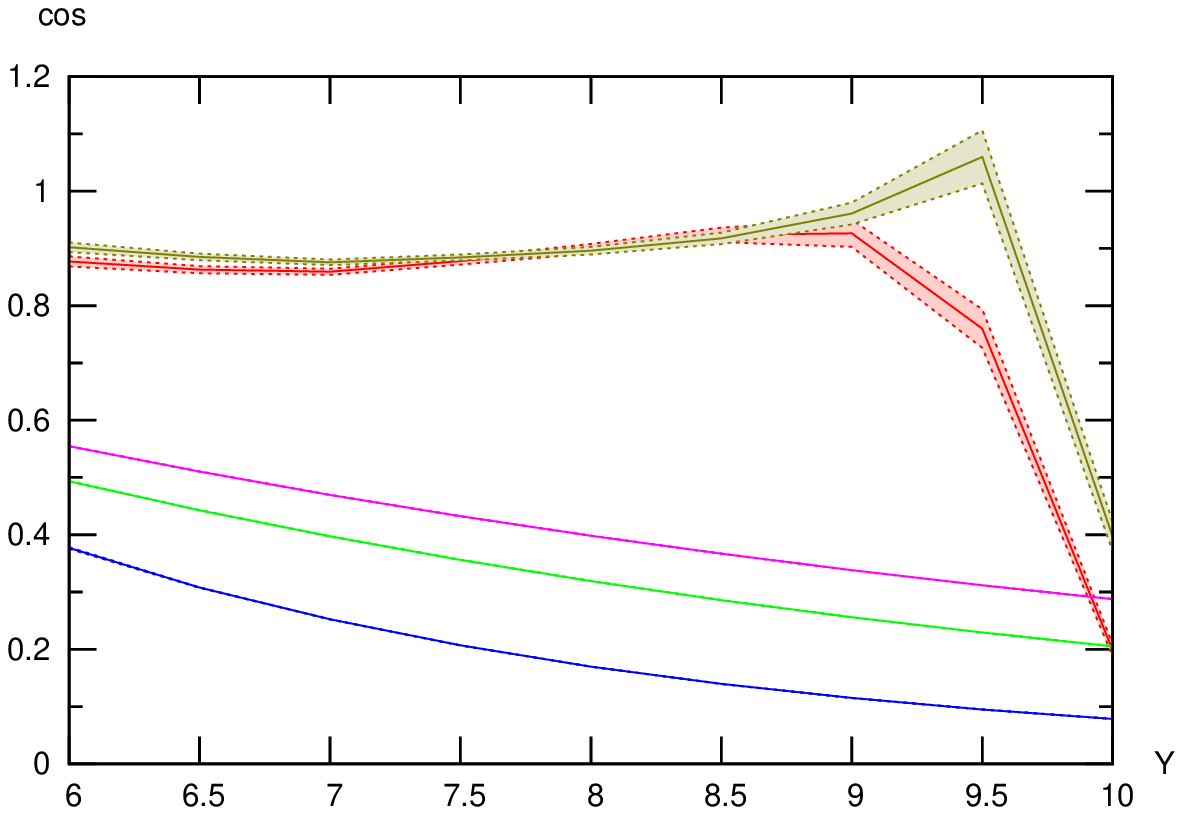}
  \end{minipage}
  \caption{Left: collinear resummation for $n=0$; right: collinear resummation for all $n$.}
  \label{Fig:cos_n0-to-ngen}
\end{figure}

The predictions of figure~\ref{Fig:cos-7-35-35} show that, contrarily to the natural expectation, the inclusion of mini-jets between the two tagged jets, when performed at full NLL, does not break the very high correlation between these two jets. Moreover, the obtained decorrelation effect is very flat with respect to $Y$. This is an effect which takes origin from the NLL corrections to vertices.

 We now study the stability of this result with respect to changes of parameters, within the pure NLL approximation. The variation due to 
change of the scales $s_0$ and $\mu_F$ is shown in figure~\ref{Fig:cos-7-35-35-parameters}~(L). The effect is sizeable, 
but does not change the conclusion that the decorrelation remains much smaller than in the pure LL or mixed LL+NLL approaches.
The second effect, due to the dependency on the set of PDFs, is shown in figure~\ref{Fig:cos-7-35-35-parameters}~(R). This dependency is very weak, much smaller than for the cross-section, see figure~\ref{Fig:sigma-7-35-35-parameters}~(R). Note that this PDF dependency does not exist when using LL jet vertices.

 \def\scalecos{.8}


\def\scalecos{.9}
\def\sca{.6}

\psfrag{central}[l][r][\sca]{\hspace{-.8cm}pure NLL}
\psfrag{muchange_0.5}[l][r][\sca]{\hspace{-1.8cm}\footnotesize $\mu_F \to \mu_F/2$}
\psfrag{muchange_2.0}[l][r][\sca]{\hspace{-1.95cm} \footnotesize $\mu_F \to2 \mu_F$}
\psfrag{s0change_0.5}[l][r][\sca]{\hspace{-2.02cm} \footnotesize $\sqrt{s_0} \to \sqrt{s_0}/2$}
\psfrag{s0change_2.0}[l][r][\sca]{\hspace{-2.02cm} \footnotesize $\sqrt{s_0} \to 2 \sqrt{s_0}$}
\begin{figure}[htbp]
\vspace{.45cm}
\psfrag{cos}{\raisebox{0cm}{\scalebox{\scalecos}{$\overline{\langle \cos \varphi \rangle}_{\rm bin}$}}}
 \begin{minipage}{0.49\textwidth}
      \includegraphics[width=7.5cm]{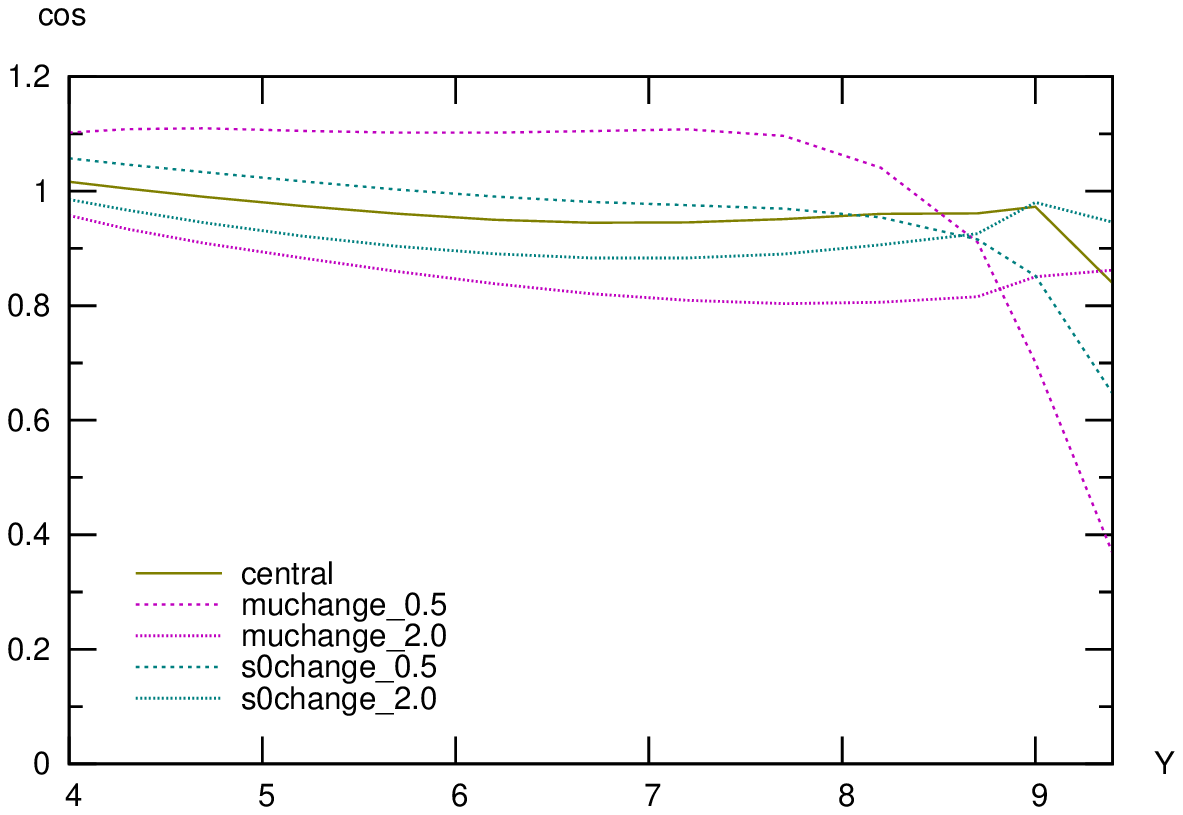}
 \end{minipage}
 \begin{minipage}{0.49\textwidth}
      \psfrag{deltacos}{\raisebox{0.3cm}{$\frac{\Delta\overline{\langle \cos \varphi \rangle}_{\rm bin}}{\overline{\langle \cos \varphi \rangle}_{\rm bin}}$}}
      \psfrag{Y}{\scalebox{0.9}{$Y$}}
      \includegraphics[width=7.5cm]{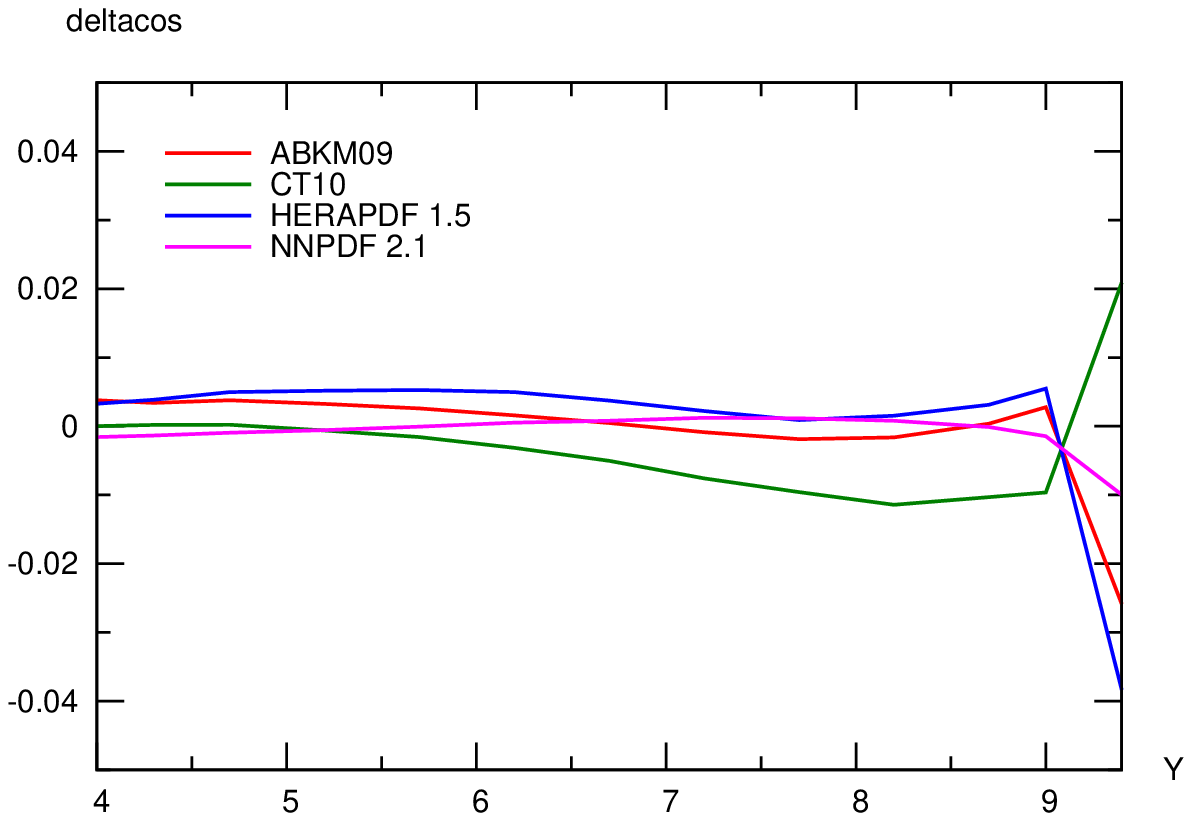}
 \end{minipage}
\caption{
    Left: 
   Variation of $\overline{\langle \cos \varphi \rangle}_{\rm bin}$  when varying $\sqrt{s_0}$ and $\mu_F$ with a factor 2.
    Right: Relative variation of $\overline{\langle \cos \varphi \rangle}_{\rm bin}$ with respect to MSTW PDFs due to the replacement by other sets of PDFs, as indicated.}
\label{Fig:cos-7-35-35-parameters}
\end{figure}

\begin{figure}[htbp]
  \begin{minipage}{0.49\textwidth}
      \includegraphics[width=7.5cm]{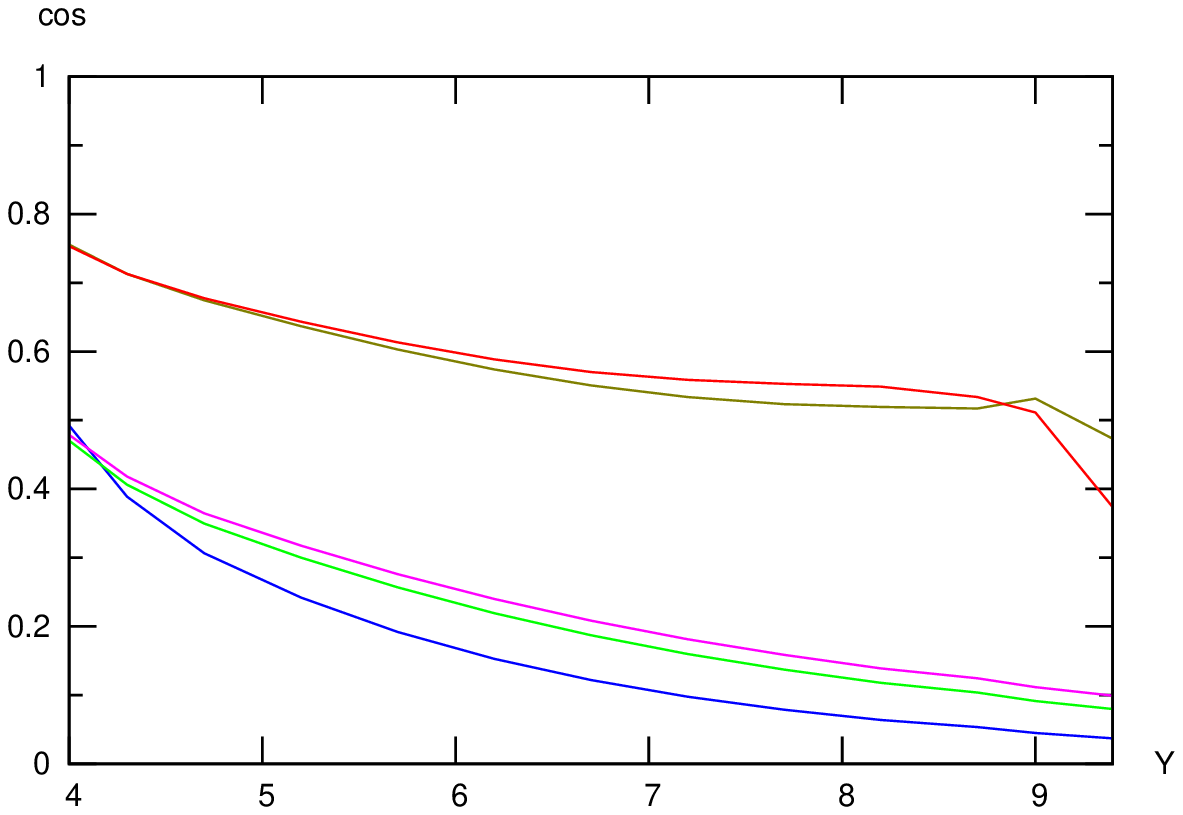}
  \end{minipage}
  \begin{minipage}{0.49\textwidth}
      \includegraphics[width=7.5cm]{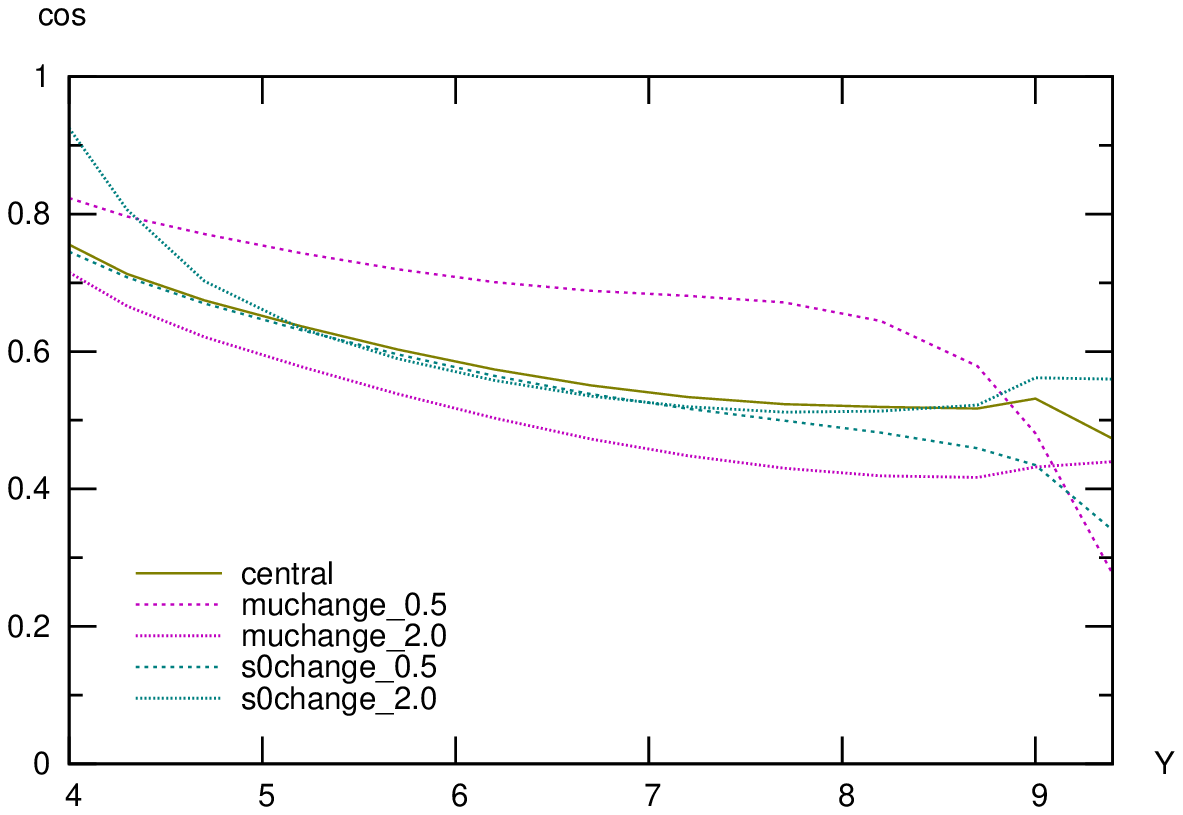}
  \end{minipage}
  \caption{Left: The bin averaged $\overline{\langle \cos (2 \varphi) \rangle}_{\rm bin}$ as a function of the jet rapidity separation $Y$, integrated over bins $35\,{\rm GeV} < |\veckjone|,|\veckjtwo| < 60 \,{\rm GeV}$ and $0 < y_1, \, y_2 < 4.7$, for the 5 scenarios described in the text, see (\ref{def:colors}). Right: Variation of $\overline{\langle \cos (2 \varphi) \rangle}_{\rm bin}$  when varying $\sqrt{s_0}$ and $\mu_F$ with a factor 2.}
 \label{Fig:cos2-7-35-35}
\end{figure}

Let us now consider the observable $\overline{\langle \cos (2 \varphi) \rangle}_{\rm bin}$. The results based on the 5 approaches (\ref{def:colors}) are displayed in figure~\ref{Fig:cos2-7-35-35}~(L).
Similar conclusions  as for  $\overline{\langle \cos \varphi \rangle}_{\rm bin}$ can be drawn. Indeed, an even more dramatic effect due to the NLL corrections to the jet vertices is observed. On the other hand, the difference between the pure NLL and the collinear improved NLL treatments is very small, this time of the same order of magnitude as the one observed between a mixed LL jet with pure NLL Green's function and the mixed LL jet with collinear improved NLL Green's function approaches.

Again, when including full NLL corrections, the decorrelation 
effect is rather small, and much smaller than the one obtained in non full NLL treatments, and the dependency with respect to $Y$ becomes much more flattish. 
 
In figure~\ref{Fig:cos2-7-35-35}~(R), we show the  
dependency of our full NLL prediction 
 with respect to changes of the scales $s_0$ and $\mu_F$. The variation due to $s_0$ changes is very small, almost negligible in comparison with the same dependency for $\overline{\langle \cos \varphi \rangle}_{\rm bin}$
 (see figure~\ref{Fig:cos-7-35-35-parameters}~(L)). Besides, the dependency with respect to $\mu_F$ remains sizeable, although a bit smaller in absolute magnitude than in $\overline{\langle \cos \varphi \rangle}_{\rm bin}$ (but comparable in relative magnitude).

 Again, these dependencies 
 do not change the conclusion that the decorrelation remains much smaller than in the pure LL or mixed LL+NLL approaches.
The second effect, due to the dependency on the set of PDFs, is very weak, similar to the one shown in figure~\ref{Fig:cos-7-35-35-parameters}~(R), and will not be displayed here.


\def\scalecos{.9}
  \psfrag{cos}{\raisebox{0cm}{\scalebox{\scalecos}{$\overline{\langle \cos (2 \varphi) \rangle}_{\rm bin}$}}}

\begin{figure}[htbp]
\vspace{.2cm}
  \begin{minipage}{0.49\textwidth}
      \includegraphics[width=7.5cm]{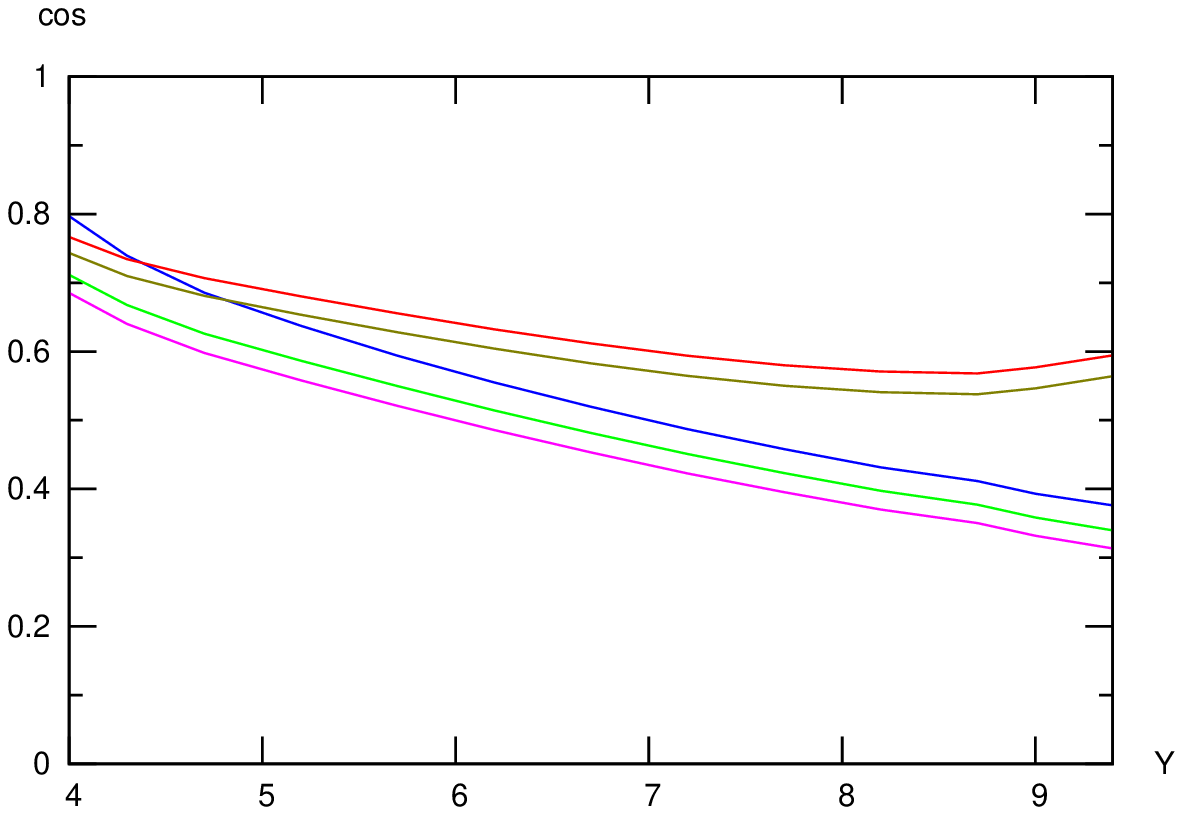}
  \end{minipage}
  \begin{minipage}{0.49\textwidth}
      \includegraphics[width=7.5cm]{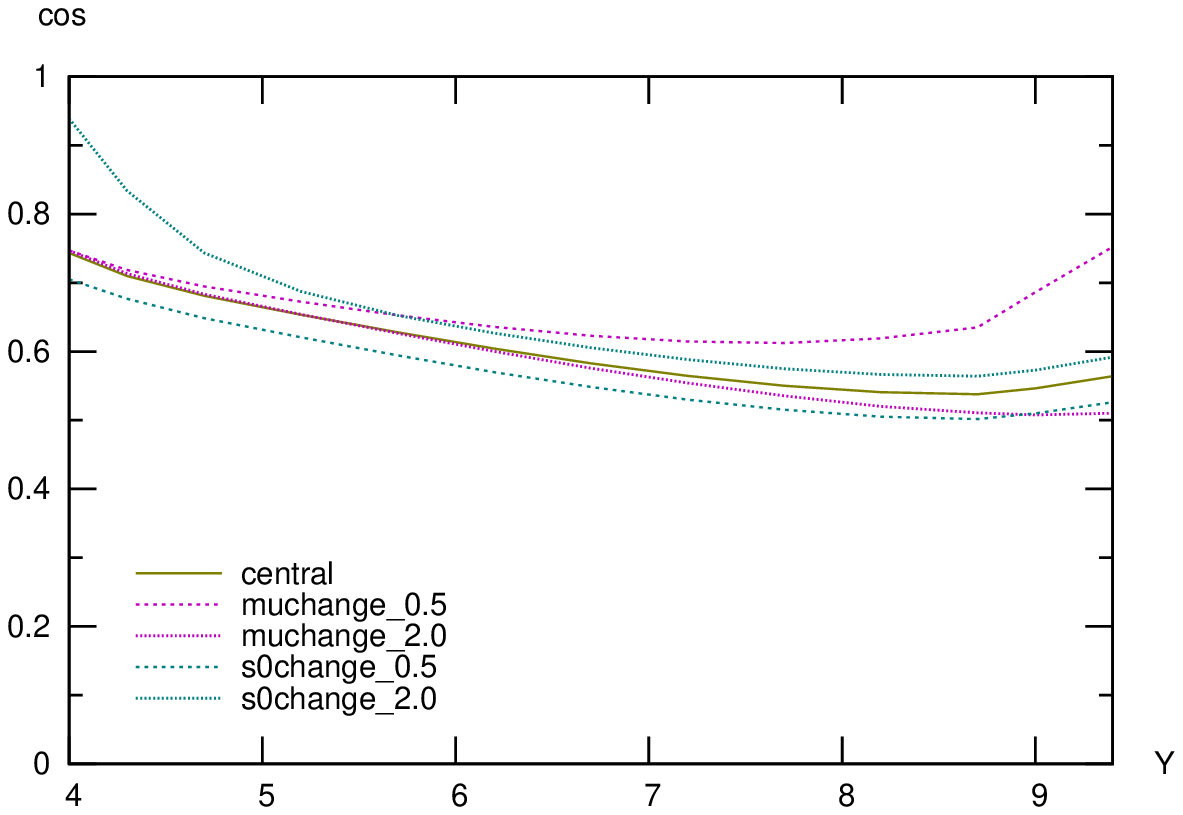}
  \end{minipage}
  \caption{Left: The bin averaged $\overline{\langle \cos (2 \varphi) \rangle}_{\rm bin}/
\overline{\langle \cos \varphi \rangle}_{\rm bin}
$
as a function of the jet rapidity separation $Y$, integrated over bins $35\,{\rm GeV} < |\veckjone|,|\veckjtwo| < 60 \,{\rm GeV}$ and $0 < y_1, \, y_2 < 4.7$, for the 5 scenarios described in the text, see (\ref{def:colors}). Right: Variation of $\overline{\langle \cos (2 \varphi) \rangle}_{\rm bin}/
\overline{\langle \cos \varphi \rangle}_{\rm bin}
$ when varying $\sqrt{s_0}$ and $\mu_F$ with a factor 2.}
  \label{Fig:cos2cos-7-35-35}
\end{figure}

It turns out that this remaining dependency with respect to $\mu_F$
is much reduced when considering the observable 
$\overline{\langle \cos (2\varphi) \rangle}_{\rm bin}/\overline{\langle \cos \varphi \rangle}_{\rm bin}\,.$ This observable is shown in figure~\ref{Fig:cos2cos-7-35-35}. In figure~\ref{Fig:cos2cos-7-35-35}~(L) we display our prediction based on the 5 approaches (\ref{def:colors}), while in
figure~\ref{Fig:cos2cos-7-35-35}~(R) we show the 
$\sqrt{s_0}$ and $\mu_F$ dependency. The difference between the full NLL prediction (either collinearly improved or not) and the non-full NLL ones is sizeable  for $Y \gtrsim 6\,,$ and the 
figure~\ref{Fig:cos2cos-7-35-35}~(R) explicitly shows that this remains valid when taking into account $\sqrt{s_0}$ and $\mu_F$ dependencies.

\def\scalecos{1}
\def\sca{.6}

\psfrag{central}[l][r][\sca]{\hspace{-.8cm}pure NLL}
\psfrag{muchange_0.5}[l][r][\sca]{\hspace{-1.8cm}\footnotesize $\mu_F \to \mu_F/2$}
\psfrag{muchange_2.0}[l][r][\sca]{\hspace{-1.95cm} \footnotesize $\mu_F \to2 \mu_F$}
\psfrag{s0change_0.5}[l][r][\sca]{\hspace{-2.0cm} \footnotesize $\sqrt{s_0} \to \sqrt{s_0}/2$}
\psfrag{s0change_2.0}[l][r][\sca]{\hspace{-2.0cm} \footnotesize $\sqrt{s_0} \to 2 \sqrt{s_0}$}

\psfrag{cos}{\raisebox{0.2cm}{\scalebox{\scalecos}{$\frac{\overline{\langle \cos (2 \varphi) \rangle}_{\rm bin}}
{\overline{\langle \cos \varphi \rangle}_{\rm bin}}
$}}}

The extraction of higher harmonics
can be as well experimentally performed.
We show in figure~\ref{Fig:cos3-7-35-35}~(L) our predictions for 
$\overline{\langle \cos (3 \varphi) \rangle}_{\rm bin}$
based on the 5 different treatments (\ref{def:colors}), and the
corresponding sensitivity with respect to $\sqrt{s_0}$ and $\mu_F$
in
figure~\ref{Fig:cos3-7-35-35}~(R). In comparison with 
$\overline{\langle \cos (2 \varphi) \rangle}_{\rm bin}\,,$
again the effect of NLL corrections in jet vertices is very important, leading to a much smaller decorrelation. The dependency with respect to $s_0$ is similarly small, while the $\mu_F$ dependency is still sizeable. It is smaller in absolute magnitude than the one for $\overline{\langle \cos (2 \varphi) \rangle}_{\rm bin}\,,$ although comparable in relative magnitude.

\def\scalecos{.9}
\def\sca{.6}

\psfrag{central}[l][r][\sca]{\hspace{-.8cm}pure NLL}
\psfrag{muchange_0.5}[l][r][\sca]{\hspace{-1.8cm}\footnotesize $\mu_F \to \mu_F/2$}
\psfrag{muchange_2.0}[l][r][\sca]{\hspace{-1.95cm} \footnotesize $\mu_F \to2 \mu_F$}
\psfrag{s0change_0.5}[l][r][\sca]{\hspace{-2.12cm} \footnotesize $\sqrt{s_0} \to \sqrt{s_0}/2$}
\psfrag{s0change_2.0}[l][r][\sca]{\hspace{-2.12cm} \footnotesize $\sqrt{s_0} \to 2 \sqrt{s_0}$}

\psfrag{cos}{\raisebox{0cm}{\scalebox{\scalecos}{$\overline{\langle \cos (3 \varphi) \rangle}_{\rm bin}$}}}
\begin{figure}[htbp]
  \begin{minipage}{0.49\textwidth}
      \includegraphics[width=7.5cm]{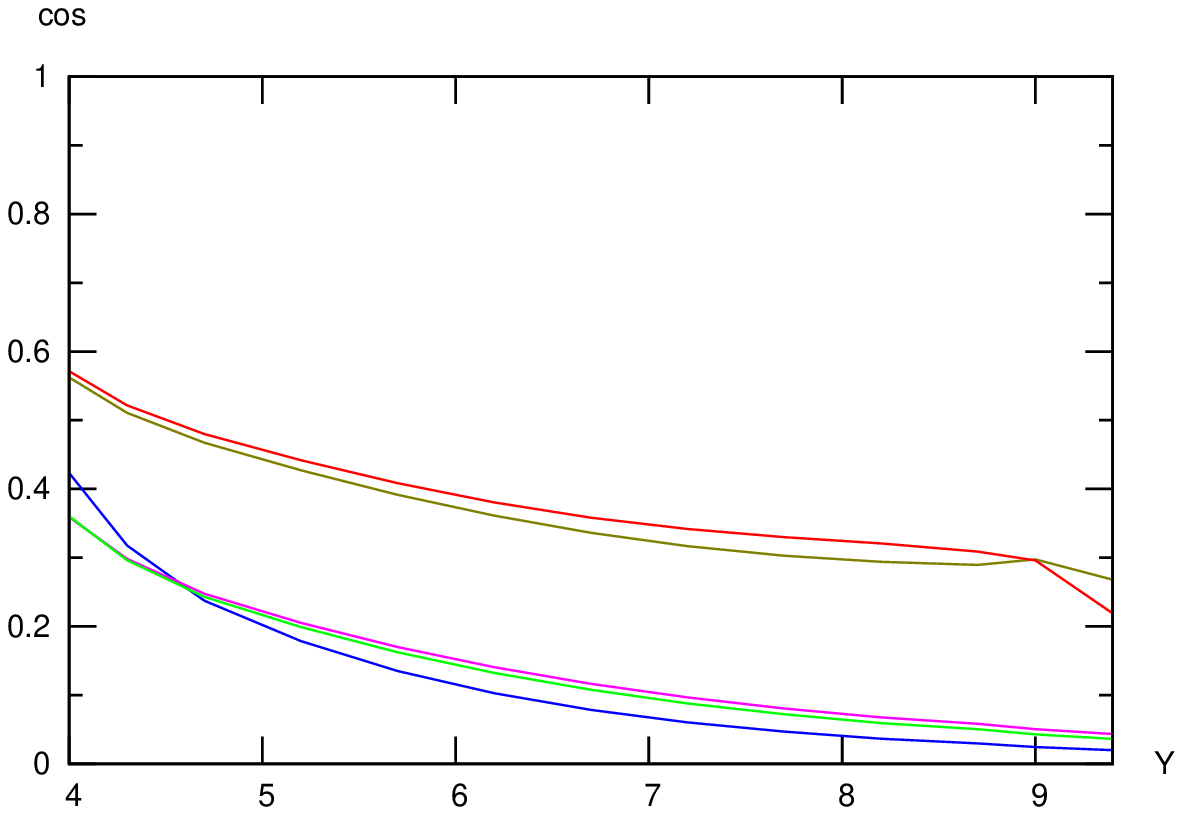}
  \end{minipage}
  \begin{minipage}{0.49\textwidth}
      \psfrag{central}[l][r][\sca]{\hspace{-1.8cm}pure NLL}
      \includegraphics[width=7.5cm]{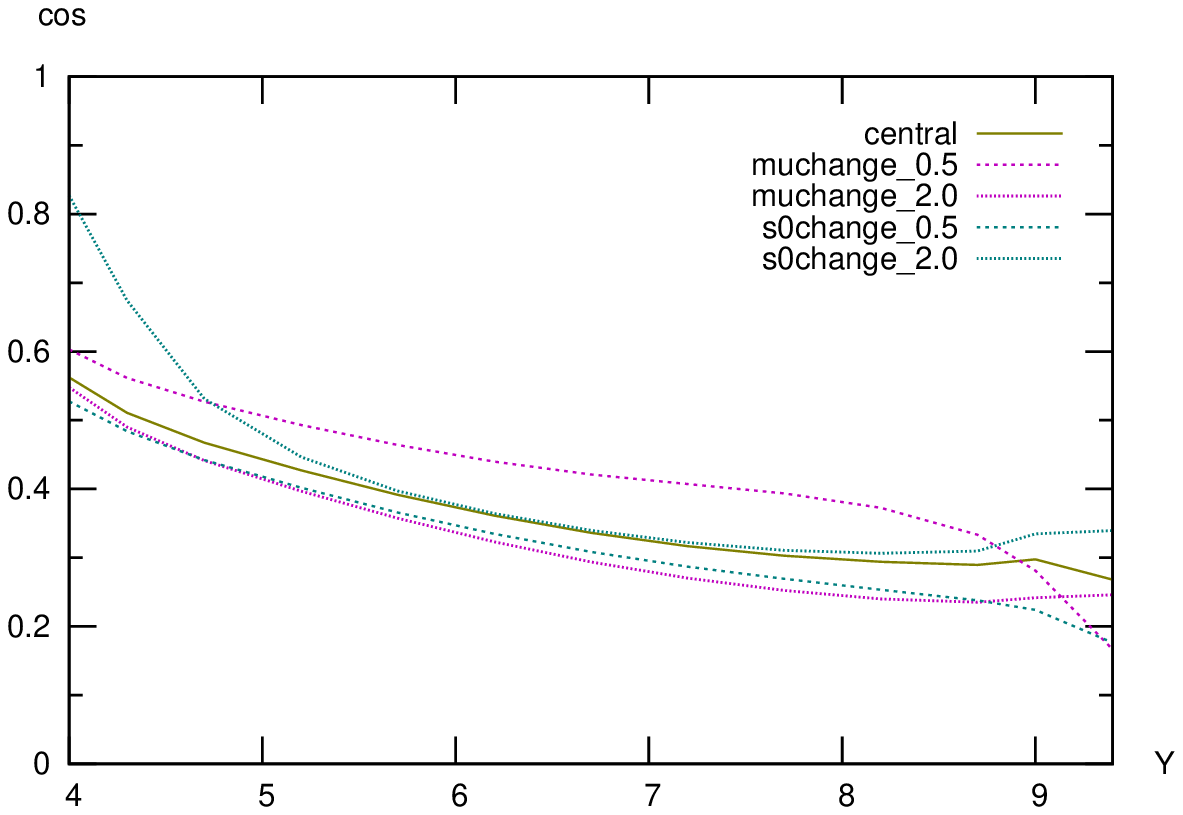}
  \end{minipage}
  \caption{Left: The bin averaged $\overline{\langle \cos (3 \varphi) \rangle}_{\rm bin}$ as a function of the jet rapidity separation $Y$, integrated over bins $35\,{\rm GeV} < |\veckjone|,|\veckjtwo| < 60 \,{\rm GeV}$ and $0 < y_1, \, y_2 < 4.7$, for the 5 scenarios described in the text, see (\ref{def:colors}). Right: Variation of $\overline{\langle \cos (3 \varphi) \rangle}_{\rm bin}$  when varying $\sqrt{s_0}$ and $\mu_F$ with a factor 2.}
  \label{Fig:cos3-7-35-35}
\end{figure}

\begin{figure}[htbp]
\vspace{.2cm}
  \begin{minipage}{0.49\textwidth}
      \includegraphics[width=7cm]{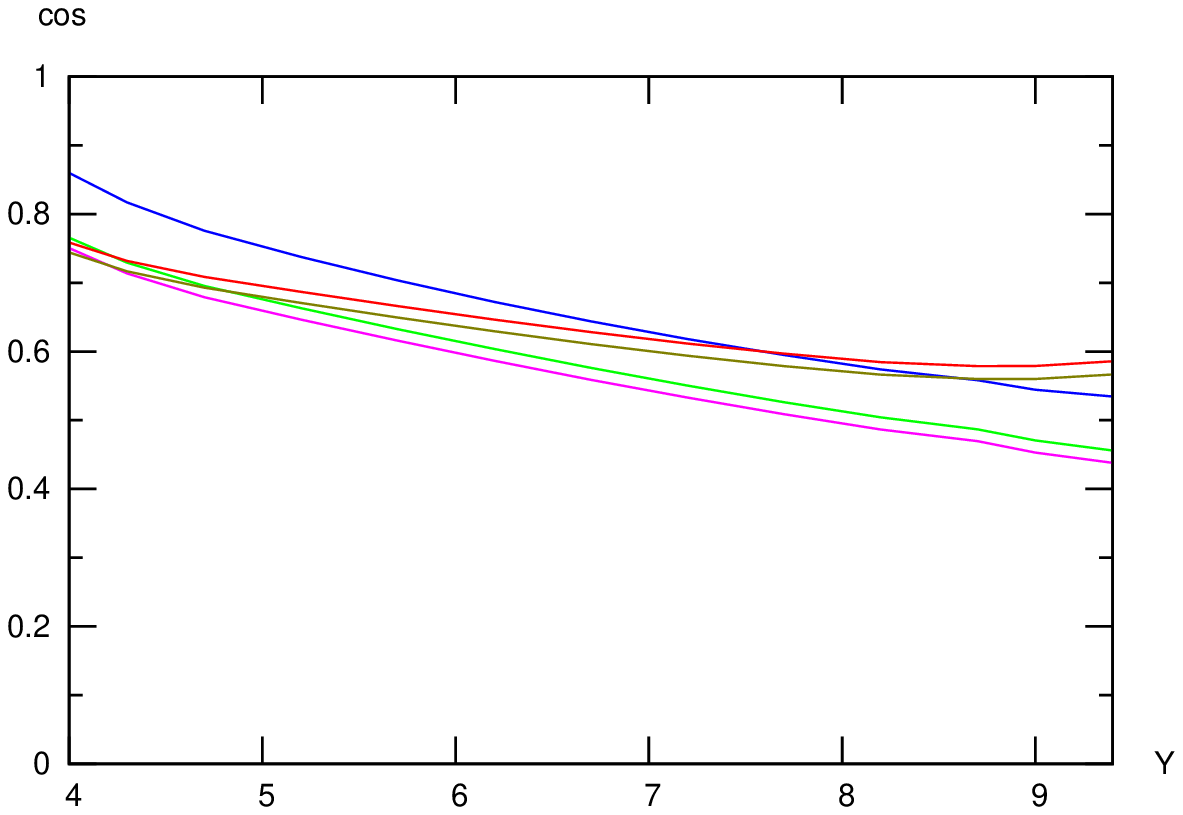}
  \end{minipage}
  \begin{minipage}{0.49\textwidth}
      \includegraphics[width=7cm]{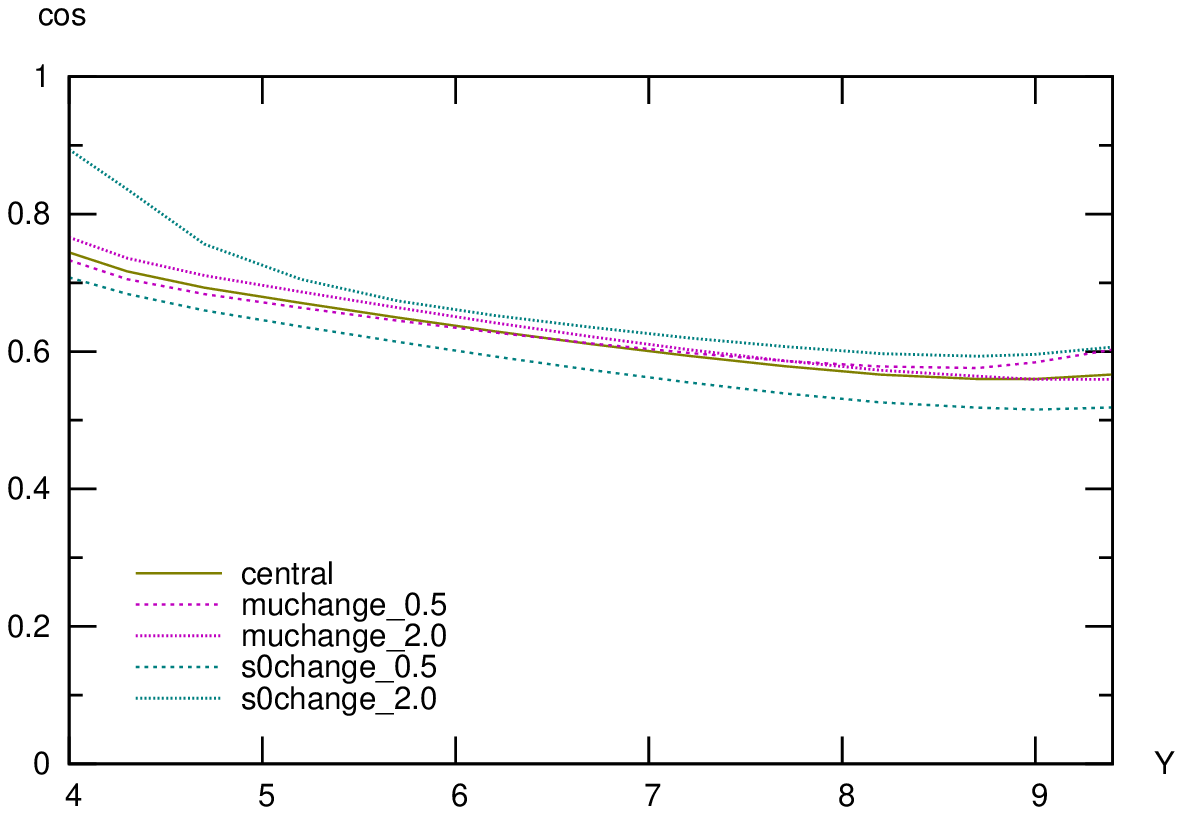}
  \end{minipage}
  \caption{Left: The bin averaged $\overline{\langle \cos (3 \varphi) \rangle}_{\rm bin}/
\overline{\langle \cos (2 \varphi) \rangle}_{\rm bin}
$
as a function of the jet rapidity separation $Y$, integrated over bins $35\,{\rm GeV} < |\veckjone|,|\veckjtwo| < 60 \,{\rm GeV}$ and $0 < y_1, \, y_2 < 4.7$, for the 5 scenarios described in the text, see (\ref{def:colors}). Right: Variation of $\overline{\langle \cos (3 \varphi) \rangle}_{\rm bin}/
\overline{\langle \cos (2 \varphi) \rangle}_{\rm bin}
$ when varying $\sqrt{s_0}$ and $\mu_F$ with a factor 2.}
  \label{Fig:cos3cos2-7-35-35}
\end{figure}

In figure~\ref{Fig:cos3cos2-7-35-35}, we show predictions for the observable $\overline{\langle \cos (3\varphi) \rangle}_{\rm bin}/\overline{\langle \cos (2 \varphi) \rangle}_{\rm bin}\,$, which is less sensitive to changes of factorization scale $\mu_F$.
In figure~\ref{Fig:cos3cos2-7-35-35}~(L) we see that the differences between the approaches (\ref{def:colors}) are not sizeable.
We see in figure ~\ref{Fig:cos3cos2-7-35-35}~(R) that this remains true when taking into account the $\sqrt{s_0}$ and $\mu_F$ dependency.

\def\scalecos{1}
\psfrag{cos}{\raisebox{0.2cm}{\scalebox{\scalecos}{$\frac{\overline{\langle \cos (3 \varphi) \rangle}_{\rm bin}}
{\overline{\langle \cos (2 \varphi) \rangle}_{\rm bin}}
$}}}

\subsection{Azimuthal distribution}
\label{SubSec:azimuthal_dist}

In practice, the observable which is most directly accessible in experiments is the azimuthal distribution of the two jets, defined as
\beq
\frac{1}{{\sigma}}\frac{d{\sigma}}{d \varphi}
~=~ \frac{1}{2\pi}
\left\{1+2 \sum_{n=1}^\infty \cos{\left(n \varphi\right)}
\left<\cos{\left( n \varphi \right)}\right>\right\}\,.
\label{dist-ang}
\eq
It is shown in figure~\ref{Fig:dphi-dist-7-35-35-LL-and-mixed-LL-Y} for LL, mixed jet LL with NLL Green's function, and mixed jet LL with collinear improved NLL Green's function, and
in figure~\ref{Fig:dphi-dist-7-35-35-NLL-Y} for pure NLL and  collinear improved NLL approaches.

The  figure~\ref{Fig:dphi-dist-7-35-35-LL-and-mixed-LL-Y} shows that the inclusion of NLL corrections to the Green's function leads to a smaller decorrelation compared to a pure LL treatment. Comparing  figure~\ref{Fig:dphi-dist-7-35-35-LL-and-mixed-LL-Y} with figure~\ref{Fig:dphi-dist-7-35-35-NLL-Y}, we see that the NLL corrections to the jet vertices lead to an even larger correlation, at fixed $Y$.
When increasing $Y\,,$ we can see on these plots that the decorrelation effect is slower, as expected in BFKL picture.

\def\scat{.8}
\def\scaf{1}

\psfrag{dist}{$\frac{1}{{\sigma}}\frac{d{\sigma}}{d \varphi}$}
\psfrag{phi}{$\varphi$}
\psfrag{Y_6.2}{\hspace{-.5cm}\scalebox{.52}{$Y = 6.2$}}
\psfrag{Y_7.2}{\hspace{-.5cm}\scalebox{.52}{$Y = 7.2$}}
\psfrag{Y_8.2}{\hspace{-.5cm}\scalebox{.52}{$Y = 8.2$}}
\begin{figure}[htbp]
\centerline{
\begin{tabular}{cc}
\hspace{-.1cm}
\psfrag{central}[r][l][\scat]{\footnotesize {pure LL}}
\scalebox{\scaf}{\includegraphics[width=7.5cm]{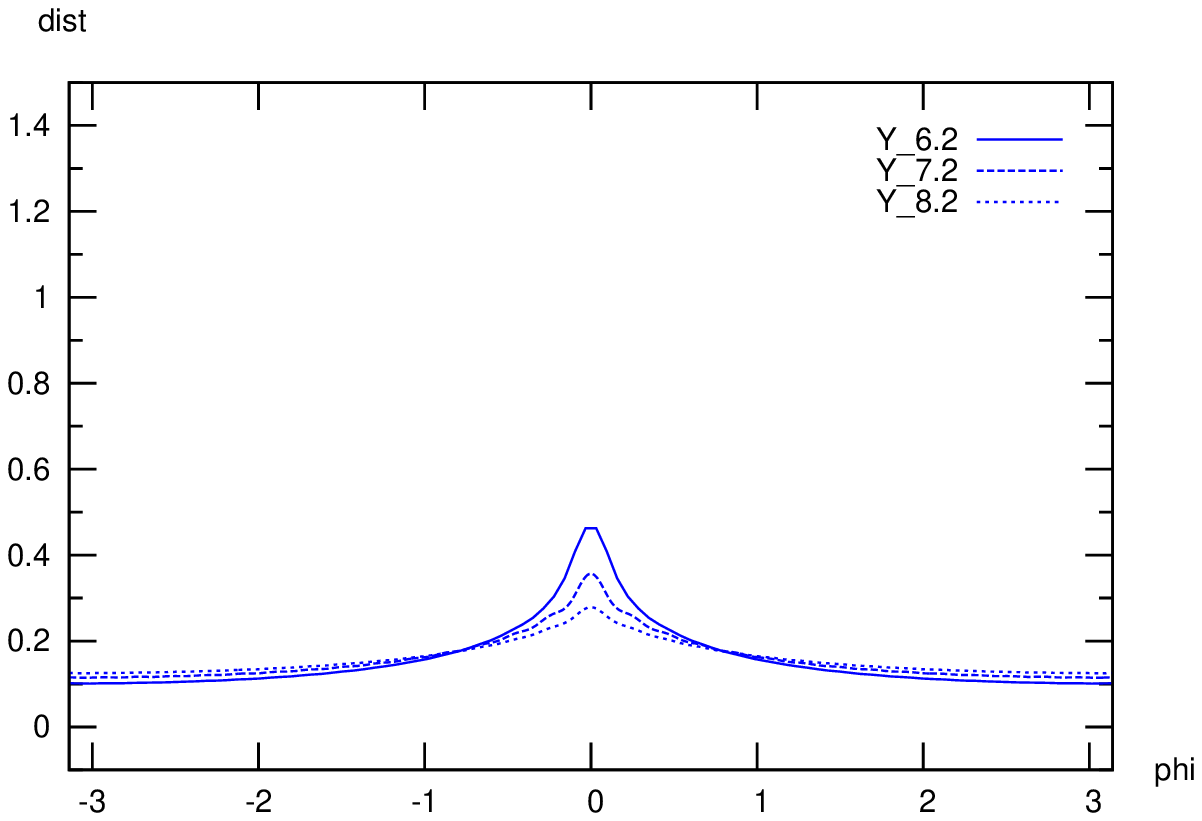}} 
& \hspace{-.8cm}
\psfrag{central}[r][l][\sca]{\footnotesize { LL vertices + NLL Green's fun.}}
\scalebox{\scaf}{\includegraphics[width=7.5cm]{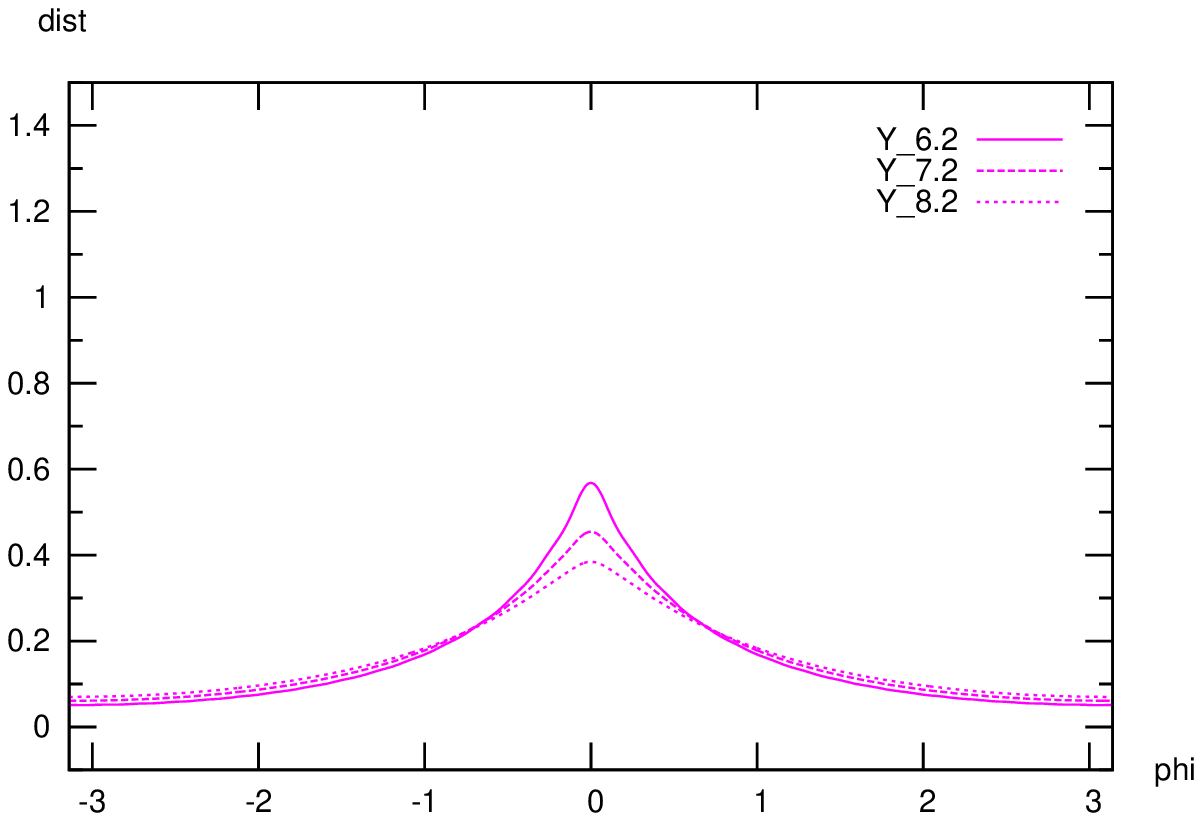}}
\\
\hspace{-.1cm}
\psfrag{central}[r][l][\scat]{\footnotesize { LL vert. + NLL resum. Green's fun.}}
\scalebox{\scaf}{\includegraphics[width=7.5cm]{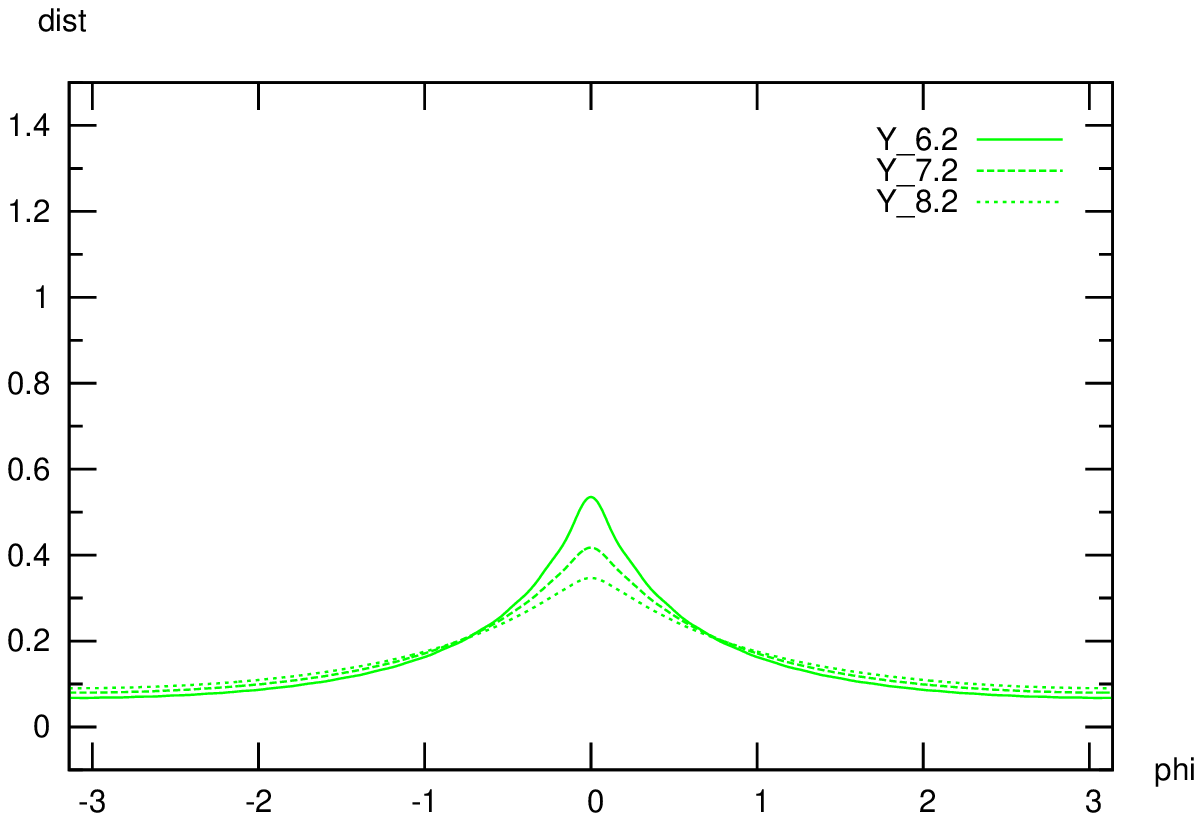}} 
\end{tabular}
}
\caption{The azimuthal distribution $\frac{1}{{\sigma}}\frac{d{\sigma}}{d \varphi}$, for the 3 first scenarios of (\ref{def:colors}), for 3 values of~$Y$.}
\label{Fig:dphi-dist-7-35-35-LL-and-mixed-LL-Y}
\end{figure}

\begin{figure}[htbp]
\centerline{
\begin{tabular}{cc}
\hspace{-.1cm}\psfrag{central}[r][l][\scat]{\footnotesize { NLL vert. + NLL Green's fun.}}
\scalebox{\scaf}{\includegraphics[width=7.5cm]{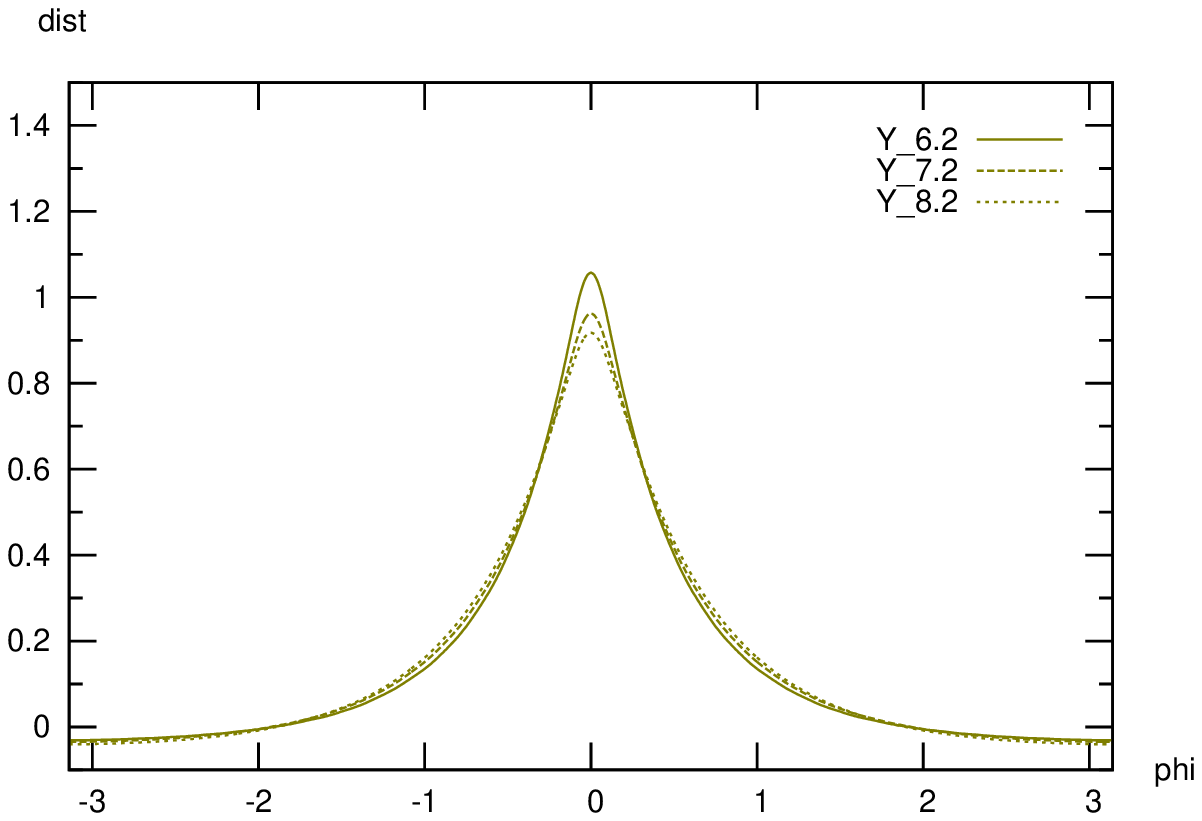}}
&\hspace{-.8cm}
\psfrag{central}[r][l][\scat]{\footnotesize { NLL vert. + NLL resum. Green's fun.}}
\scalebox{\scaf}{\includegraphics[width=7.5cm]{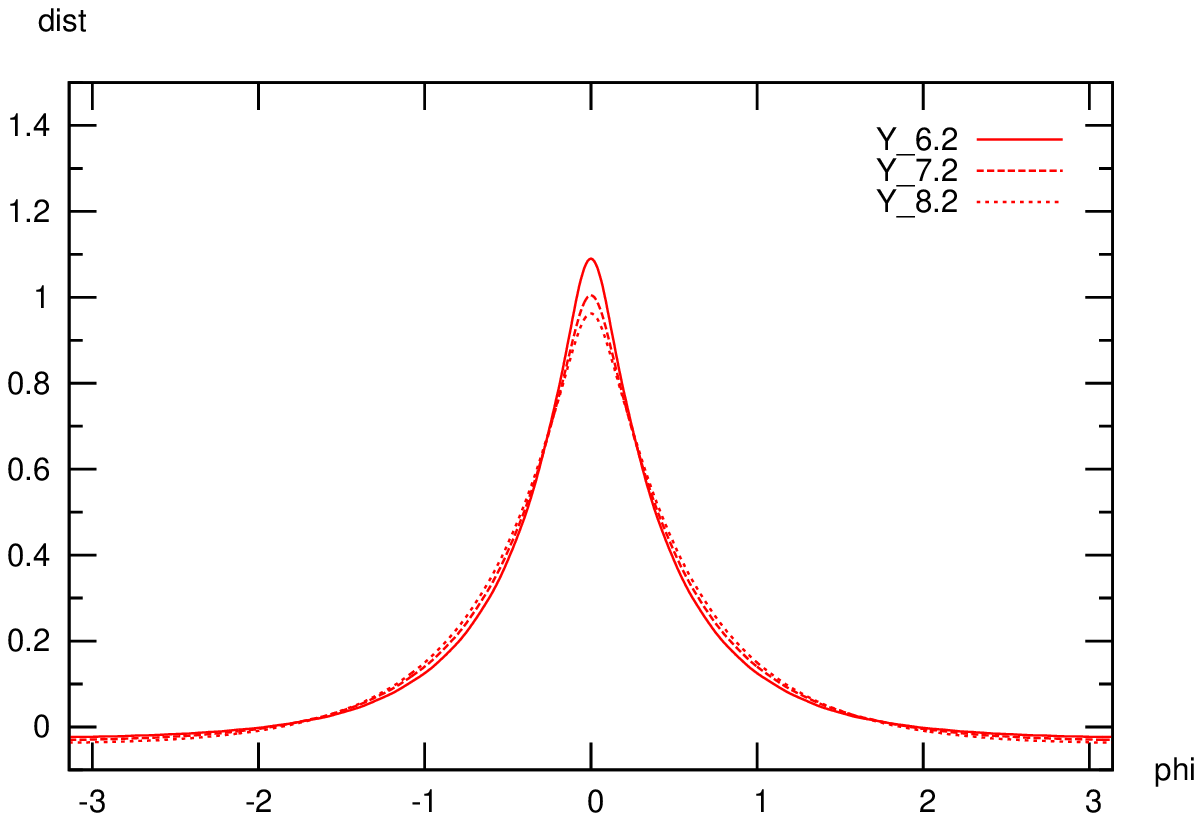}}
\end{tabular}
}
\caption{The azimuthal distribution $\frac{1}{{\sigma}}\frac{d{\sigma}}{d \varphi}$ for the 2 full NLL scenarios of (\ref{def:colors}), for 3 values of $Y$.}
\label{Fig:dphi-dist-7-35-35-NLL-Y}
\end{figure}


\psfrag{dist}[][][1]{$\frac{1}{\sigma}\frac{d \sigma}{d\varphi}$}
\psfrag{phi}[][][1]{\raisebox{-.5cm}{$\!\!\varphi$}}

\psfrag{muchange_0.5}[l][l][.6]{\footnotesize $ \mu_F \to \mu_F/2$}
\psfrag{muchange_2.0}[l][l][.6]{\footnotesize $\mu_F \to2 \mu_F$}
\psfrag{s0change_0.5}[l][l][.6]{\footnotesize \hspace{-.25cm}$\sqrt{s_0} \to \sqrt{s_0}/2$}
\psfrag{s0change_2.0}[l][l][.6]{\footnotesize \hspace{-.25cm}$\sqrt{s_0} \to 2 \sqrt{s_0}$}

\def\sca{.6}
\begin{figure}[htbp]
\centerline{
\begin{tabular}{cc}
\hspace{0cm}
\psfrag{central}[r][r][\sca]{pure LL\hbox{ \ \ \ }}
\scalebox{\scaf}{\includegraphics[width=7.5cm]{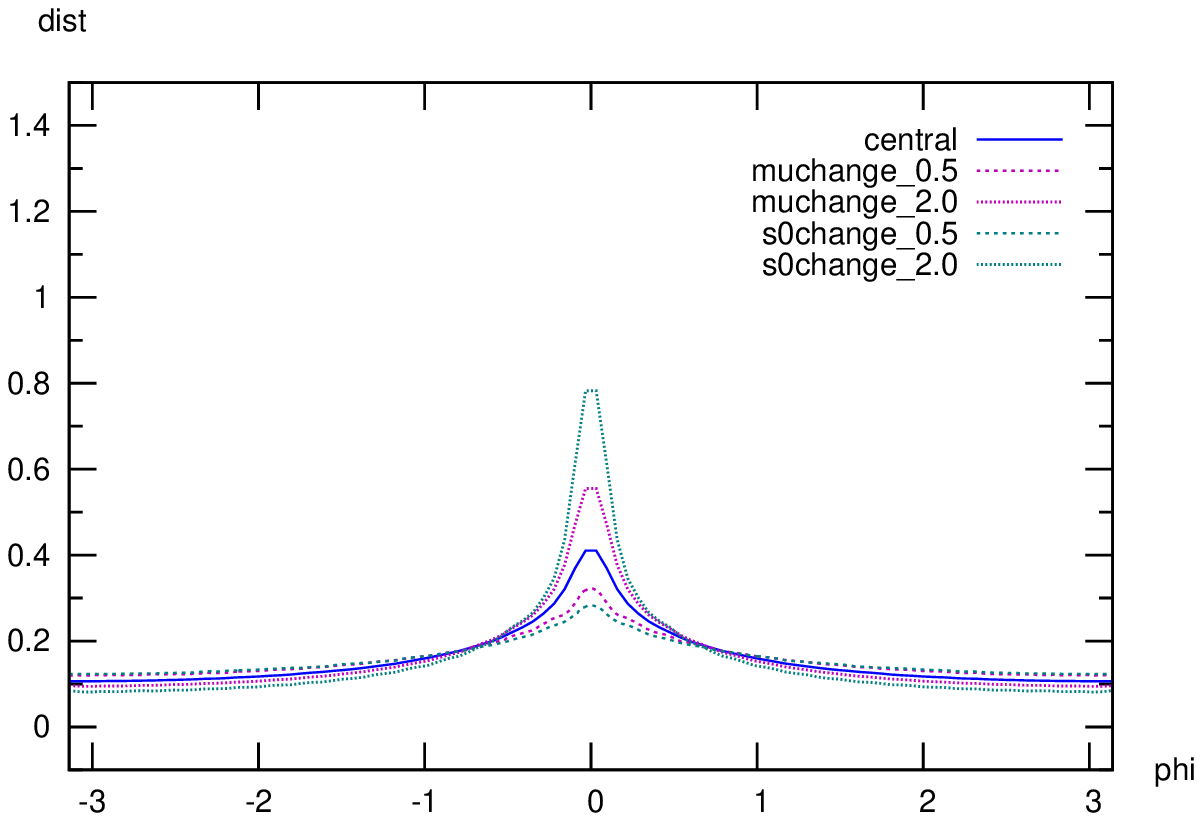}} 
& \hspace{-.7cm}
\psfrag{central}[r][r][\sca]{\hspace{.3cm}{ LL vertices + NLL Green's fun.\hbox{ \  }}}
\scalebox{\scaf}{\includegraphics[width=7.5cm]{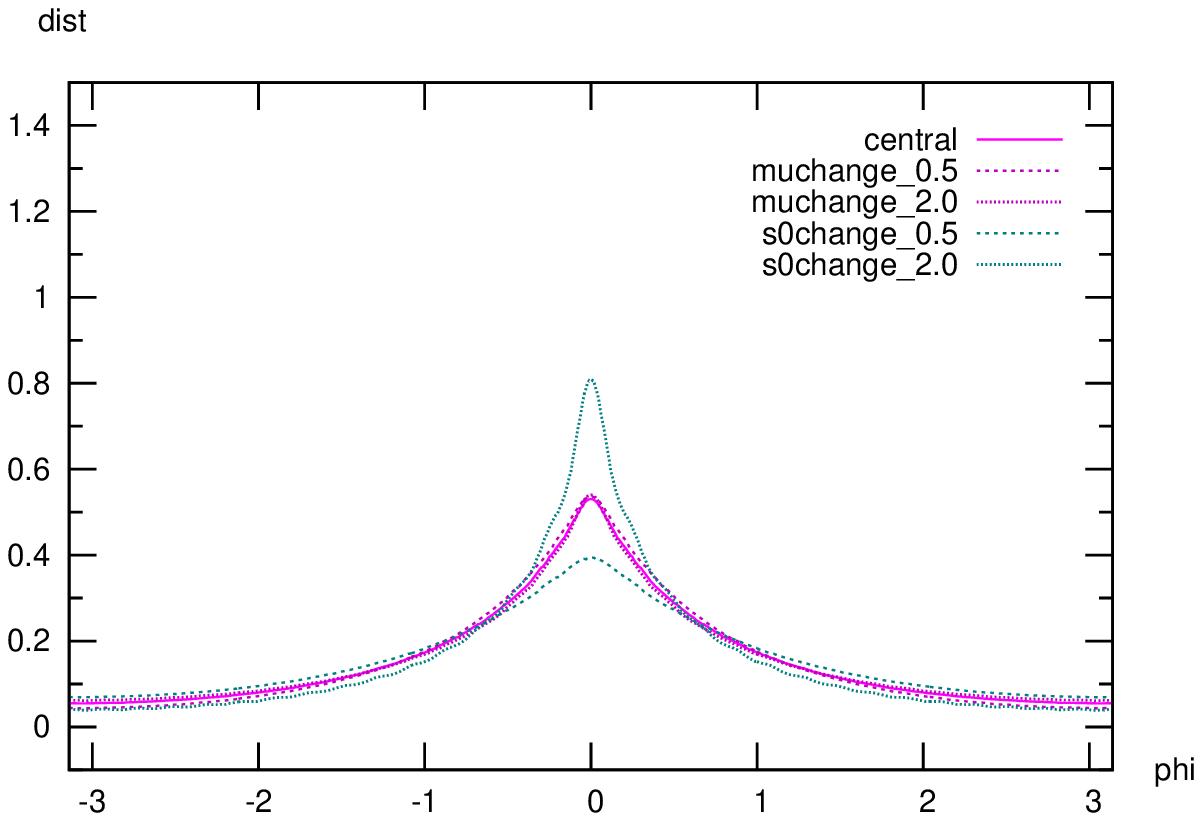}}\\
\hspace{0cm}
\psfrag{central}[r][r][\sca]{ LL vert. + NLL resum. Green's fun.\hbox{ \ }}
\scalebox{\scaf}{\includegraphics[width=7.5cm]{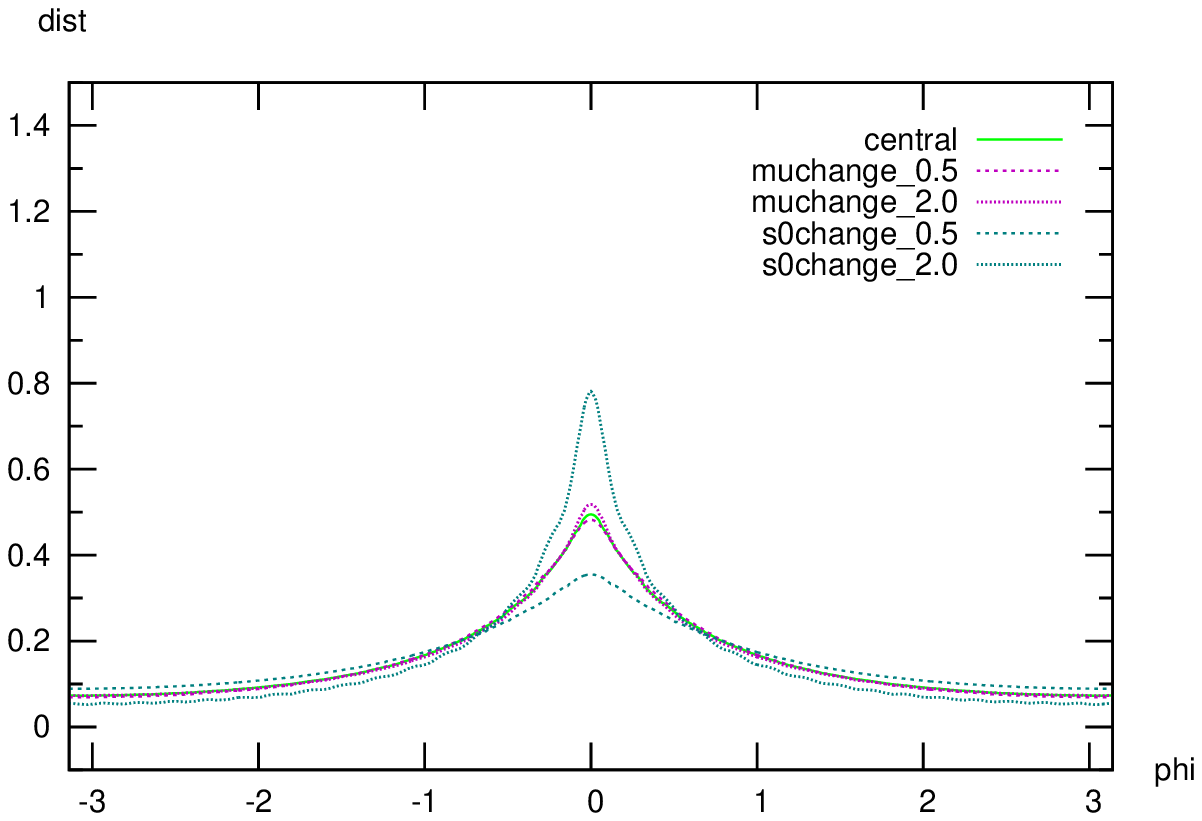}} 
\end{tabular}
}
\caption{The azimuthal distribution $\frac{1}{{\sigma}}\frac{d{\sigma}}{d \varphi}$ integrated over bins $35\,{\rm GeV} < |\veckjone|,|\veckjtwo| < 60 \,{\rm GeV}$, $0 < y_1, \, y_2 < 4.7$ and $6<Y<9.4$, for the 3 first scenarios of (\ref{def:colors}), including a variation of $\sqrt{s_0}$ and $\mu_F$ with a factor 2 with respect to the central values.}
\label{Fig:dphi-dist-7-35-35-LL-and-mixed-LL-parameters}
\end{figure}

\def\sca{.6}
\begin{figure}[htbp]
\centerline{
\begin{tabular}{cc}
\hspace{0.1cm}\psfrag{central}[r][r][\sca]{ NLL vert. + NLL Green's fun.\hbox{ \ }}
\scalebox{\scaf}{\includegraphics[width=7.5cm]{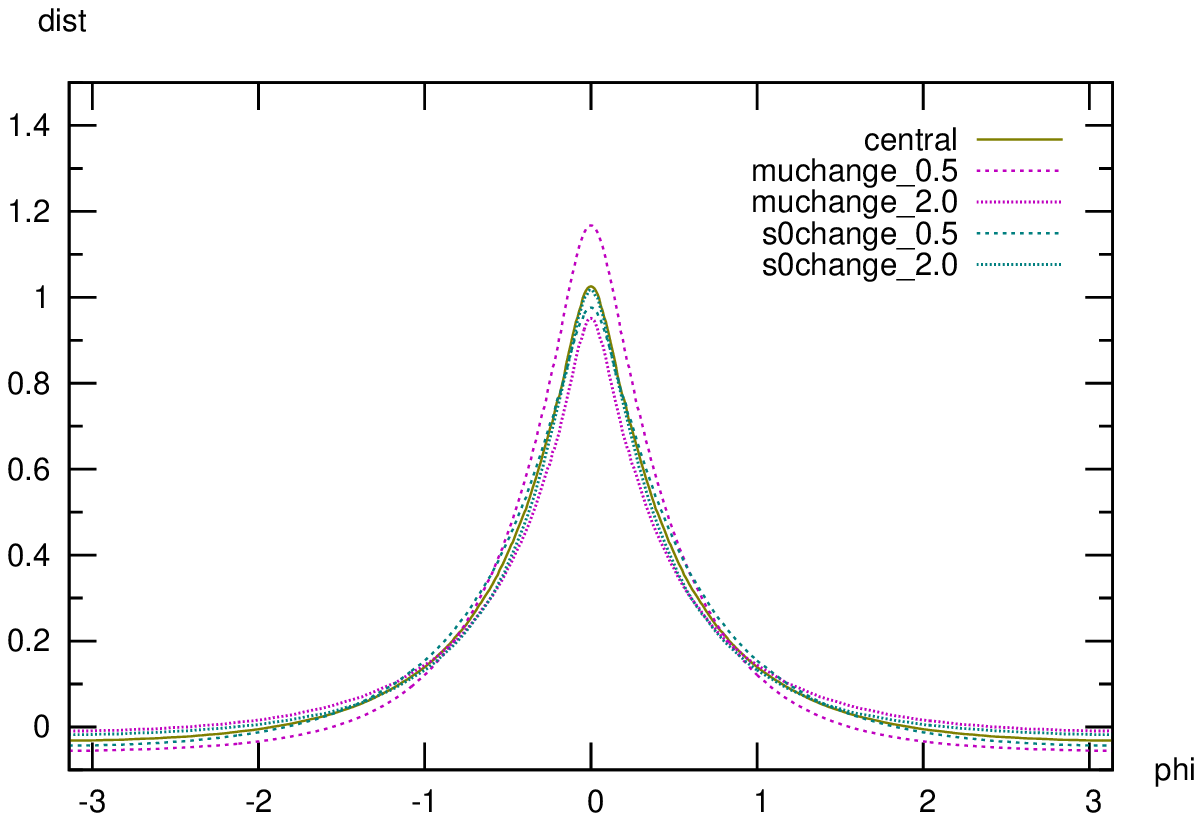}}
&\hspace{-.7cm}
\psfrag{central}[r][r][\sca]{NLL vert. + NLL resum. Green's fun.\hbox{ \ }}
\scalebox{\scaf}{\includegraphics[width=7.5cm]{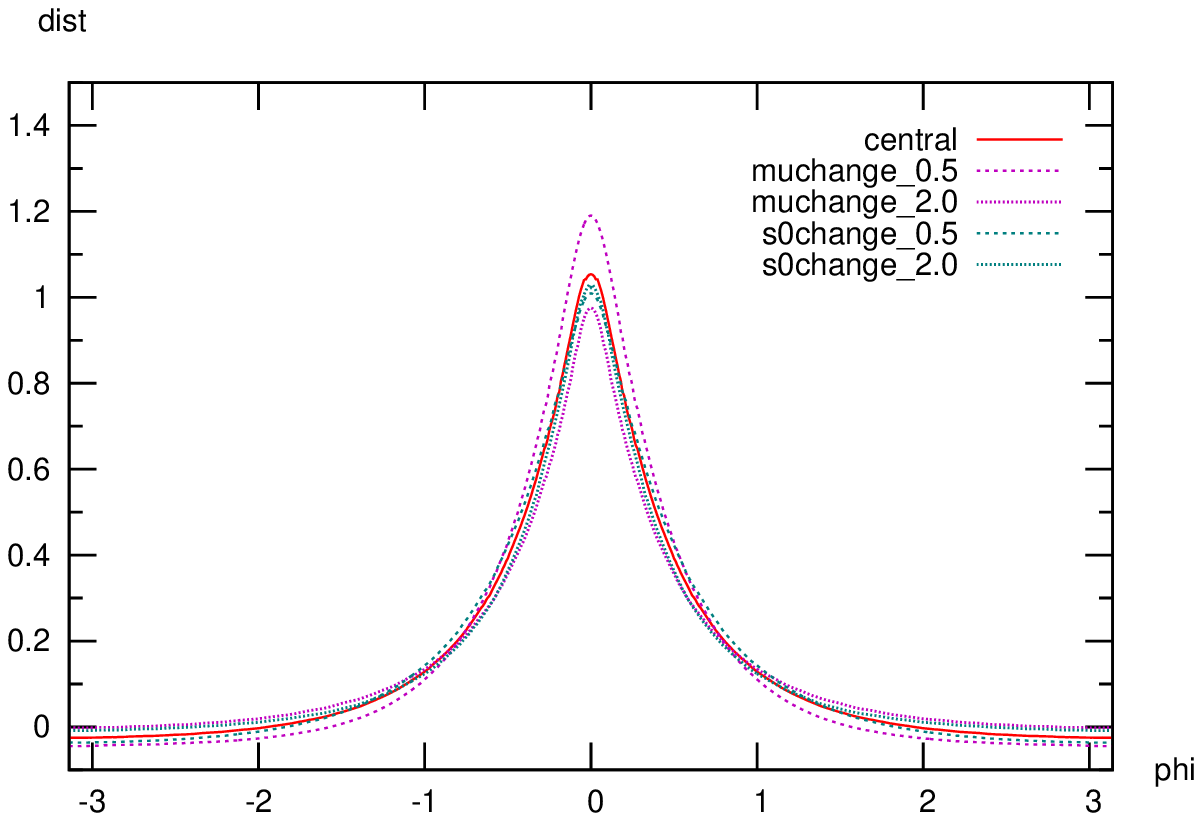}}
\end{tabular}
}
\caption{The azimuthal distribution $\frac{1}{{\sigma}}\frac{d{\sigma}}{d \varphi}$ integrated over bins $35\,{\rm GeV} < |\veckjone|,|\veckjtwo| < 60 \,{\rm GeV}$, $0 < y_1, \, y_2 < 4.7$ and $6<Y<9.4$, for the 2 full NLL scenarios of (\ref{def:colors}), including a variation of $\sqrt{s_0}$ and $\mu_F$ with a factor 2 with respect to the central values.}
\label{Fig:dphi-dist-7-35-35-NLL-parameters}
\end{figure}

We now integrate also over $Y$ in the range $6<Y<9.4$ in addition to $ |\veckjone|,|\veckjtwo|,y_1, \, y_2$.
The resulting azimuthal distribution is shown in 
 figure~\ref{Fig:dphi-dist-7-35-35-LL-and-mixed-LL-parameters} 
 for LL, mixed jet LL with NLL Green's function, and mixed jet LL with collinear improved NLL Green's function, and
in figure~\ref{Fig:dphi-dist-7-35-35-NLL-parameters} for the two full NLL scenarios of (\ref{def:colors}). In the same plots are shown the dependency with respect to $s_0$ and $\mu_F\,.$
We see that the pure LL approach is quite dependent on the scales $\sqrt{s_0}$ and $\mu_F$, whereas a mixed treatment using LL vertices with NLL Green's function shows a smaller dependency on $\mu_F$. The full NLL approaches are much more stable with respect to $\sqrt{s_0}$, while still $\mu_F$ dependent.


 \def\sca{.7}

\section{Results: asymmetric configuration}
\label{Sec:asym}

An asymmetric configuration with very different $k_{J {\rm min},\,1}$
and $k_{J {\rm min},\,2}$ allows us to compare our predictions with the ones obtained by fixed order NLO approaches, since it is known that symmetric configurations lead to unstable predictions in fixed order calculation~\cite{Andersen:2001kta,Fontannaz:2001nq}. Here we compare our predictions with the results obtained by the \textsc{Dijet} code \cite{Aurenche:2008dn}. Below we show the same observables as the ones which we considered in section~\ref{Sec:sym} for the symmetric configuration, now supplemented by a comparison with the \textsc{Dijet} predictions, for which we include a scale uncertainty on $\mu_F$ of a factor 2, for every plot\footnote{Note that the results obtained with \textsc{Dijet} use a scale $\mu_F=\frac{|\veckjone|+|\veckjtwo|}{2}$, which is very close in the domain we consider to the value $\sqrt{|\veckjone| \cdot |\veckjtwo|}$ we use in our BFKL calculation
}\!.
We consider bins with cuts
\beqa
  35\,{\rm GeV} < &|\veckjone|, |\veckjtwo| & < 60 \,{\rm GeV} \,, \nonumber\\
  50\,{\rm GeV} < &{\rm Max}(|\veckjone|, |\veckjtwo|)\,, \nonumber\\
 0 < &y_1, \, y_2& < 4.7\,.
\label{asym-cuts} 
\eqa
Since the cross-section is dominated by minimal allowed values of 
$|\veckjone|, |\veckjtwo|\,,$ such a choice of binning reduces the domain where $|\veckjone|$ and $|\veckjtwo|$ are very close to each other, for which unstable results at fixed order may be a source of worry.

As the behaviour of the BFKL results is very similar to the one for a symmetric configuration, we will mainly focus in this section on the comparison with \textsc{Dijet}.

\subsection{Cross-section}

\begin{figure}[htbp]
 \centerline{\includegraphics[width=10cm]{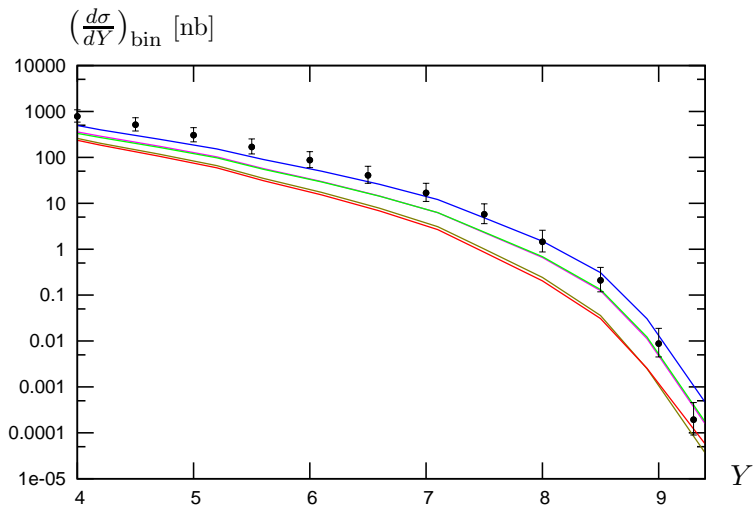}}
 \caption{Differential cross-section as a function of the jet rapidity separation $Y$, using cuts defined in (\ref{asym-cuts}). The different curves correspond to the 5 scenarios (\ref{def:colors}). The dots correspond to the predictions of the \textsc{Dijet} code.}
  \label{Fig:sigma-asym-7-35-35}
\end{figure}

On the figure~\ref{Fig:sigma-asym-7-35-35}, we show the cross-section. This figure shows surprising results: the fixed order results are above BFKL predictions, contrarily to the expectation. 
%
This difference remains valid after 
including the effect of possible variations of the parameters $s_0$ and $\mu_F\,,$ as can be seen in figure~\ref{Fig:sigma-asym-7-35-35-parameters}. 


\def\sca{.6}
\psfrag{central}[l][l][\sca]{ \hspace{-0.2cm}
pure NLL }
\psfrag{muchange_0.5}[l][l][\sca]{ $ \mu_F \to \mu_F/2$}
\psfrag{muchange_2.0}[l][l][\sca]{ $\mu_F \to2 \mu_F$}
\psfrag{s0change_0.5}[l][l][\sca]{ \hspace{-.17cm}$\sqrt{s_0} \to \sqrt{s_0}/2$}
\psfrag{s0change_2.0}[l][l][\sca]{ \hspace{-.17cm}$\sqrt{s_0} \to 2 \sqrt{s_0}$}
\begin{figure}[htbp]
\psfrag{Dijet}[l][l][\sca]{{\hspace{-.1cm} fixed order NLO}}
\centerline{\includegraphics[width=10.5cm]{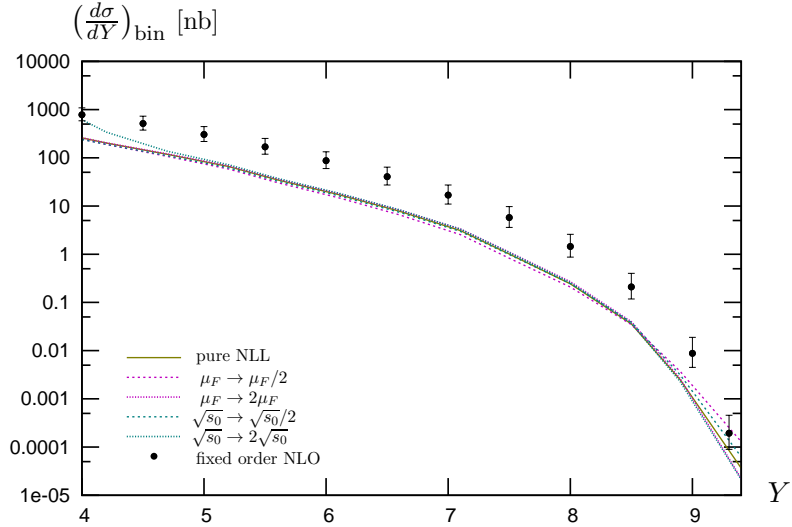}}
  \caption{Differential cross-section as a function of the jet rapidity separation $Y$, using cuts defined in (\ref{asym-cuts}), in the full NLL approximation. The dots correspond to the predictions of the \textsc{Dijet} code. We show the effect of a variation of $\sqrt{s_0}$ and $\mu_F$ with a factor 2 with respect to the central values.}
  \label{Fig:sigma-asym-7-35-35-parameters}
\end{figure}

\subsection{Azimuthal correlations}

\begin{figure}[htbp]
  \begin{minipage}{0.49\textwidth}
    \hspace{-.3cm}  \includegraphics[width=7.8cm]{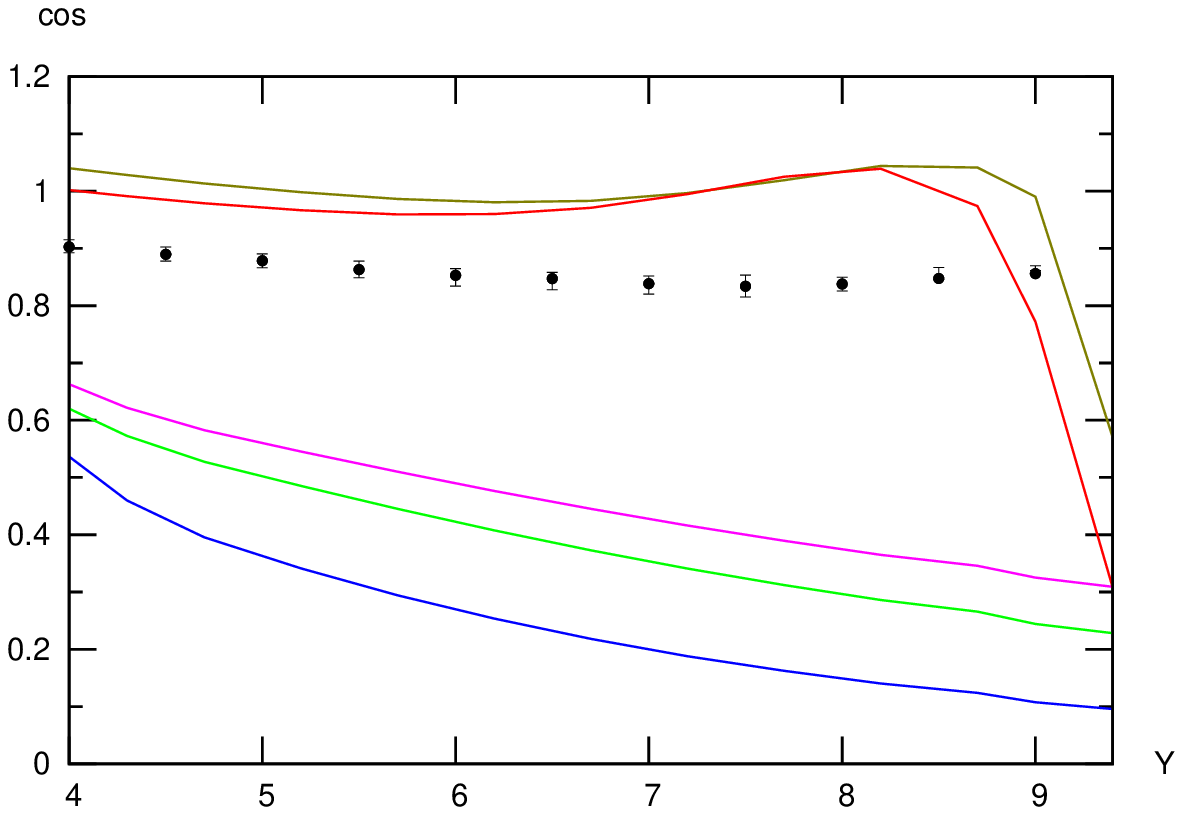}
  \end{minipage}
  \hspace{-.2cm}\begin{minipage}{0.49\textwidth}
      \includegraphics[width=7.8cm]{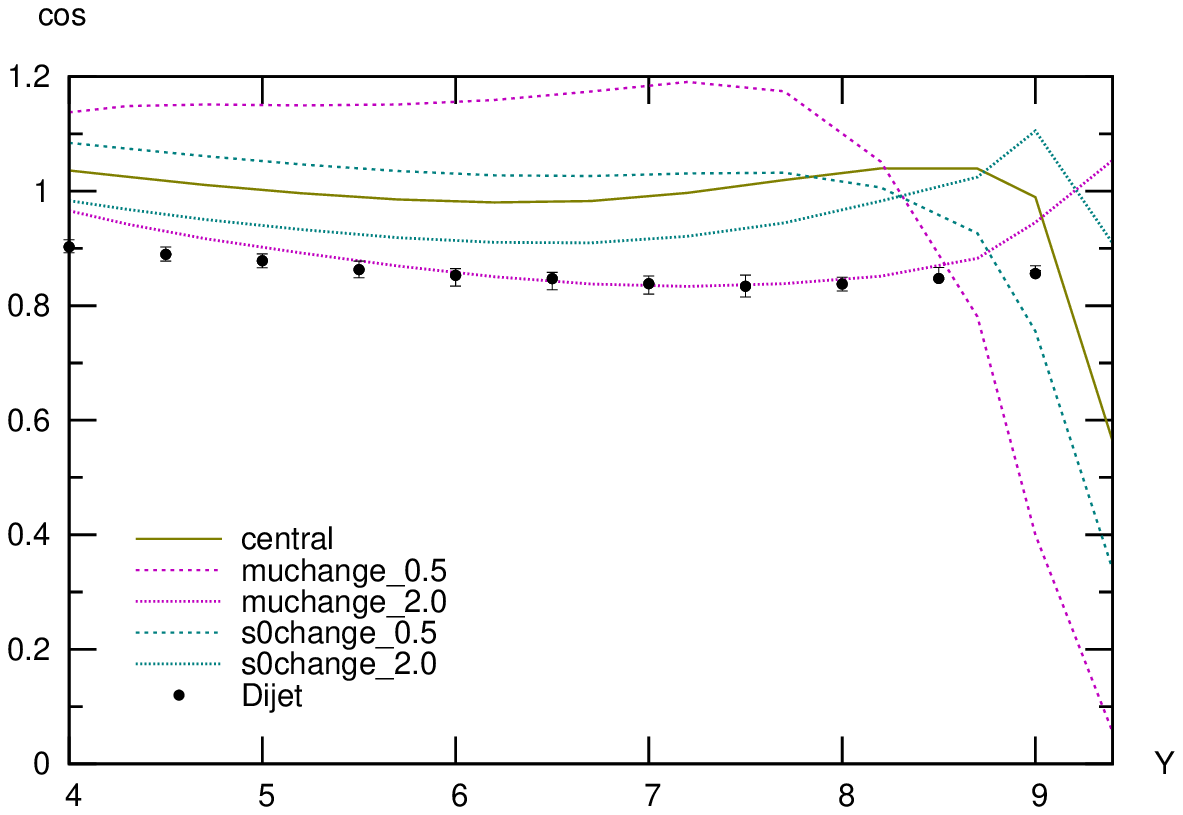}
  \end{minipage}
  \caption{Left: The bin averaged $\overline{\langle \cos \varphi \rangle}_{\rm bin}$ as a function of the jet rapidity separation $Y$, using cuts defined in (\ref{asym-cuts}), for the 5 scenarios described in the text, see (\ref{def:colors}). Right: Variation of $\overline{\langle \cos \varphi \rangle}_{\rm bin}$  when varying $\sqrt{s_0}$ and $\mu_F$ with a factor 2. The dots correspond to the predictions of the \textsc{Dijet} code.}  \label{Fig:cos-asym-7-35-35-parameters}
\end{figure}

We now consider the azimuthal correlation, for which we again compare predictions based on 5 kinds of BFKL scenarios with a  fixed order NLO prediction. The predictions for $\overline{\langle \cos \varphi \rangle}_{\rm bin}$ are displayed in figure~\ref{Fig:cos-asym-7-35-35-parameters}~(L). The two full NLL BFKL predictions (pure and collinearly improved) are noticeably above the fixed order NLO prediction. However, the figure~\ref{Fig:cos-asym-7-35-35-parameters}~(R) shows that the uncertainties with respect to $s_0$ and $\mu_F$ are quite significant, and do not allow to distinguish between the full NLL BFKL predictions and the  
fixed order NLO one.

\def\scalecos{.9}
\def\sca{.5}
\psfrag{central}[l][l][\sca]{ \hspace{-0.2cm}
pure NLL }
\psfrag{muchange_0.5}[l][l][\sca]{ $ \mu_F \to \mu_F/2$}
\psfrag{muchange_2.0}[l][l][\sca]{ $\mu_F \to2 \mu_F$}
\psfrag{s0change_0.5}[l][l][\sca]{ \hspace{-.17cm}$\sqrt{s_0} \to \sqrt{s_0}/2$}
\psfrag{s0change_2.0}[l][l][\sca]{ \hspace{-.17cm}$\sqrt{s_0} \to 2 \sqrt{s_0}$}
\psfrag{Dijet}[l][l][\sca]{ \hspace{-0.2cm}
fixed order NLO}

\psfrag{cos}{\raisebox{0cm}{\scalebox{\scalecos}{$\overline{\langle \cos \varphi \rangle}_{\rm bin}$}}}

\begin{figure}[htbp]
  \begin{minipage}{0.49\textwidth}
    \hspace{-.3cm}  \includegraphics[width=7.8cm]{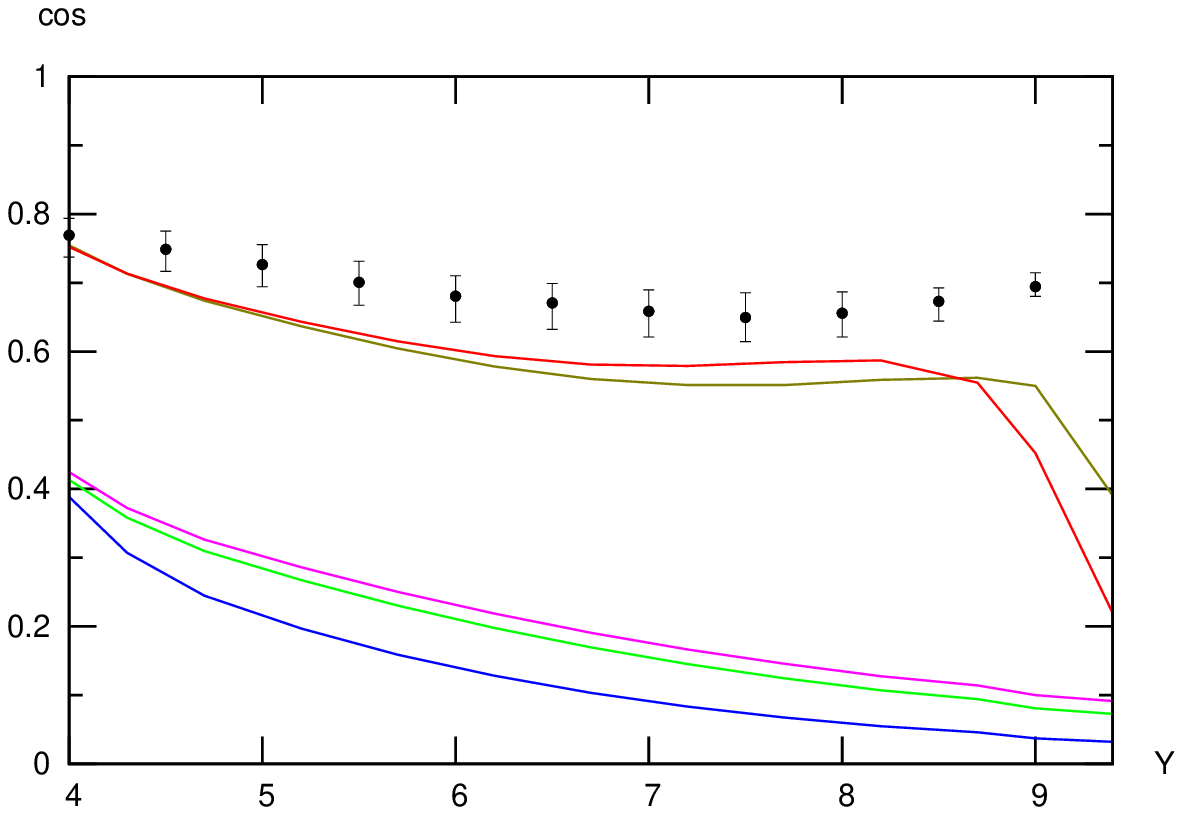}
  \end{minipage}
  \begin{minipage}{0.49\textwidth}
    \hspace{-.3cm}  \includegraphics[width=7.8cm]{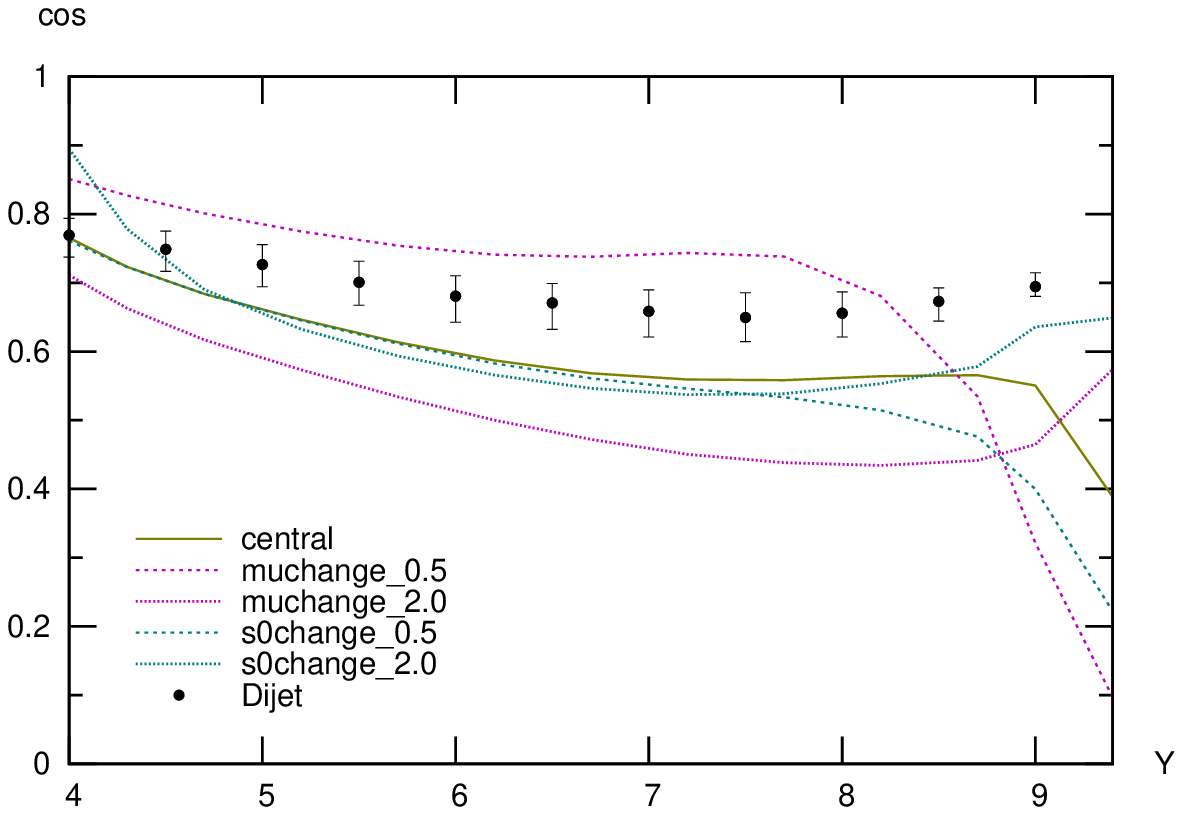}
  \end{minipage}
  \caption{Left: The bin averaged $\overline{\langle \cos (2\varphi) \rangle}_{\rm bin}$ as a function of the jet rapidity separation $Y$, using cuts defined in (\ref{asym-cuts}), for the 5 scenarios described in the text, see (\ref{def:colors}). Right: Variation of $\overline{\langle \cos (2\varphi) \rangle}_{\rm bin}$  when varying $\sqrt{s_0}$ and $\mu_F$ with a factor 2. The dots correspond to the predictions of the \textsc{Dijet} code.}
   \label{Fig:cos2-asym-7-35-35-parameters}
\end{figure}

We then display the predictions for $\overline{\langle \cos (2\varphi) \rangle}_{\rm bin}$ in figure~\ref{Fig:cos2-asym-7-35-35-parameters}~(L). The two full NLL BFKL predictions (pure and collinearly improved) are now a bit below
 the fixed order NLO prediction, and again, the figure~\ref{Fig:cos2-asym-7-35-35-parameters}~(R) shows that the uncertainties with respect to $s_0$ and $\mu_F$ are quite significant, and do not allow to distinguish between the full NLL BFKL predictions and the  
fixed order NLO one with this observable.


\def\sca{.48}
\psfrag{central}[l][l][\sca]{ \hspace{-0.2cm}
pure NLL }
\psfrag{muchange_0.5}[l][l][\sca]{ $ \mu_F \to \mu_F/2$}
\psfrag{muchange_2.0}[l][l][\sca]{ $\mu_F \to2 \mu_F$}
\psfrag{s0change_0.5}[l][l][\sca]{ \hspace{-.17cm}$\sqrt{s_0} \to \sqrt{s_0}/2$}
\psfrag{s0change_2.0}[l][l][\sca]{ \hspace{-.17cm}$\sqrt{s_0} \to 2 \sqrt{s_0}$}
\psfrag{Dijet}[l][l][\sca]{ \hspace{-0.2cm}
fixed order NLO}

\def\scalecos{.9}
\psfrag{cos}{\raisebox{0cm}{\scalebox{\scalecos}{$\overline{\langle \cos (2\varphi) \rangle}_{\rm bin}$}}}

\begin{figure}[htbp]
\vspace{.2cm}
   \begin{minipage}{0.49\textwidth}
   \hspace{-.3cm}  \includegraphics[width=7.8cm]{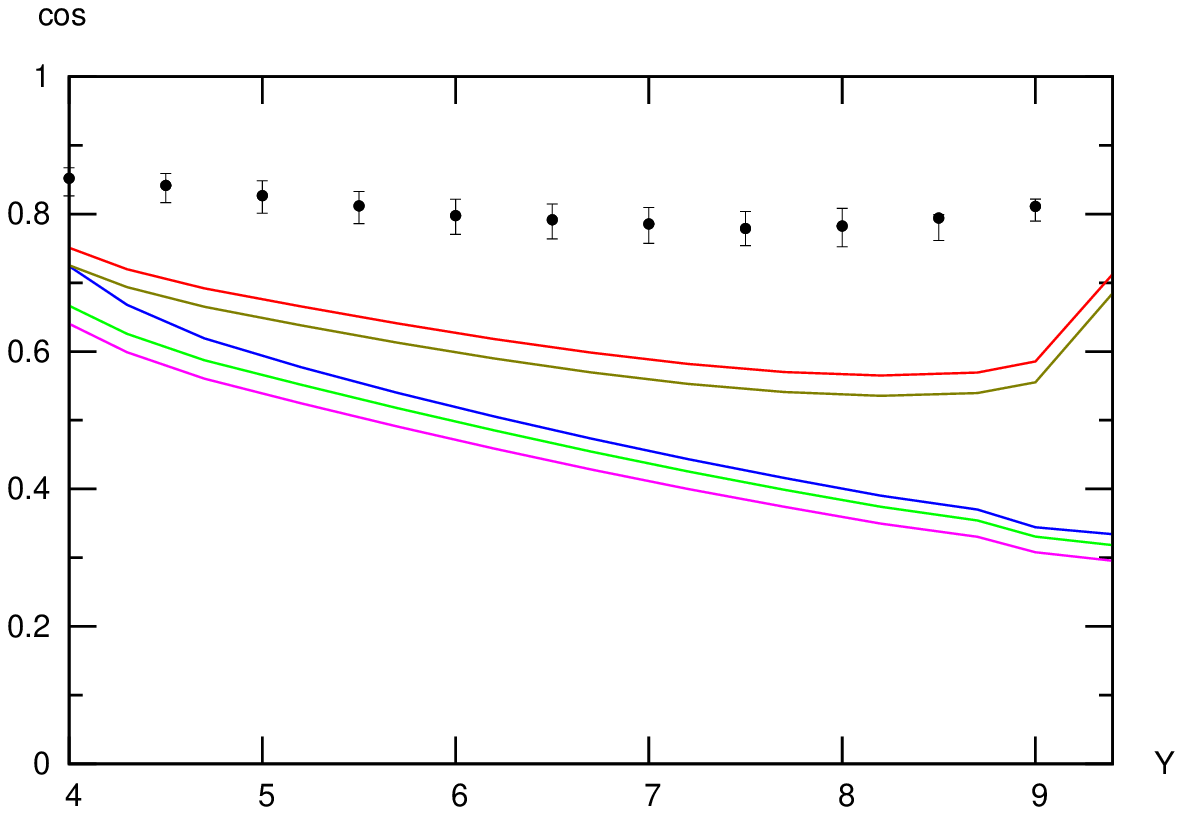}
  \end{minipage}
  \begin{minipage}{0.49\textwidth}
  \hspace{-.3cm}    \includegraphics[width=7.8cm]{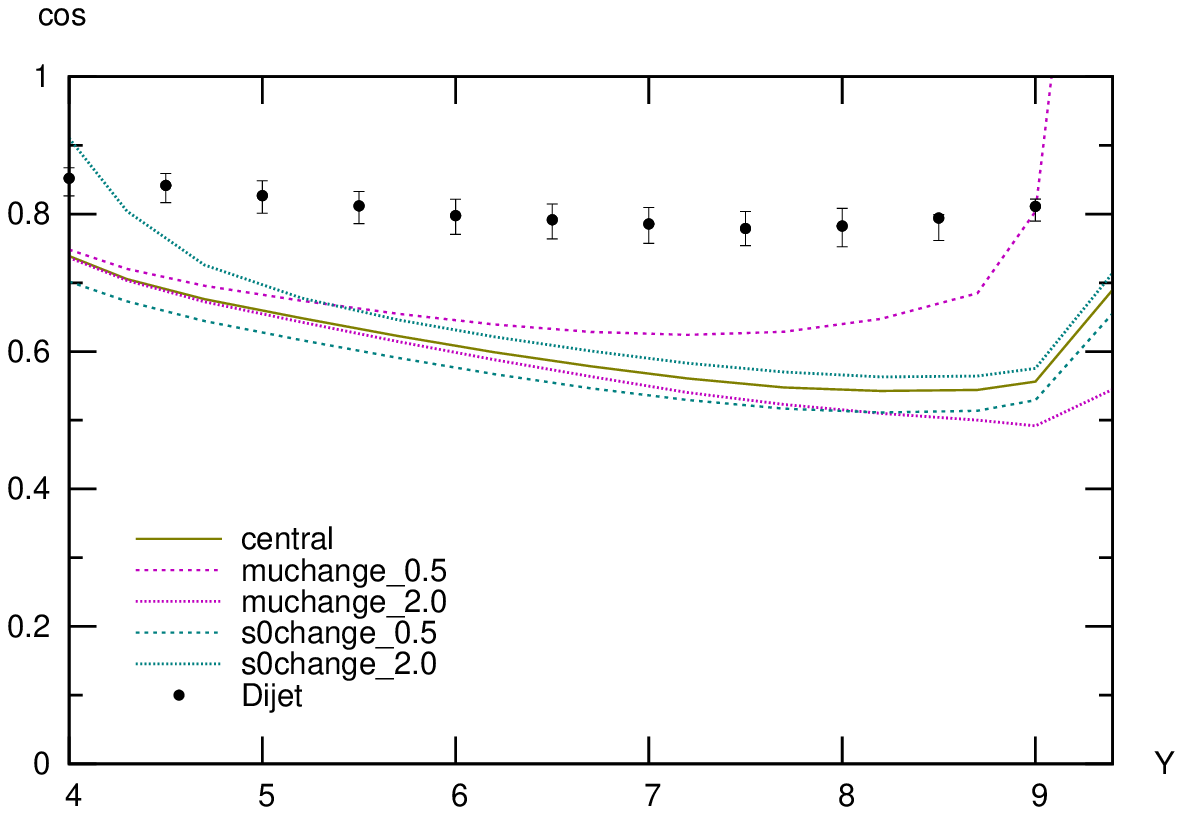}
  \end{minipage}
  \caption{Left: The bin averaged 
  $\overline{\langle \cos (2 \varphi) \rangle}_{\rm bin}/
\overline{\langle \cos \varphi \rangle}_{\rm bin}
$ as a function of the jet rapidity separation $Y$, using cuts defined in (\ref{asym-cuts}), for the 5 scenarios described in the text, see (\ref{def:colors}). Right: Variation of $\overline{\langle \cos (2 \varphi) \rangle}_{\rm bin}/
\overline{\langle \cos \varphi \rangle}_{\rm bin}
$  when varying $\sqrt{s_0}$ and $\mu_F$ with a factor 2. The dots correspond to the predictions of the \textsc{Dijet} code.}
\label{Fig:cos2cos-asym-7-35-35-parameters}
\end{figure}

Let us now study the observable $\frac{\overline{\langle \cos (2 \varphi) \rangle}_{\rm bin}}
{\overline{\langle \cos \varphi \rangle}_{\rm bin}}
\,.$ The fact that in the full NLL BFKL predictions, $\overline{\langle \cos \varphi \rangle}_{\rm bin}$  (resp.
$\overline{\langle \cos (2\varphi) \rangle}_{\rm bin}$) are above (resp. below) the fixed NLO order predictions now leads to a ratio which is very significantly, in the full NLL BFKL approximation,
under the NLO fixed order one, as can be seen from figure~\ref{Fig:cos2cos-asym-7-35-35-parameters}~(L). This difference is not washed out when including the uncertainties due to $s_0$ and $\mu_F$ variations, as shown in figure~\ref{Fig:cos2cos-asym-7-35-35-parameters}~(R). We want to emphasize the fact that
this is valid in particular in the region $Y \sim 6$, for which according to the discussion of section~\ref{SubSec:energy-momentum}, the corrections due to energy-momentum conservation are not expected to be very significant.

\def\scalecos{1}

\def\sca{.48}
\psfrag{central}[l][l][\sca]{ \hspace{-0.2cm}
pure NLL }
\psfrag{muchange_0.5}[l][l][\sca]{ $ \mu_F \to \mu_F/2$}
\psfrag{muchange_2.0}[l][l][\sca]{ $\mu_F \to2 \mu_F$}
\psfrag{s0change_0.5}[l][l][\sca]{ \hspace{-.17cm}$\sqrt{s_0} \to \sqrt{s_0}/2$}
\psfrag{s0change_2.0}[l][l][\sca]{ \hspace{-.17cm}$\sqrt{s_0} \to 2 \sqrt{s_0}$}
\psfrag{Dijet}[l][l][\sca]{ \hspace{-0.2cm}
fixed order NLO}
\psfrag{cos}{\raisebox{0.2cm}{\scalebox{\scalecos}{$\frac{\overline{\langle \cos (2 \varphi) \rangle}_{\rm bin}}
{\overline{\langle \cos \varphi \rangle}_{\rm bin}}
$}}}

\begin{figure}[htbp]
  \begin{minipage}{0.49\textwidth}
 \hspace{-.3cm}     \includegraphics[width=7.8cm]{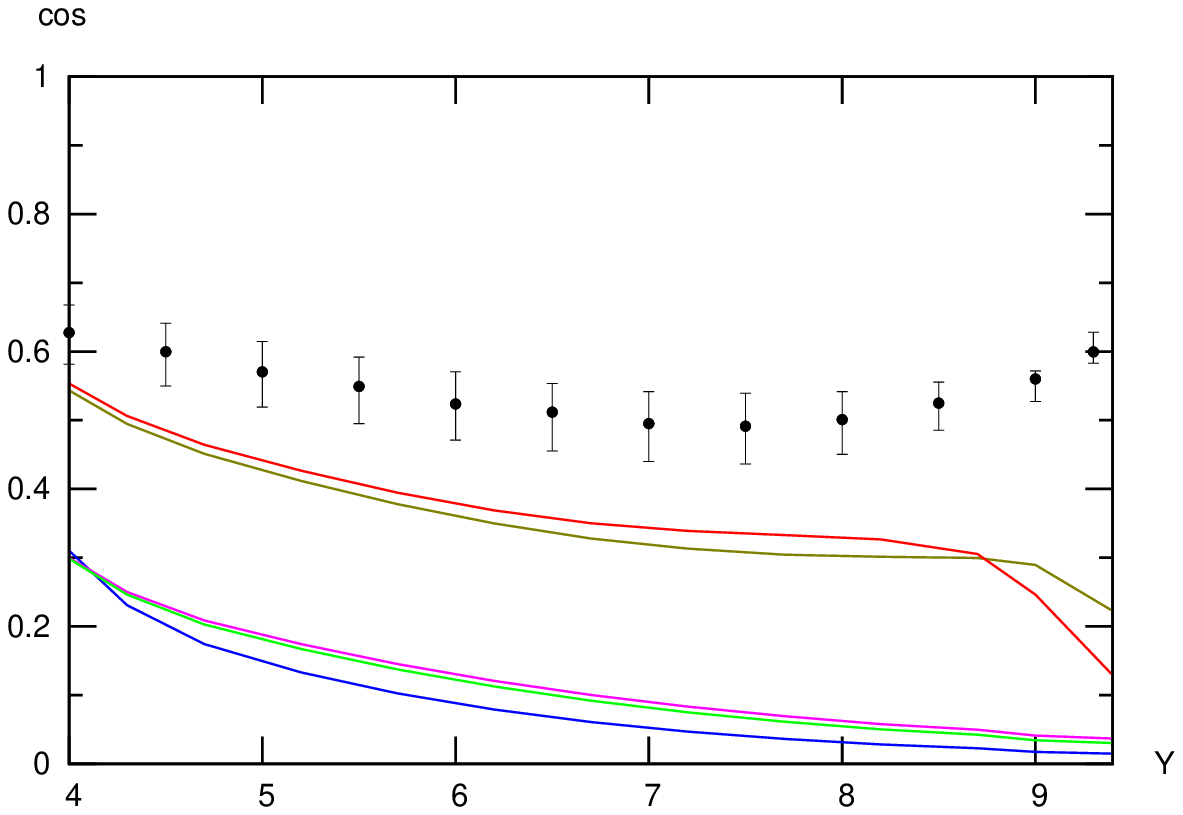}
  \end{minipage}
  \begin{minipage}{0.49\textwidth}
   \hspace{-.3cm}    \includegraphics[width=7.8cm]{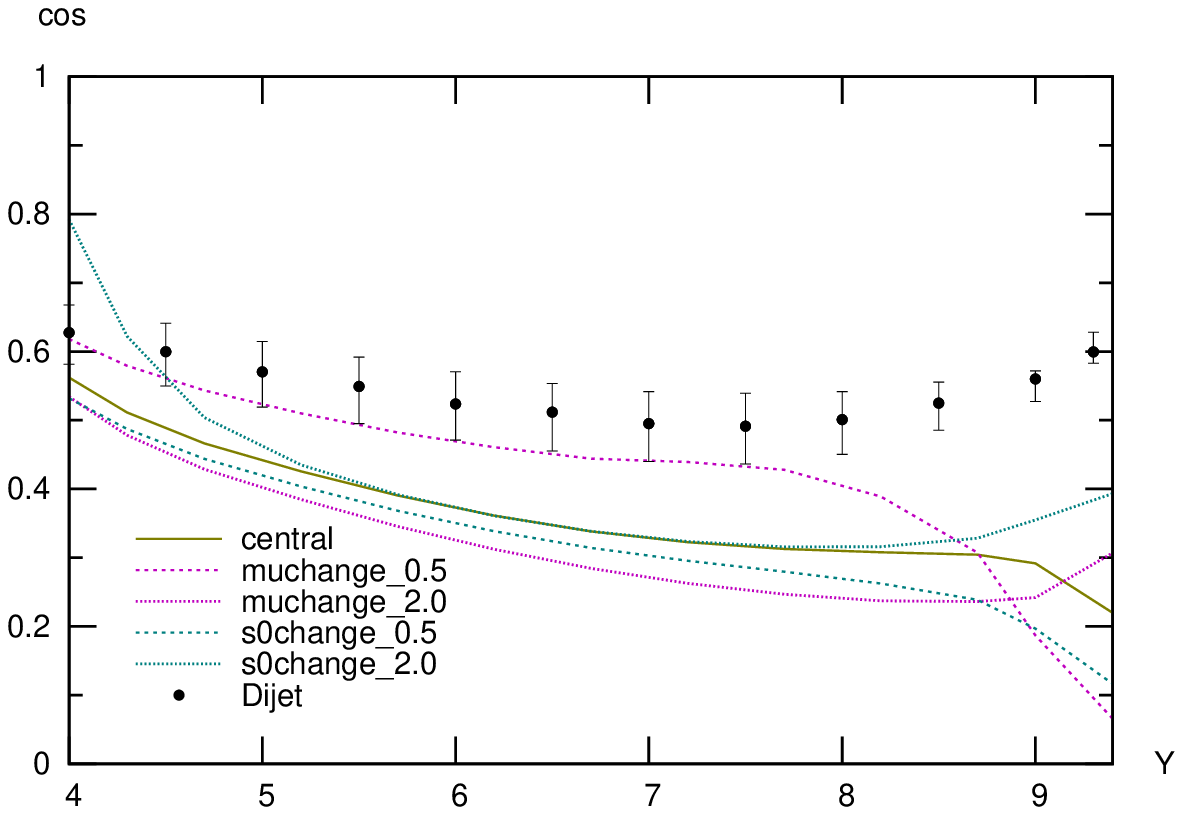}
  \end{minipage}
  \caption{Left: The bin averaged $\overline{\langle \cos (3\varphi) \rangle}_{\rm bin}$ as a function of the jet rapidity separation $Y$, using cuts defined in (\ref{asym-cuts}), for the 5 scenarios described in the text, see (\ref{def:colors}). Right: Variation of $\overline{\langle \cos (3\varphi) \rangle}_{\rm bin}$  when varying $\sqrt{s_0}$ and $\mu_F$ with a factor 2. The dots correspond to the predictions of the \textsc{Dijet} code.}
   \label{Fig:cos3-asym-7-35-35-parameters}
\end{figure}

We finally consider  the observable $\overline{\langle \cos (3\varphi) \rangle}_{\rm bin}$ in figure~\ref{Fig:cos3-asym-7-35-35-parameters}~(L). The two full NLL BFKL predictions (pure and collinearly improved) are now significantly below
 the fixed order NLO prediction. The figure~\ref{Fig:cos3-asym-7-35-35-parameters}~(R) shows that the uncertainties with respect to $s_0$ and $\mu_F$ are quite significant, and can marginally alter the possibility of distinguishing the two types of scenarios, mainly due to the $\mu_F$ uncertainty. Considering
 the ratio $\frac{\overline{\langle \cos (3 \varphi) \rangle}_{\rm bin}}
{\overline{\langle \cos (2 \varphi) \rangle}_{\rm bin}}
\,,$ we observe that this latter observable is much more favorable, as can be seen from figure~\ref{Fig:cos3-asym-7-35-35-parameters}.
  

\def\sca{.48}
\psfrag{central}[l][l][\sca]{ \hspace{-0.2cm}
pure NLL }
\psfrag{muchange_0.5}[l][l][\sca]{ $ \mu_F \to \mu_F/2$}
\psfrag{muchange_2.0}[l][l][\sca]{ $\mu_F \to2 \mu_F$}
\psfrag{s0change_0.5}[l][l][\sca]{ \hspace{-.17cm}$\sqrt{s_0} \to \sqrt{s_0}/2$}
\psfrag{s0change_2.0}[l][l][\sca]{ \hspace{-.17cm}$\sqrt{s_0} \to 2 \sqrt{s_0}$}
\psfrag{Dijet}[l][l][\sca]{ \hspace{-0.2cm}
fixed order NLO}

\def\scalecos{.9}
\psfrag{cos}{\raisebox{0cm}{\scalebox{\scalecos}{$\overline{\langle \cos (3\varphi) \rangle}_{\rm bin}$}}}


\def\sca{.48}
\psfrag{central}[l][l][\sca]{ \hspace{-0.2cm}
pure NLL }
\psfrag{muchange_0.5}[l][l][\sca]{ $ \mu_F \to \mu_F/2$}
\psfrag{muchange_2.0}[l][l][\sca]{ $\mu_F \to2 \mu_F$}
\psfrag{s0change_0.5}[l][l][\sca]{ \hspace{-.17cm}$\sqrt{s_0} \to \sqrt{s_0}/2$}
\psfrag{s0change_2.0}[l][l][\sca]{ \hspace{-.17cm}$\sqrt{s_0} \to 2 \sqrt{s_0}$}
\psfrag{Dijet}[l][l][\sca]{ \hspace{-0.2cm}
fixed order NLO}

\def\scalecos{1}
\psfrag{cos}{\raisebox{0.2cm}{\scalebox{\scalecos}{$\frac{\overline{\langle \cos (3 \varphi) \rangle}_{\rm bin}}
{\overline{\langle \cos (2 \varphi) \rangle}_{\rm bin}}
$}}}
\begin{figure}[htbp]
\vspace{.2cm}
  \begin{minipage}{0.49\textwidth}
 \hspace{-.3cm}     \includegraphics[width=7.8cm]{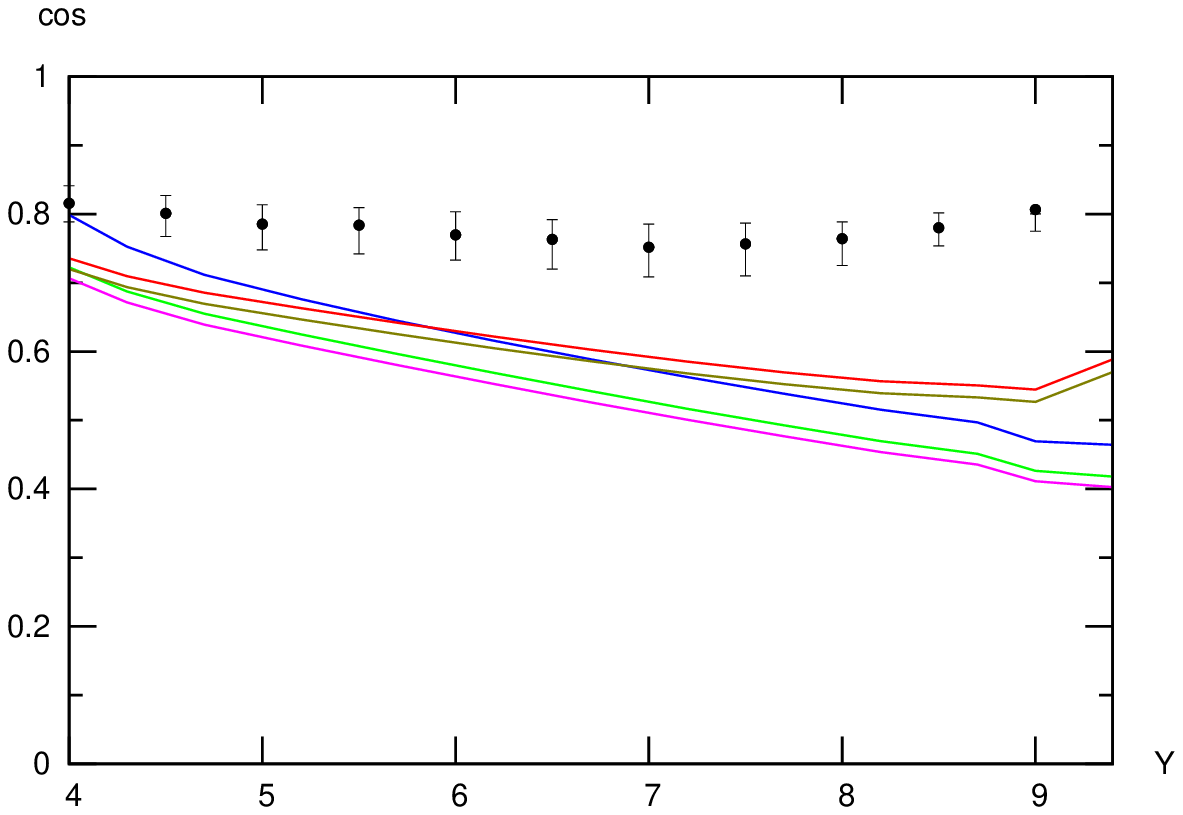}
  \end{minipage}
  \begin{minipage}{0.49\textwidth}
  \hspace{-.3cm}     \includegraphics[width=7.8cm]{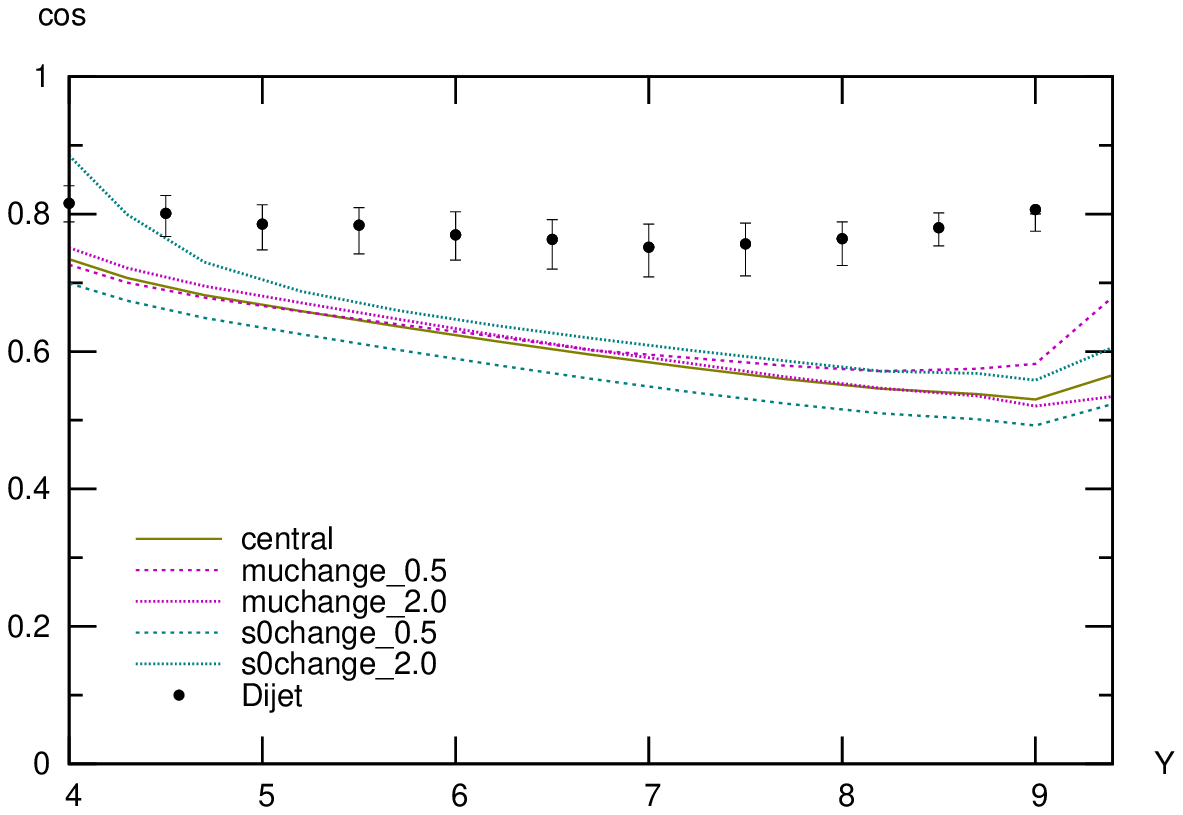}
  \end{minipage}
  \caption{Left: The bin averaged 
  $\overline{\langle \cos (3 \varphi) \rangle}_{\rm bin}/
\overline{\langle \cos (2 \varphi) \rangle}_{\rm bin}
$ as a function of the jet rapidity separation $Y$, using cuts defined in (\ref{asym-cuts}), for the 5 scenarios described in the text, see (\ref{def:colors}). Right: Variation of $\overline{\langle \cos (3 \varphi) \rangle}_{\rm bin}/
\overline{\langle \cos (2 \varphi) \rangle}_{\rm bin}
$  when varying $\sqrt{s_0}$ and $\mu_F$ with a factor 2. The dots correspond to the predictions of the \textsc{Dijet} code.}
\label{Fig:cos3cos2-asym-7-35-35-parameters}
\end{figure}
We do not study here the azimuthal distribution as we have done
in the symmetric case of section~\ref{SubSec:azimuthal_dist}. Indeed it is not possible to confront our BFKL predictions with the fixed order NLO predictions of the \textsc{Dijet} code, due to instabilities when evaluating the higher harmonics, which are necessary 
for a precise study of $\varphi$ distribution, using the \textsc{Dijet} code.

\section{Limit of small-R cone}

A detailed study, based on the work of ref.~\cite{Ivanov:2012ms} where the jet vertices were computed in an approximated small $R$ treatment, shows that the difference between an exact treatment, as used in the present work, and that small $R$ approximation is small. This is illustrated in figure~\ref{Fig:R-0.3} for $R=0.3$ and in figure~\ref{Fig:R-0.5} for $R=0.5\,,$ 
and shows explicitly the consistency of the two approaches. This small $R$ limit has been used in ref.~\cite{Caporale:2012ih} for phenomenological studies\footnote{We thank the authors of ref.~\cite{Caporale:2012ih} for pointing out that the values for $\mathcal{C}_1$ and $\mathcal{C}_2$ given in ref.~\cite{Colferai:2010wu} should be multiplied by a factor of 2 to get the proper normalization.}\!. In this paper, it is stated that  sizeable differences are obtained when comparing with the results we got in ref.~\cite{Colferai:2010wu}, with the same set of parameters. We believe that this is mainly due to the way NNLL corrections, which are beyond the precision of both studies, are treated. Indeed, when convoluting jet vertices with the Green's function, there is a freedom to neglect terms of magnitude $\alpha_s^3 \, Y\,.$ A close inspection on the way both papers deal with such contributions shows that in ref.~\cite{Caporale:2012ih}, eq.~(42), terms involving the product of NLL corrections in both vertices 
are 
explicitly neglected, while they are kept in ref.~\cite{Colferai:2010wu} and in the present study.  

\def\scalecos{1}

\def\scalecos{1}
\begin{figure}[htbp]
  \begin{minipage}{0.49\textwidth}
      \psfrag{deltasigma}{\scalebox{\scalecos}{\raisebox{0.1cm}{$\frac{\Delta \sigma}{\sigma}$}}}
      \psfrag{Y}{\scalebox{0.9}{$Y$}}
     \includegraphics[width=7cm]{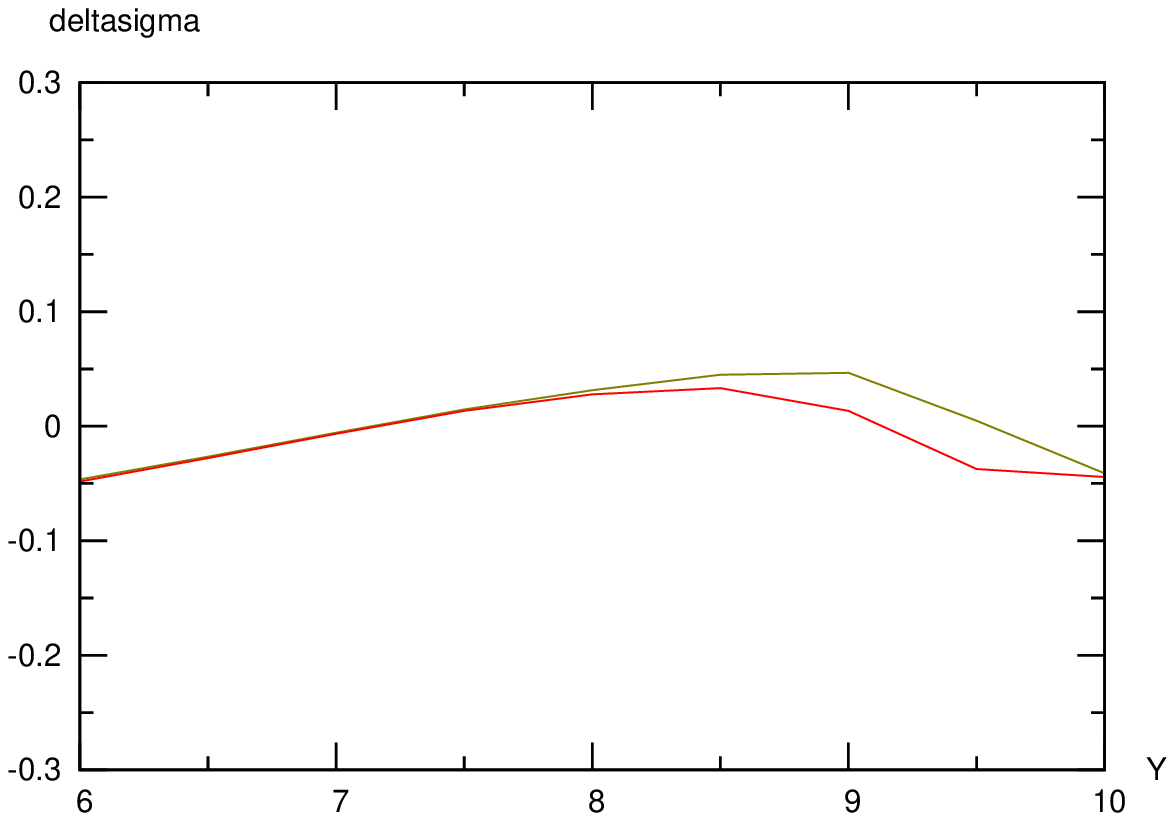}
  \end{minipage}
  \begin{minipage}{0.49\textwidth}
      \psfrag{deltacos}{\scalebox{\scalecos}{\raisebox{0.1cm}{$\frac{\Delta \langle \cos \varphi \rangle}{\langle \cos \varphi\rangle}$}}}
      \psfrag{Y}{\scalebox{0.9}{$Y$}}
     \raisebox{0cm}{\includegraphics[width=7cm]{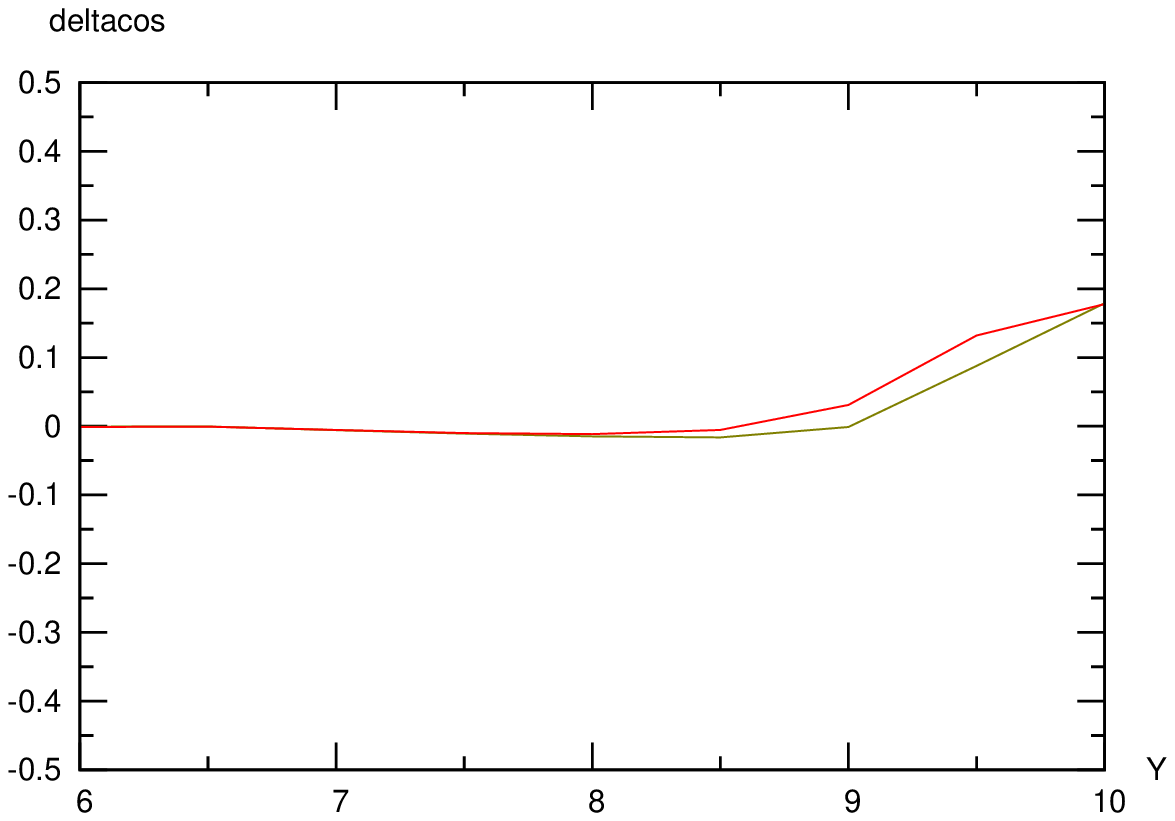}}
  \end{minipage}
  \caption{Relative difference between an exact treatment of the cone size and the small cone approximation for $R=0.3$ and 
  $|\veckjone|=|\veckjtwo|=35$~GeV\,, in the two full NLL BFKL scenarios.}
  \label{Fig:R-0.3}
\end{figure}


\begin{figure}[htbp]
  \begin{minipage}{0.49\textwidth}
   \psfrag{deltasigma}{\scalebox{\scalecos}{\raisebox{0.1cm}{$\frac{\Delta \sigma}{\sigma}$}}}
      \psfrag{Y}{\scalebox{0.9}{$Y$}}
    \includegraphics[width=7cm]{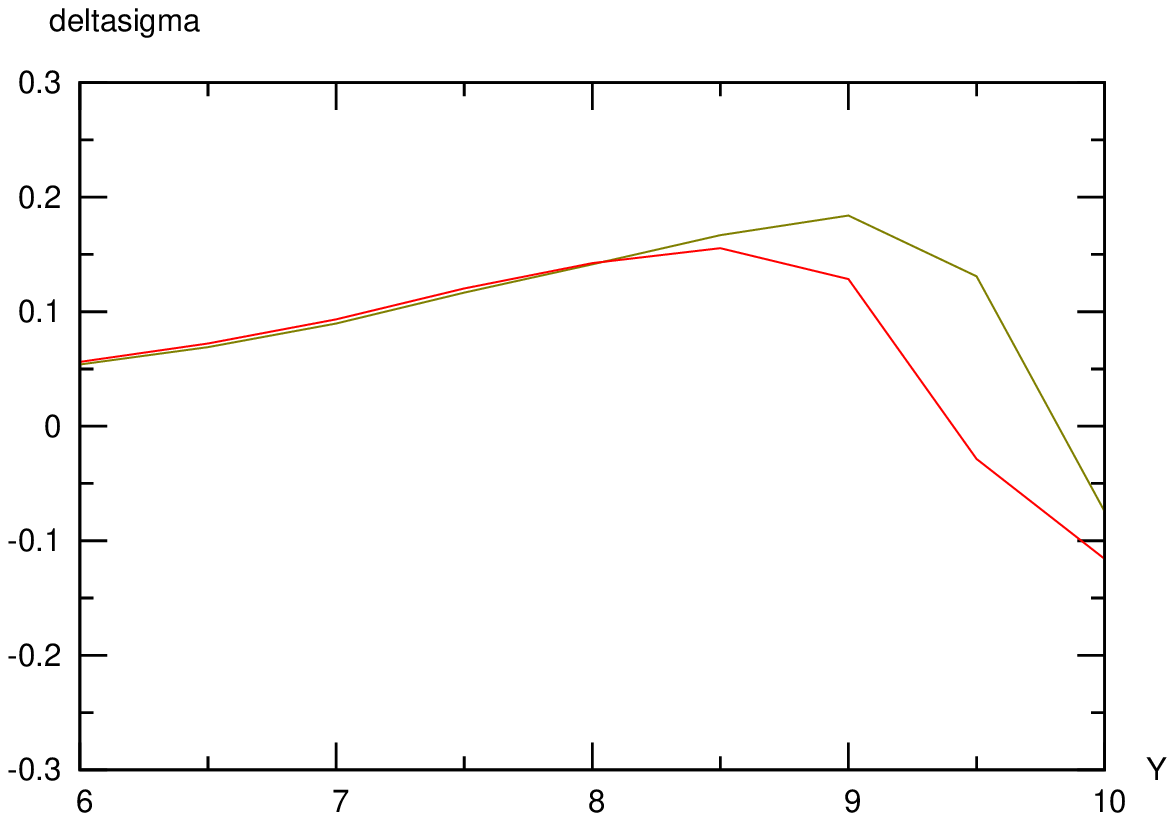}
  \end{minipage}
  \begin{minipage}{0.49\textwidth}
      \psfrag{deltacos}{\scalebox{\scalecos}{\raisebox{0.1cm}{$\frac{\Delta \langle \cos \varphi \rangle}{\langle \cos \varphi\rangle}$}}}
      \psfrag{Y}{\scalebox{0.9}{$Y$}}
    \includegraphics[width=7cm]{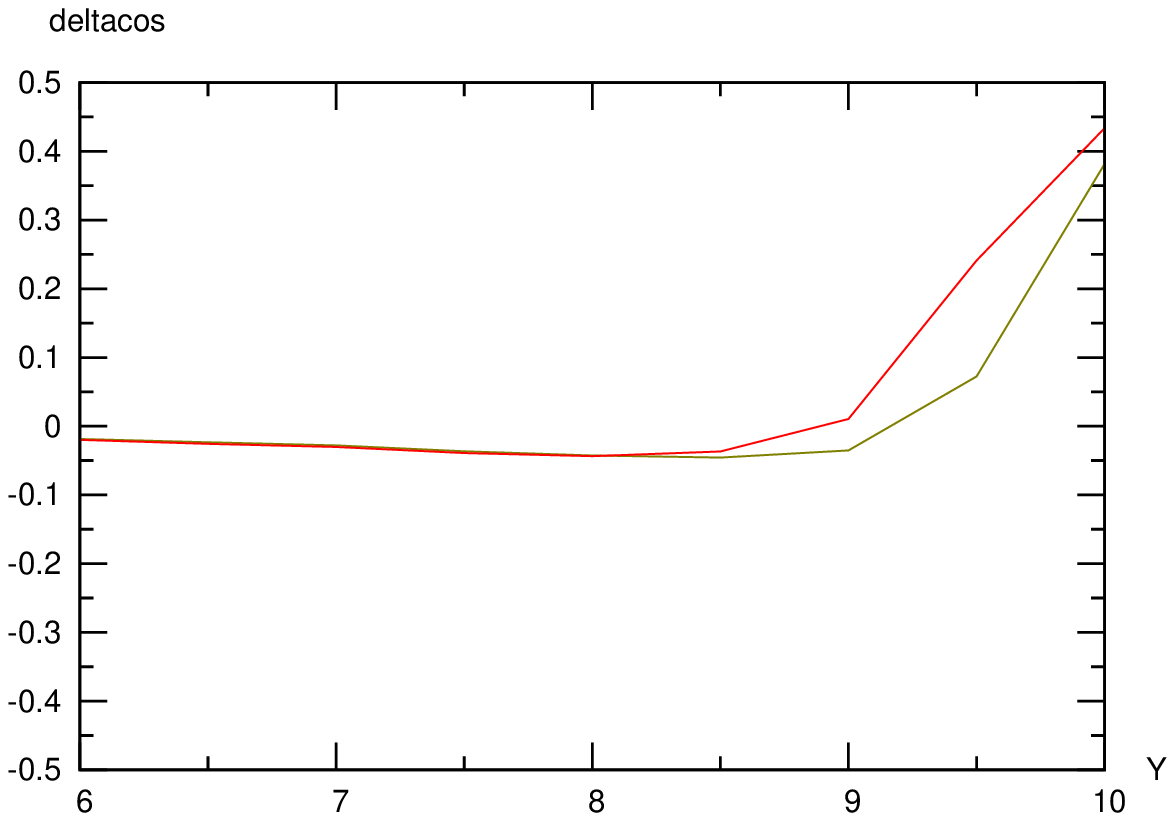}
  \end{minipage}
  \caption{Relative difference between an exact treatment of the cone size and the small cone approximation for $R=0.5$ and 
  $|\veckjone|=|\veckjtwo|=35$~GeV\,, in the two full NLL BFKL scenarios.}
  \label{Fig:R-0.5}
\end{figure}

\section{Conclusions}

In this paper, we have continued our NLL analysis of  Mueller-Navelet jets, at the partonic level, concentrating on the kinematical conditions of ongoing experiments ATLAS and CMS at LHC. We have made a detailed study of the azimuthal distributions for the first time at full NLL BFKL accuracy. Finally, we confronted 
our predictions with the fixed order predictions based on the \textsc{Dijet} code. Our analysis does not take into account hadronization effects. We did not estimate the importance of potentially  competing production mechanisms involving multiparton interactions. The evaluation of the importance of these effects is left for future analysis.

The predictions of the present study confirm the main result of ref.~\cite{Colferai:2010wu} that the effect of NLL corrections to jets vertices is dramatically large, similar in magnitude to the one due to the NLL Green's function corrections.

We have investigated the stability of our predictions with respect to changes of factorization scale $\mu_F$, of scale $s_0$  and of sets of PDFs.
For the cross-section, in comparison with 
scenarios using LL jet vertices,
the predictions are much more stable with respect to variation of $\mu_F$ and $s_0$, and of similar small order of magnitude for PDFs variations.
Our full NLL BFKL predictions are surprisingly
sizeably below the fixed order NLO prediction. 
	
For the decorrelation effect, the full NLL BFKL predictions and 
  fixed order NLO  one are very close for $\langle \cos \varphi\rangle$ and $\langle \cos 2 \varphi\rangle$.
  They are very flat in rapidity $Y$, but still rather dependent on  $s_0$, and specially on $\mu_F\,,$ while weakly dependent on PDFs.
	
We have taken into account the effect of 
collinearly improved NLL  BFKL Green's function. Including these effects for non zero conformal spins $n$  has an important impact on our predictions for azimuthal decorrelation. It 
leads to results which are very close to the pure NLL BFKL treatment. 

The angular $\varphi$ distribution which we predict at full NLL BFKL is very strongly peaked around $0$ and does not evolve strongly with respect to $Y\,.$ 
This prediction significantly differs from the ones based on LL jet vertices, and is stable when changing $\mu_F\,,$ $s_0$ and PDFs.

Finally, we have shown that
for the ratios $\langle \cos 2\varphi\rangle / \langle \cos \varphi\rangle$ and  $\langle \cos 3\varphi\rangle / \langle \cos 2 \varphi\rangle$ the differences between NLL  BFKL and fixed order NLO are sizeable, and stable with respect to scale choices.

To conclude, our analysis suggests that the ratios of harmonics are most suitable observables to distinguish between full NLL BFKL  predictions and fixed order NLO ones.

\section*{Acknowledgements}

\noindent

We acknowledge the  collaboration
with
Florian Schwennsen and Dimitri Colferai, which 
was the starting basis of the present study.

We are very grateful to 
 Michel Fontannaz, Cyrille Marquet and Christophe Royon for providing their codes.

We strongly thank Grzegorz Brona, David d'Enterria,
Hannes Jung, Victor Kim and  Maciej Misiura for many discussions and fruitful suggestions on the experimental aspects of this study.

S.~W. thanks the participants of MPI@TAU for discussions, and Halina Abramowicz for support. L.~S. and S.~W. thank the participants of the CMS small-$x$ and forward physics working group for discussions, and G.~Salam for support.

This work is partly supported by the French-Polish collaboration agreement
Polonium, the Polish Grant NCN
No. DEC-2011/01/B/ST2/03915, the P2IO Labex 
and the Joint Research Activity Study of Strongly 
Interacting Matter (acronym HadronPhysics3, Grant Agreement n.283286) under the Seventh
Framework
Programme of the European Community.


\providecommand{\href}[2]{#2}\begingroup\raggedright\endgroup

\end{document}